\documentclass[12pt]{JHEP3}
\usepackage[centertags]{amsmath}
\usepackage{amssymb}
\usepackage{epsfig}
\usepackage{amsmath}
\usepackage{graphicx}
\usepackage{afterpage}

\def\beq{\begin{equation}}
\def\eeq{\end{equation}}
\def\bea{\begin{eqnarray}}
\def\eea{\end{eqnaray}}
\def\nn{\nonumber}
\def\nl{\nonumber\\}
\def\sss{\scriptscriptstyle}

\def\roughly#1{\mathrel{\raise.3ex\hbox
{$#1$\kern-.75em\lower1ex\hbox{$\sim$}}}}

\def\gsim{\roughly>}
\def\lesssim{\mathrel{\hbox{\rlap{\hbox{\lower4pt\hbox{$\sim$}}}\hbox{$<$}}}}
\def\gtrsim{\mathrel{\hbox{\rlap{\hbox{\lower4pt\hbox{$\sim$}}}\hbox{$>$}}}}

\def\sla#1{\raise.15ex\hbox{$/$}\kern-.57em #1}

\def\ks{K_{\sss S}}

\def\Kbar{\bar K}
\def\bd{B_d^0}
\def\bs{B_s^0}
\def\bdbar{{\bar B}_d^0}
\def\bsbar{{\bar B}_s^0}

\def\btos{ b \to  s}
\def\bsmumu{ b \to  s \mu^+ \mu^-}

\def\AFB{A_{FB}}
\def\btopik{B \to \pi K}
\def \kstar{{\bar{K}^*}}
\def\Bsmumu{\bsbar \to \mu^+ \mu^-}
\def\Bsmumugamma{\bsbar \to \mu^+ \mu^- \gamma}
\def\BKmumu{\bdbar \to {\bar K} \mu^+ \mu^-}
\def\BKstarmumu{\bdbar\to \kstar \mu^+ \mu^-}
\def\BXsmumu{\bdbar \to X_s \mu^+ \mu^-}
\def\Bsdecay{\bsbar\to J/\psi \phi}

\def\fTfL{f_{\sss T}/f_{\sss L}}
\def\gev{{\rm GeV}}

\def\barr{\begin{eqnarray}}
\def\earr{\end{eqnarray}}
\def\beast{\begin{eqnarray*}}
\def\eeast{\end{eqnarray*}}

\def\be{\begin{equation}}
\def\ee{\end{equation}}
\def\bea{\begin{eqnarray}}
\def\eea{\end{eqnarray}}
\def\mc{\hat{{m}}_{{c}}}

\def\pan#1#2#3{{\it Phys.\ Atom.\ Nucl.} {\bf #1} (#2) #3}

\title{\boldmath New Physics in $\bsmumu$: \\
~~~CP-Conserving Observables}

\author{
Ashutosh Kumar Alok$^a$,
Alakabha Datta$^b$,
Amol Dighe$^c$,
Murugeswaran Duraisamy$^b$,
Diptimoy Ghosh$^c$ and
David London$^a$
\\
$^a$ Physique des Particules, Universit\'e de Montr\'eal,
\\ ~~C.P. 6128, succ. centre-ville, Montr\'eal, QC, Canada H3C 3J7 \\
$^b$ Department of Physics and Astronomy, 108 Lewis Hall,
\\ ~~University of Mississippi, Oxford, MS 38677-1848, USA \\
$^c$ Tata Institute of Fundamental Research, Homi Bhabha Road,
\\ ~~Mumbai 400005, India 
\\
E-mail:
\email{alok@lps.umontreal.ca},
\email{datta@phy.olemiss.edu},
\email{amol@theory.tifr.res.in},
\email{duraism@phy.olemiss.edu},
\email{diptimoyghosh@theory.tifr.res.in},
\email{london@lps.umontreal.ca}
}

\abstract{ We perform a comprehensive study of the impact of
  new-physics operators with different Lorentz structures on decays
  involving the $b \to s \mu^+ \mu^-$ transition.  We examine the
  effects of new vector-axial vector (VA), scalar-pseudoscalar (SP)
  ``` and tensor (T) interactions on the differential branching ratios
  and forward-backward asymmetries ($A_{FB}$'s) of $\Bsmumu$,
  $\BXsmumu$, $\Bsmumugamma$, $\BKmumu$, and $\BKstarmumu$, taking the
  new-physics couplings to be real.  In $\BKstarmumu$, we further
  explore the polarization fraction $f_L$, the angular asymmetry
  $A_T^{(2)}$, and the longitudinal-transverse asymmetry $A_{LT}$.  We
  identify the Lorentz structures that would significantly impact
  these observables, providing analytical arguments in terms of the
  contributions from the individual operators and their interference
  terms.  In particular, we show that while the new VA operators can
  significantly enhance most of the asymmetries beyond the Standard
  Model predictions, the SP and T operators can do this only for
  $A_{FB}$ in $\BKmumu$.  }

\keywords{$B$ Physics, Beyond Standard Model}

\preprint{UdeM-GPP-TH-10-191, UMISS-HEP-2010-01, TIFR/TH/10-23}

\begin{document}

\section{Introduction}

In recent years, there have been quite a few measurements of
quantities in $B$ decays which differ from the predictions of the
Standard Model (SM) by $\sim2\sigma$. For example, in $\btopik$, the
SM has some difficulty in accounting for all the experimental
measurements \cite{piKupdate}.  The measured indirect (mixing-induced)
CP asymmetry in some $\btos$ penguin decays is found not to be
identical to that in $\bd\to J/\psi\ks$ \cite{btos-1,btos-2,btos-3},
counter to the expectations of the SM.  While the SM predicts that the
indirect CP asymmetry in $\Bsdecay$ should be $\simeq 0$, the
measurement of this quantity by the CDF and D\O\ collaborations shows
a deviation from the SM \cite{cdf-d0-note}.  One naively expects the
ratio of transverse and longitudinal polarizations of the decay
products in $B\to\phi K^*$ to be $\fTfL \ll 1$, but it is observed
that $\fTfL \simeq 1$ \cite{phiK*-babar,phiK*-belle}.  It may be
possible to explain this value of $\fTfL$ within the SM, but this is
not certain.  Finally, the recent observation of the anomalous dimuon
charge asymmetry by the D\O\ collaboration \cite{D0-dimuon} also
points towards some new physics in $B_s$ mixing that affects the
lifetime difference and mixing phase involved therein (for example,
see Ref.~\cite{dgs-rc}).  Though none of the measurements above show a
strong enough deviation from the SM to claim positive evidence for new
physics (NP), they are intriguing since (i) the effects are seen in
several different $B$ decay channels, (ii) use a number of independent
observables, and (iii) all involve $b \to s$ transitions.

A further hint has recently been seen in the leptonic decay channel:
in the exclusive decay $\BKstarmumu$, the forward-backward asymmetry
($\AFB$) has been found to deviate somewhat from the predictions of
the SM
\cite{Belle-oldKstar,Belle-newKstar,BaBar-Kmumu,BaBar-Kstarmumu}.
This is interesting since it is a CP-conserving process, whereas most
of the other effects involve CP violation.  Motivated by this
tantalizing hint of NP in $\BKstarmumu$, we explore the
consequences of such NP in related decays.  We do not restrict
ourselves to any particular model, but work in the framework of
effective operators with different Lorentz structures.

If NP affects $\BKstarmumu$, it must be present in the decay
$\bsmumu$, and will affect the related decays $\Bsmumu$, $\BXsmumu$,
$\Bsmumugamma$, and $\BKmumu$.  The analyses of these decays in the
context of the SM as well as in some NP models have been performed in
the literature: $\Bsmumu$ \cite{Skiba:1992mg, Choudhury:1998ze,
  Huang:2000sm, Bobeth:2001sq, Huang:2002ni, Chankowski:2003wz,
  Alok:2005ep, blanke, Alok:2009wk, Buras:2010pi, Golowich:2011cx},
$\BXsmumu$
\cite{Ali:1991is,Buras:1994dj,Ali:1996bm,Huang:1998vb,Fukae:1998qy,bmu,Ali:2002jg,
  Huber:2007vv,Lee:2006gs,Ligeti:2007sn}, $\Bsmumugamma$
\cite{Eilam:1996vg,Aliev:1996ud,Geng:2000fs, Dincer:2001hu,
  Kruger:2002gf, Melikhov:2004mk, Melikhov:2004fs,
  Alok:2006gp,Balakireva:2009kn}, $\BKmumu$
\cite{Ali:2002jg,ali-00,Aliev:2001pq,Bensalem:2002ni, Bobeth:2007dw,
  Alok:2008wp,Charles,Beneke},
$\BKstarmumu$\cite{Kruger:2000zg,Beneke:2001at, Yan:2000dc,
  Aliev:2003fy,Aliev:2004hi,kruger-matias,Lunghi:2006hc,HHM,Egede:2008uy,
  Altmannshofer:2008dz,AFBNP,Soni:2010xh,hill1,Lunghi:2010tr,hill2,aoife}.
Correlations between some of these modes have been studied in
Refs.~\cite{Hiller:2003js,Alok:2008aa,Alok:2008hh}.

In this paper, we consider the addition of NP vector-axial vector
(VA), scalar-pseudoscalar (SP), and tensor (T) operators that
contribute to $\bsmumu$, and compute their effects on the above
decays.  Our aim here is not to obtain precise predictions, but rather
to obtain an understanding of how the NP affects the observables, and
to establish which Lorentz structure(s) can provide large deviations
from the SM predictions.  Some of these effects have already been
examined by some of us: for example, new VA and SP operators in
$\Bsmumu$ \cite{Alok:2005ep}, new VA and SP operators in
$\Bsmumugamma$ \cite{Alok:2006gp}, the correlation between $\Bsmumu$
and $\BKmumu$ with SP operators \cite{Alok:2008aa,Alok:2008hh}, large
forward-backward asymmetry in $\BKmumu$ from T operators
\cite{Alok:2008wp}, and the contribution of all Lorentz structures to
$\BKstarmumu$, with a possible explanation of the $\AFB$ anomaly
\cite{AFBNP}.  Here we perform a combined study of all of these decay
modes with all the Lorentz structures, consolidating and updating some
of the earlier conclusions, and adding many new results and insights.
Such a combined analysis, performed here for the first time, is
crucial for obtaining a consistent picture of the bounds on NP and the
possible effect of NP on the observables of interest.  While
observables like the differential branching ratio (DBR) and
$A_{FB}(q^2)$ by themselves are sensitive to NP, we also examine the
correlations between them in the context of NP Lorentz structures.

A full angular distribution of $\BKstarmumu$ allows us access to many
independent observables, and hence to multiple avenues for probing NP.
We present here for the first time the full angular distribution,
including all the NP Lorentz structures, for this decay mode.  This
leads to the identification of observables that could be significantly
influenced by specific Lorentz structures of NP.  In addition to the
DBR and $A_{FB}$, we also examine the longitudinal polarization
fraction $f_L$ and the angular asymmetry $A_T^{(2)}$, introduced
recently in Ref.~\cite{kruger-matias}.  We further analyze the
longitudinal-transverse asymmetry $A_{LT}$, which, as we will argue,
has very small hadronic uncertainties.

Hadronic uncertainties often are the main source of error in the
calculation of SM predictions of a quantity, and make the positive
identification of NP rather difficult.  In this paper, for $\BKmumu$
we use the form factors from light-cone sum rules. For $\BKstarmumu$,
we use the form factors obtained from QCD factorization at low $q^2$,
and those from light-cone sum rules at high $q^2$. The latest
next-to-leading order (NLO QCD) corrections \cite{Beylich:2011aq} have
not been included. These corrections would affect the central values
of the SM predictions to a small extent, while also decreasing the
renormalization-scale uncertainty. However, since our primary interest
is looking for observables for which the NP effects are large, a LO
analysis is sufficient at this stage.  In our figures, we display
bands for the SM predictions that include the form-factor
uncertainties as claimed by the respective authors.

In addition to the form-factor uncertainties, the SM prediction bands
also include the uncertainties due to quark masses,
Cabibbo-Kobayashi-Maskawa (CKM) matrix elements and meson decay
constants.  In our figures, these bands are overlaid with some
examples of the allowed values of these observables when NP
contributions are included.  This allows the scaling of these
uncertainties to be easily visualized.  It turns out that in many
cases, the results with the NP can be significantly different from
those without the NP, even taking into account inflated values for the
hadronic uncertainties.  We identify and emphasize such observables.
We also show that the hadronic uncertainties in several of these
observables are under control, especially when the invariant mass of
the muon pair is small and one can use the limit of large-energy
effective theory (LEET).  This makes such observables excellent probes
of new physics.  Also, since all the observables are shown as
functions of $q^2$, we have the information not just about the
magnitudes of the observables, but also about their shape as a
function of $q^2$, where some of the uncertainties are expected to
cancel out.

In this paper, we restrict ourselves to real values for all the NP
couplings, and study only the CP-conserving observables\footnote{The
  CP-violating observables, with complex values of the couplings, are
  treated in the companion paper \cite{CPviol}.}.  In
section~\ref{bsmumuops}, we examine the various SM and NP $\bsmumu$
operators, and give the current constraints on the NP couplings.  The
effects of the NP operators on the observables of the decays are
discussed in the following sections: $\Bsmumu$ (Sec.~\ref{Bsmumu}),
$\BXsmumu$ (Sec.~\ref{BXsmumu}), $\Bsmumugamma$
(Sec.~\ref{Bsmumugamma}), $\BKmumu$ (Sec.~\ref{BKmumu}), and
$\BKstarmumu$ (Sec.~\ref{BKstarmumu}).  Our notation in these sections
clearly distinguishes the contributions from VA, SP and T operators
and their interference terms, which offers many insights into their
impact on modifying the observables.  We give the details of the
calculations involved in sections \ref{BXsmumu}--\ref{BKstarmumu} in
the appendices \ref{app-bxsmumu}-\ref{app-bkstarmumu}, respectively,
for the sake of completeness and in order to have a clear consistent
notation for this combined analysis.  In Sec.~\ref{summary}, we
summarize our findings and discuss their implications.  In particular,
we point out the measurements which will allow one to distinguish
among the different classes of NP operators, and thus clearly identify
which type of new physics is present.

\section{\boldmath $\bsmumu$ Operators
\label{bsmumuops}}

\subsection{Standard Model and New Physics: effective Hamiltonians}

Within the SM, the effective Hamiltonian for the quark-level
transition $\bsmumu$ is
\bea
{\cal H}_{\rm eff}^{SM} &=& -\frac{4 G_F}{\sqrt{2}}
\, V_{ts}^* V_{tb} \, \Bigl\{ \sum_{i=1}^{6} {C}_i (\mu) {\cal O}_i (\mu)
+ C_7 \,\frac{e}{16 \pi^2}\, [\bar{s}
  \sigma_{\mu\nu} (m_s P_L + m_b P_R) b] \,
F^{\mu \nu} \nn \\
&& +\, C_9 \,\frac{\alpha_{em}}{4 \pi}\, (\bar{s}
\gamma^\mu P_L b) \, \bar{\mu} \gamma_\mu \mu + C_{10}
\,\frac{\alpha_{em}}{4 \pi}\, (\bar{s} \gamma^\mu P_L b) \, \bar{\mu}
\gamma_\mu \gamma_5
\mu  \, \Bigr\} ~,
\label{HSM}
\eea
where $P_{L,R} = (1 \mp \gamma_5)/2$. The operators ${\cal O}_i$
($i=1,..6$) correspond to the $P_i$ in Ref.~\cite{bmu}, and $m_b =
m_b(\mu)$ is the running $b$-quark mass in the $\overline{\rm MS}$
scheme.  We use the SM Wilson coefficients as given in
Ref.~\cite{Altmannshofer:2008dz}.  In the magnetic dipole operator
with the coefficient $C_7$, we neglect the term proportional to $m_s$.

The operators $O_i$, $i=1$-6, can contribute indirectly to $\bsmumu$
and their effects can be included in an effective Wilson coeficient as
\cite{Altmannshofer:2008dz}
\begin{eqnarray}
\label{effecWC1}
C^{\rm eff}_9 &\!=\!& C_9(m_b) + h(z,\hat{m_c}) \left(\frac{4}{3} C_1 + C_2 + 6\, C_3 + 60\, 
C_5 \right) \nn\\
&&-~\frac{1}{2} h(z,\hat{m_b}) \left(7 C_3 + \frac{4}{3} C_4 + \, 76 
C_5 + 
\frac{64}{3} C_6 \right) \\
&&-~\frac{1}{2} h(z,0) \left( C_3 + \frac{4}{3} C_4 + 16\, C_5 + 
\frac{64}{3} C_6 
\right) + \frac{4}{3} C_3 + \frac{64}{9} C_5 + \frac{64}{27} C_6 ~. \nn
\end{eqnarray}
Here $z \equiv q^2/m_b^2$, and $\hat{m}_q \equiv m_q/m_b$ for all
quarks $q$.  The function $h(z,\hat m)$ represents the one-loop
correction to the four-quark operators $O_1$-$O_6$ and is given by
\cite{Buras:1994dj,Altmannshofer:2008dz}
 \begin{eqnarray}
 \label{effecWC}
 h(z,\hat m) &  = & -\frac{8}{9}\ln\frac{m_b}{\mu_b} - \frac{8}{9}\ln \hat m +
 \frac{8}{27} + \frac{4}{9} x \\
 & & - \frac{2}{9} (2+x) |1-x|^{1/2} 
 \left\{\begin{array}{ll}
 \left( \ln\left| \frac{\sqrt{1-x} + 1}{\sqrt{1-x} - 1}\right| - i\pi \right), &
 \mbox{for } x \leq 1 ~, \nonumber \\
 2 \arctan \frac{1}{\sqrt{x-1}}, & \mbox{for } x > 1 ~,
 \end{array}
 \right.\ 
 \end{eqnarray}
where $x \equiv {4\hat m^2}/{z}$.  In the numerical analysis, the
renormalization scale $\mu_b$ is varied between $m_b/2$ and
$2m_b$. Note that in the high-$q^2$ region one can perform an operator
product expansion (OPE) in $ 1/Q$ with $Q=(m_b \sqrt{q^2})$
\cite{grin1, grin2}. Numerically the results of Refs.~\cite{grin1,
  grin2} differ little from those in Eq.~(\ref{effecWC1}) and so we
use the above expression for the entire range of $q^2$.  An analysis
of $\bsmumu$ where the OPE in the high-$q^2$ region is used can be
found in Refs.~\cite{hill1, hill2}.

We now add new physics to the effective Hamiltonian for $\bsmumu$,
so that it becomes
\beq
{\cal H}_{\rm eff}(\bsmumu) = {\cal
H}_{\rm eff}^{SM} + {\cal H}_{\rm eff}^{VA} + {\cal H}_{\rm eff}^{SP} +
{\cal H}_{\rm eff}^{T} ~,
\label{NP:effHam}
\eeq
where ${\cal H}_{\rm eff}^{SM}$ is given by Eq.~(\ref{HSM}), while
\bea
{\cal H}_{\rm eff}^{VA} &=& - \frac{4 G_F}{\sqrt{2}} \,
\frac{\alpha_{em}}{4\pi} \, V_{ts}^* V_{tb} \,
\Bigl\{ R_V \, (\bar{s} \gamma^\mu P_L b)
\, \bar{\mu} \gamma_\mu \mu + R_A \, (\bar{s} \gamma^\mu P_L b)
\, \bar{\mu} \gamma_\mu \gamma_5 \mu \nn \\
&& \hskip3.0 truecm +~R'_V \, (\bar{s} \gamma^\mu P_R b) \,
\bar{\mu} \gamma_\mu \mu + R'_A \, (\bar{s} \gamma^\mu P_R b)
\, \bar{\mu} \gamma_\mu \gamma_5\mu \Bigr\} ~, \\
{\cal H}_{\rm eff}^{SP} &=& - \frac{4G_F}{\sqrt{2}} \,
\frac{\alpha_{em}}{4\pi}\, V_{ts}^* V_{tb} \,
\Bigl\{R_S ~(\bar{s} P_R b) ~\bar{\mu}\mu +
R_P ~(\bar{s} P_R b) ~ \bar{\mu}\gamma_5 \mu \nn\\
&& \hskip3.0 truecm +~R'_S ~(\bar{s} P_L b) ~\bar{\mu}\mu +
R'_P ~(\bar{s} P_L b) ~ \bar{\mu}\gamma_5 \mu \Bigr\} \;, \\
{\cal H}_{\rm eff}^{T} &=& -\frac{4 G_F}{\sqrt{2}} \,
\frac{\alpha_{em}}{4\pi}\, V_{ts}^* V_{tb} \,
\Bigl\{C_T (\bar{s} \sigma_{\mu \nu } b)
\bar{\mu} \sigma^{\mu\nu}\mu + i C_{TE} (\bar{s} \sigma_{\mu \nu } b)
\bar{\mu} \sigma_{\alpha \beta } \mu ~\epsilon^{\mu
\nu \alpha \beta} \Bigr\}
\eea
are the new contributions.  Here, $R_V, R_{A}, R_V', R_A', R_S, R_P,
R_S', R_P', C_{T}$ and $C_{TE}$ are the NP effective couplings.  We do
not consider NP in the form of the $O_7 =\bar s\sigma^{\alpha\beta}
P_R b \, F_{\alpha\beta}$ operator or its chirally-flipped counterpart
$O_7^\prime= \bar s\sigma^{\alpha\beta} P_L b \,F_{\alpha\beta}$.
This is because there has been no hint of NP in the radiative decays
${\bar B} \to X_s \gamma, {\bar K}^{(*)} \gamma$ \cite{ali-00}, which
imposes strong constraints on $|C_7^{\rm eff}|$.  This by itself does
not rule out the possibility of a flipped-sign $C_7^{\rm eff}$
scenario.  However this solution can be ruled out at 3$\sigma$ from
the decay rate of ${\bar B}\to X_s\ell^+\ell^-$ if there are no NP
effects in $C_9$ and $C_{10}$~\cite{GHM}.  Thus, NP effects
exclusively in $C_7$ cannot provide large deviations from the SM.  The
impact of $O_7^\prime$ on the forward-backward asymmetry in
$\BKstarmumu$, together with other observables, was studied in
Ref.~\cite{Egede:2008uy}.

Note that the operators with coefficients $R_V$ and $R_A$ have the same
Lorentz structure as those in the SM involving $C_9$ and $C_{10}$,
respectively [see Eq.~(\ref{HSM})], so that any measurement will be
sensitive only to the combinations $(C_9+R_V)$ or $(C_{10}+R_A)$.  For
simplicity, in our numerical analysis of the observables of various
decays, these couplings are taken to be real. As a consequence, 
the results in this paper would be the same if the
corresponding CP-conjugate decays were considered.
However, for completeness, the expressions allow for a 
complex-coupling analysis.

When calculating the transition amplitudes, for the leptonic part
we use the notation
\beq
\begin{tabular}{ll}
$L^\mu  \equiv  \langle \mu^+(p_+) \mu^-(p_-) |
\bar\mu \gamma^\mu \mu | 0 \rangle$ , &
$L^{\mu 5}  \equiv  \langle \mu^+(p_+) \mu^-(p_-) |
\bar\mu \gamma^\mu \gamma^5 \mu | 0 \rangle$ , \\
$L  \equiv  \langle \mu^+(p_+) \mu^-(p_-) |
\bar\mu \mu | 0 \rangle$ ,  &
$L^{5}  \equiv  \langle \mu^+(p_+) \mu^-(p_-) |
\bar\mu \gamma^5 \mu | 0 \rangle$ ,  \\
$L^{\mu\nu}  \equiv  \langle \mu^+(p_+) \mu^-(p_-) |
\bar\mu \sigma^{\mu\nu} \mu | 0 \rangle$ . & \\
\end{tabular}
\label{Ldefs}
\eeq

\subsection{Constraints on NP couplings}
\label{constraints}

The constraints on the NP couplings in $\bsmumu$ come mainly from the
upper bound on the branching ratio $B(\Bsmumu)$ and the measurements
of the total branching ratios $B(\BXsmumu)$ and $B(\BKmumu)$
\cite{:2007kv,pdg,Barberio:2008fa,Aubert:2004it,Iwasaki:2005sy}:
\barr
B(\Bsmumu) & < & 3.60 \times 10^{-8} \quad \mbox{(90\% C.L.)} \; , \\
B(\BXsmumu) & = & \left\{ \begin{array}{ll}
\left( 1.60 \pm 0.50 \right) \times 10^{-6} & (\mbox{low } q^2)  \\
\left( 0.44 \pm 0.12 \right) \times 10^{-6} & (\mbox{high } q^2)  \\
\end{array} \right. \; , \\
B(\BKmumu) & = & \left(4.5^{+1.2}_{-1.0} \right) \times 10^{-7} \; ,
\earr
where the low-$q^2$ and high-$q^2$ regions correspond to 1 GeV$^2 \le
q^2 \le 6$ GeV$^2$ and $q^2 \ge 14.4$ GeV$^2$, respectively, where
$q^2$ is the invariant mass squared of the two muons.  The constraints
from the first two quantities above have been derived in
Ref.~\cite{AFBNP}.  Here we also include the additional constraints
from $B(\BKmumu)$.  The three decays above provide complementary
information about the NP operators.  For the SM predictions here, we
use the latest NNLO calculations.  Note that the measurements for
$B(\BKstarmumu)$ are also available
\cite{Belle-newKstar,CDF-Kstar}. However, the form-factor
uncertainties in $\BKstarmumu$ are rather large, and as a result the
constraints due to this decay mode are subsumed in those from the
other three modes.

\FIGURE[t]{
\includegraphics[width=0.4\linewidth]{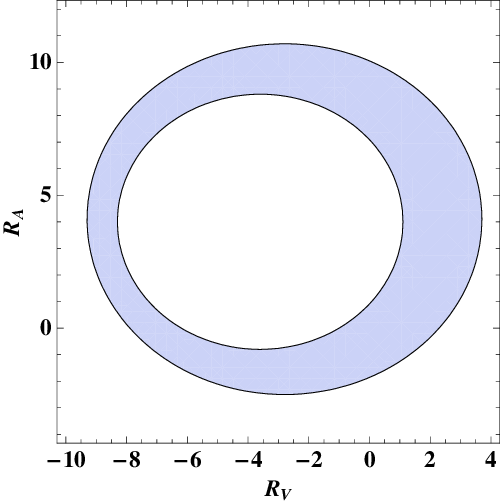}
~~~~~\includegraphics[width=0.4\linewidth]{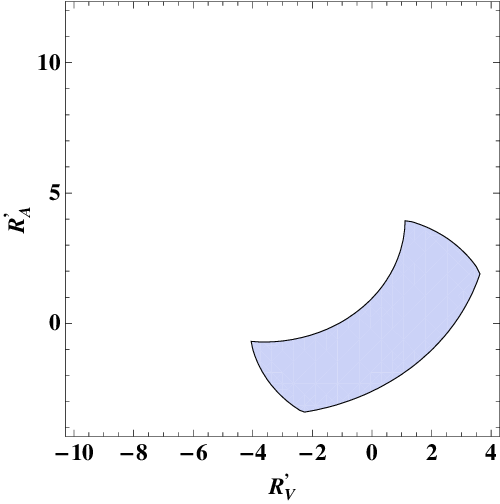}
\caption{The constraints on the couplings $R_V,R_A$ (left panel)
and $R'_V,R'_A$ (right panel) when only primed or unprimed
couplings are present.
\label{VA-constraints}}
}

The constraints on the new VA couplings come mainly from $B(\BXsmumu)$
and $B(\BKmumu)$.  Their precise values depend on which NP operators
are assumed to be present. For example, if only $R_{V,A}$ or only
$R'_{V,A}$ couplings are present, the constraints on these couplings
take the form shown in Fig.~\ref{VA-constraints}.
For $R_{V,A}$, the allowed parameter space is the region between
two ellipses:
\beq
1.0 \lesssim \frac{|R_V + 3.6|^2}{(4.7)^2} + \frac{|R_A -4.0|^2}{(4.8)^2} \; ,
\quad \frac{|R_V + 2.8|^2}{(6.5)^2} + \frac{|R_A -4.1|^2}{(6.6)^2} \lesssim 1 ~,
\eeq
while for $R'_{V,A}$, the allowed region is the intersection of
an annulus and a circle:
\beq
22.2 \lesssim
|R'_V + 3.6|^2 + |R'_A - 4.0|^2 \lesssim 56.6\; , \quad |R'_V|^2 + |R'_A|^2 \lesssim 17 ~.
\eeq
If both $R_{V,A}$ and $R'_{V,A}$ are present, the constraints on them get
individually weakened to
\beq
\frac{|R_V + 2.8|^2}{(6.5)^2} + \frac{|R_A -4.1|^2}{(6.6)^2} \lesssim 1 ~,
\label{RVAconstraints}
\eeq
and
\beq
|R'_V|^2 + |R'_A|^2 \lesssim 40 ~,
\eeq
respectively\footnote{Note: the constraints on $R_{V,A}$ obtained here
  are milder than those obtained in Ref.~\cite{Alok:2006gp} using
  $B(\bdbar \to ({\bar K}\,,{\bar K}^*)\, \mu^+ \, \mu^-)$.  This is
  because Ref.~\cite{Alok:2006gp} had neglected the interference terms
  between the SM and new physics VA operators. Their inclusion relaxes
  the stringent constraints therein.}.

For the SP operators, the present upper bound on $B(\Bsmumu)$ provides
the limit
\begin{equation}
|R_S - R'_S|^2 + |R_P - R'_P|^2 \lesssim 0.44 ~,
\label{SP-constraints}
\end{equation}
where we have used $f_{B_s}=(238.8 \pm 9.5)\,{\rm MeV}$
\cite{Laiho:2009eu} and $|V_{ts}^* V_{tb}|=0.0407\pm 0.0010$
\cite{pdg}.  This constitutes a severe constraint on the NP couplings
if only $R_{S,P}$ or $R'_{S,P}$ are present. However, if both types of
operators are present, these bounds can be evaded due to cancellations
between the $R_{S,P}$ and $R'_{S,P}$.  In that case, $B(\BXsmumu)$ and
$B(\BKmumu)$ can still bound these couplings. The stronger bound is
obtained from the measurement of the latter quantity, which yields
\beq
|R_S|^2 + |R_P|^2 \lesssim 9 \; ,
\quad R_S \approx R'_S \; , \quad R_P \approx R'_P \; .
\eeq

Finally, the constraints on the NP tensor operators come entirely from
$B(\BXsmumu)$.  When only the T operators are present,
\begin{equation}
|C_T|^2 +4 |C_{TE}|^2 \lesssim 1.0 ~.
\end{equation}
 
Although the bounds presented in this section for VA, SP and T
couplings are obtained by taking one kind of Lorentz structure at a
time, in our numerical analysis for scenarious where we consider
combinations of two or more kinds of Lorentz structres, we use the
allowed parameter space obtained by considering the corresponding
combined Lorentz structures.

We now analyze the $\bsmumu$ modes in detail and present our results.
As explained in the Introduction, the figures have the SM prediction
bands overlaid with the predictions for specific allowed values of NP
couplings.  The SM band is generated by varying the form factors
within their ranges as predicted by the respective authors, while the
CKM matrix elements, quark masses and meson decay constants are varied
within their $1.6 \sigma$ allowed values.

\section{\boldmath $\Bsmumu$
\label{Bsmumu}}

In this section we examine the NP contributions to $\Bsmumu$.  Within
the SM, $\Bsmumu$ is chirally suppressed.  The SM prediction for the
branching ratio is $B(\Bsmumu) = (3.35\pm 0.32)\times 10^{-9}$
\cite{blanke}.  The Tevatron gives an upper bound on its branching
ratio (BR) of $3.6 \times 10^{-8}$ at 90\%
C.L. \cite{:2007kv,pdg,Barberio:2008fa}.  This decay can be observed
at the Tevatron only if NP enhances its BR above $10^{-8}$. LHCb is
the only experiment which will probe $B(\Bsmumu)$ down to its SM
value. It has the potential for a $3 \sigma$ observation ($5 \sigma$
discovery) of $\Bsmumu$ with $\sim 2\, {\rm fb}^{-1}$ ($\sim 6\, {\rm
  fb}^{-1}$) of data \cite{Lenzi:2007nq}.  LHCb therefore has the
potential to observe either an enhancement or a suppression of
$B(\Bsmumu)$.  It can observe $\Bsmumu$ as long as its BR is above
$1.0 \times 10^{-9}$.

\subsection{Branching ratio}

The transition amplitude for $\bsbar \to \mu^+ \mu^-$ is given by
\barr
i{\cal M}(\bsbar \to \mu^+ \mu^-)  & = &
(-i)\frac{1}{2}\Bigg[-\frac{4 G_F}{\sqrt{2}} \frac{\alpha_{em}}{4 \pi}
(V_{ts}^* V_{tb})\Bigg] \times \nn \\
 && \hskip-1truecm \Bigg\{
\left< 0 \left|\bar{s}\gamma_{\mu}\gamma_{5}b\right|\bsbar(p)\right>
(-C_{10}^{\rm eff} - R_A + R'_A) L^{5\mu} \nn \\
 & + & \left< 0 \left|\bar{s} \gamma_5 b\right|\bsbar(p)\right>
\left[ (R_S - R'_S) L + (R_P - R'_P) L^5  \right] \Bigg\} \; ,
\earr
where $L^{5\mu}$, $L$ and $L^5$ are defined in
Eq.~(\ref{Ldefs}). Using the matrix elements \cite{Skiba:1992mg}
\beq
\left< 0 \left|\bar{s}\gamma_{\mu}\gamma_{5}b\right|\bsbar(p)\right> =
i\,p_\mu\,f_{B_s}\;, \quad
\left< 0 \left|\bar{s}\gamma_{5}b\right|\bsbar(p)\right> = -
i\,f_{B_s}\frac{m_{B_s}^2}{m_b + m_s}\;,
\eeq
the calculation of the BR gives
\begin{eqnarray}
B({\bsbar} \to \mu^+ \, \mu^-) & =& \frac{G^2_F \alpha_{em}^2
m^5_{B_s} f_{B_s}^2 \tau_{B_s}}{64 \pi^3}
|V_{tb}^{}V_{ts}^{\ast}|^2 \sqrt{1 - \frac{4 m_\mu^2}{m_{B_s}^2}}\times
\nn\\
&& \hskip-2truecm  \Bigg\{
\Bigg(1 - \frac{4m_\mu^2}{m_{B_s}^2} \Bigg) \Bigg|
\frac{R_S - R'_S}{m_b + m_s}\Bigg|^2
+ \Bigg|\frac{R_P - R'_P}{m_b + m_s} + \frac{2 m_\mu}{m^2_{B_s}} (C_{10}+R_A-R'_A)\Bigg|^2 \Bigg\}. \phantom{space}
\label{bmumu-BR}
\end{eqnarray}
Clearly, NP in the form of tensor operators does not contribute to
$\Bsmumu$.  From Eq.~(\ref{bmumu-BR}) and the constraints on NP
couplings obtained in Sec.~\ref{constraints}, one can study the effect
of new VA and SP couplings.

Since the NP contribution from VA operators is suppressed by a factor
of $\sim m_\mu/m_b$ compared to that from the SP operators, the effect
of SP operators dominates.  Both VA and SP operators can suppress
$B(\Bsmumu)$ significantly below the SM prediction.  However while VA
operators can only marginally enhance $B(\Bsmumu)$ above $10^{-8}$,
making the decay accessible at the Tevatron in an optimistic scenario,
the SP operators can enhance the branching ratio even up to the
present experimental bound.  Indeed, the strongest limit on the SP
couplings comes from this decay.  This strong limit prevents the SP
operators from expressing themselves in many other observables, as we
shall see later in this paper.

\subsection{Muon polarization asymmetry}
\label{ALP}

The longitudinal polarization asymmetry of muons in $\Bsmumu$ is defined as
\beq
 A_{LP} = \frac{N_R-N_L}{N_R+N_L}\;,
\eeq
where $N_R\,(N_L)$ is the number of $\mu^{-}$'s emerging with positive
(negative) helicity. $A_{LP}$ is a clean observable that is not
suppressed by $m_\mu/m_{B_s}$
only if the NP contribution is in the form of SP operators, such as in
an extended Higgs sector.

$A_{LP}$ for the most general NP is \cite{Alok:2008hh}
\beq
A_{LP} = \frac{2\sqrt{1 - \frac{4m_\mu^2}{m_{B_s}^2}}\;{\rm Re}\Bigg[\Big(
\frac{R_S - R'_S}{m_b + m_s}\Big)\Big(\frac{R_P - R'_P}{m_b + m_s} + \frac{2 m_\mu}{m^2_{B_s}} (C_{10}+R_A-R'_A)\Big)\Bigg]}
{\Big(1 - \frac{4m_\mu^2}{m_{B_s}^2} \Big) \Bigg|
\frac{R_S - R'_S}{m_b + m_s}\Bigg|^2
+ \Bigg|\frac{R_P - R'_P}{m_b + m_s} + \frac{2 m_\mu}{m^2_{B_s}} (C_{10}+R_A-R'_A)\Bigg|^2} ~.
\eeq
From the above equation, we see that $A_{LP}$ can be nonzero if and
only if $R_S-R'_S\neq 0$, i.e.\ there must be a contribution from NP
SP operators. (Within the SM, SP couplings are negligibly small, so
that $A_{LP} \simeq 0$.)

The present upper bound on $B(\Bsmumu)$ puts no constraint on
$A_{LP}$, and it can be as large as $100\%$ \cite{Alok:2008hh}.
$A_{LP}$ can be maximal even if $B(\Bsmumu)$ is close to its SM
prediction. Therefore, in principle $A_{LP}$ can serve as an important
tool to probe NP of the SP form.  However, in order to measure its
polarization, the muon must decay within the detector. This is not
possible due to the long muon lifetime ($c\tau$ for the muon is 659
m). Hence in practice, this quantity is not measurable at current
detectors.

\section{\boldmath $\BXsmumu$
\label{BXsmumu}}

The BR of $\BXsmumu$ in the low-${q}^2$ and high-${q}^2$ regions has
been measured to be \cite{Aubert:2004it,Iwasaki:2005sy}
\bea
{B} ( {\bar {\text{B}}} \to {X}_{s} \ell^+ \ell^-)_{{\rm low}~{q}^2} ~=~
\left\{ \begin{array}{ll}
\left( 1.49 \pm 0.50^{+0.41}_{-0.32} \right) \times
10^{-6}~, & (\rm Belle)~, \\
\left( 1.8 \pm 0.7 \pm 0.5 \right) \times 10^{-6} ~, & (\rm
BaBar)~, \\
\left( 1.60 \pm 0.50 \right) \times 10^{-6}~, & (\rm Average)
~. \\
\end{array} \right. \\
{B} ( {\bar B} \to X_s \ell^+ \ell^-)_{{\rm high}~{q}^2} ~=~
\left\{ \begin{array}{ll}
\left( 0.42 \pm 0.12^{+0.06}_{-0.07} \right) \times
10^{-6} ~, & (\rm Belle) ~, \\
\left( 0.50 \pm 0.25 ^{+0.08}_{-0.07} \right) \times 10^{-6}
~, & (\rm BaBar) ~, \\
\left( 0.44 \pm 0.12 \right) \times 10^{-6} ~,
& (\rm Average) ~. \\
\end{array} \right.
\eea
The SM predictions for ${B} ( {\bar {{B}}} \to {X}_{s} \, \mu^+ \,
\mu^-)$ in the low-${q}^2$ and high-${q}^2$ regions are $(1.59\pm0.11)
\times 10^{-6}$ and $(0.24 \pm 0.07) \times 10^{-6}$, respectively
\cite{Huber:2007vv}.

Apart from the measurement of the total BR of $\BXsmumu$, which has
already been used to restrict the VA and T operators in
Sec.~\ref{constraints}, the differential branching ratio (DBR) as a
function of $q^2$ also contains valuable information that can help us
detect NP.  In particular, the SM predicts a positive zero crossing
for $\AFB$ in $\BXsmumu$ in the low-$q^2$ region, i.e.\ for $q^2$ less
than (greater than) the crossing point, the value of $\AFB$ is
negative (positive).  This zero crossing is sufficiently away from the
charm resonances so that its value can be determined perturbatively to
an accuracy of $\sim 5\%$. The NNLO prediction \cite{Huber:2007vv} for
the zero of $A_{FB}(q^2)$ is (taking $m_b = 4.8$ GeV)
\begin{equation}
(q^2)_0= (3.5 \pm 0.12)\, {\rm GeV}^2 \,.
\end{equation}
This quantity has not yet been measured. However, estimates show that
a precision of about $5\%$ could be obtained at a Super-$B$ factory
\cite{Browder:2008em}.  A deviation from the zero crossing point
predicted above will be a clear signal of NP.

\subsection{Differential branching ratio and forward-backward asymmetry}

After including all the NP interactions, and neglecting terms
suppressed by $m_\mu/m_b$ and $m_s/m_b$, the total differential
branching ratio $\text{d}{B}/{\text{d}z}$ is given by
\bea
\Bigg(\frac{\text{d}{B}}{\text{d}z}\Bigg)_{\text{Total}}~=~
\Bigg(\frac{\text{d}{B}}{\text{d}z}\Bigg)_{\text{SM}} + {B}_0
\Bigg[{B}_{SM{\hbox{-}}VA} +  {B}_{VA} +  {B}_{SP} + {B}_{T}\Bigg] \; ,
\eea
where the quantities $B$ depend on the SM and NP couplings and
kinematic variables.  The complete expressions for these quantities
are given in Appendix~\ref{app-bxsmumu}.  The subscripts denote the
Lorentz structure(s) contributing to that term.

The forward-backward asymmetry in $\BXsmumu$ is
\bea
\label{AFB-Xsmumu-1}
A_{FB}(q^2) &=&\frac{\int^1_0 d\cos{\theta_\mu}
\frac{d^2B}{dq^2d\cos{\theta_\mu}  }-\int^0_{-1}
d\cos{\theta_\mu}\frac{d^2B}{dq^2d\cos{\theta_\mu}  }}
{\int^1_0 d\cos{\theta_\mu} \frac{d^2B}{dq^2d\cos{\theta_\mu}  }
+\int^0_{-1} d\cos{\theta_\mu}\frac{d^2B}{dq^2d\cos{\theta_\mu}  }} \; ,
\eea
where $\theta_\mu$ is the angle between the $\mu^+$ and the
$\bar{B}^0$ in the dimuon center-of-mass frame.  We can write $\AFB$
in the form
\begin{equation}
A_{FB}(q^2)=\frac{N(z)}{\text{d} {B}/ \text{d}z}\;,
\end{equation}
where the numerator is given by
\bea
N(z) &~=~& B_0
\Bigg[N_{SM} + N_{SM{\hbox{-}}VA} +  N_{VA} + N_{SP{\hbox{-}}T}
\Bigg] \;.
\label{N-Xsmumu-1}
\eea
The terms suppressed by $m_\mu/m_b$ and $m_s/m_b$ have been neglected
as before.  Again for the detailed expressions, we refer the reader to
Appendix~\ref{app-bxsmumu}.

\FIGURE[t]{
\includegraphics[width=0.4\linewidth]{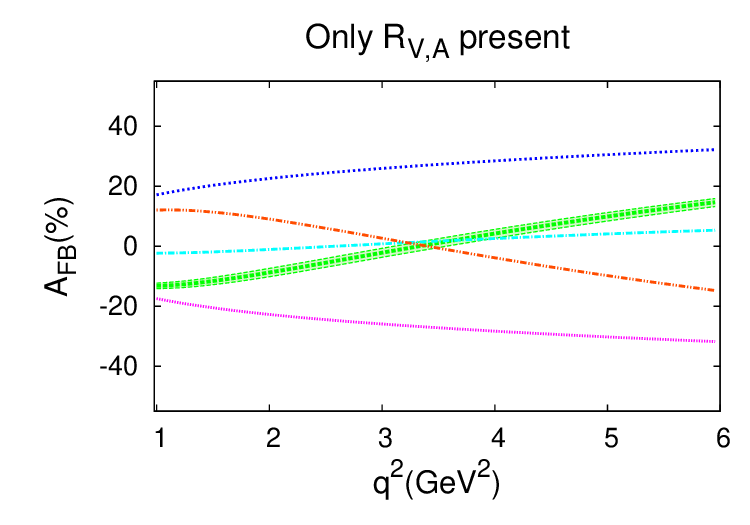}
\includegraphics[width=0.4\linewidth]{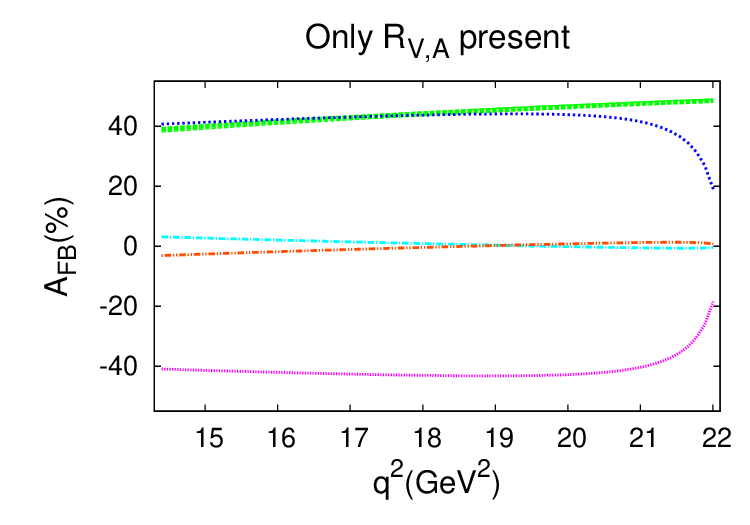} \\
  \includegraphics[width=0.4\linewidth]{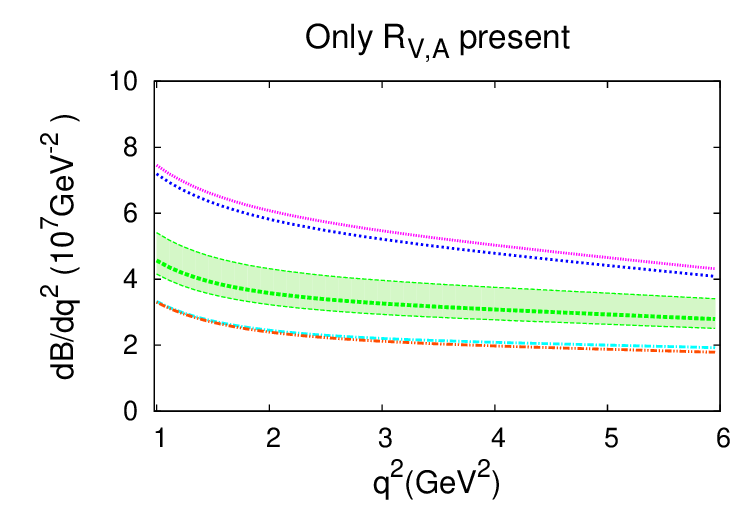}
 \includegraphics[width=0.4\linewidth]{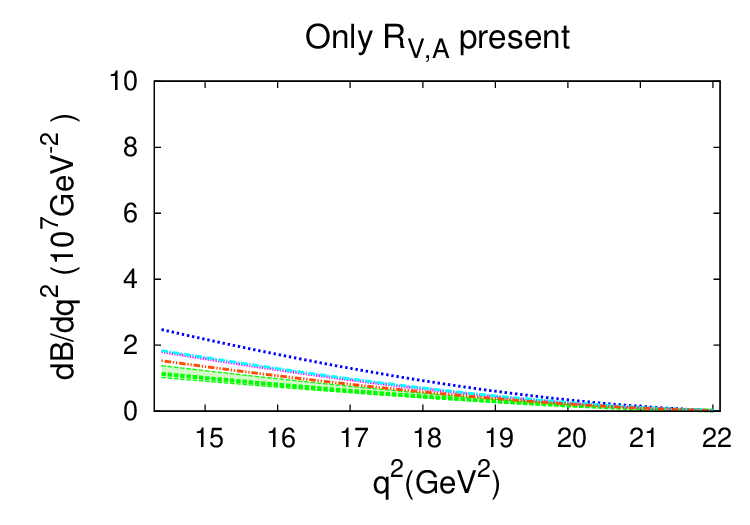}
\caption{The left (right) panels of the figure show $\AFB$ and DBR for
  $\BXsmumu$ in the low-$q^2$ (high-$q^2$) region, in the scenario
  where only $(R_V, R_A)$ terms are present.  The band corresponds to
  the SM prediction and its uncertainties; the lines show predictions
  for some representative values of NP parameters $(R_V, R_A)$. For
  example, the blue curves in the low-$q^2$ and high-$q^2$ regions
  correspond to $(-6.85,8.64)$ and $(-9.34,8.85)$, respectively.
\label{fig:incl-va}}
}

Fig.~\ref{fig:incl-va} shows $A_{FB}(q^2)$ and the DBR for $\BXsmumu$
in the presence of NP in the form of $R_{V,A}$ couplings, which are
the ones that can most influence these observables.  Enhancement or
suppression of the DBR by a factor of 2 is possible.  The NP couplings
can enhance $\AFB$ up to 30\% at low $q^2$, make it have either sign,
and even make the zero crossing disappear altogether.  At high $q^2$,
however, $A_{FB}$ can only be suppressed.  The $R'_{V,A}$ couplings
can only affect these observables mildly: a 50\% enhancement in DBR is
possible (no suppression), but $A_{FB}$ can only be marginally
enhanced and a positive zero crossing in the $q^2= 2$-4 GeV$^2$ region
is maintained.  The mild effect of $R'_{V,A}$ couplings as compared to
the $R_{V,A}$ couplings is a generic feature for almost all
observables. This may be attributed to the bounds on the magnitudes of
these couplings: from Sec.~\ref{constraints}, while $|R_{V,A}|<10$,
the values of $|R'_{V,A}| <5$.

Eq.~(\ref{N-Xsmumu-1}) shows that if SP or T couplings are
individually present, their contribution to $A_{FB}$ is either absent
or suppressed by $m_\mu/m_b$. In such a case, though they can enhance
the DBR (marginally for SP, by up to a factor of 2 for T), $A_{FB}$ is
suppressed in general (marginally for SP, significantly for T).
However if both SP and T operators are present, their interference
term is not suppressed and some enhancement of $A_{FB}$ is possible.
This still is not significant, since the magnitude of the SP couplings
is highly constrained from $\Bsmumu$ measurements.  A positive zero
crossing in the low-$q^2$ region is always maintained.  This may be
seen in Fig.~\ref{fig:incl-spsppt}.

\FIGURE[t]{
\includegraphics[width=0.4\linewidth]{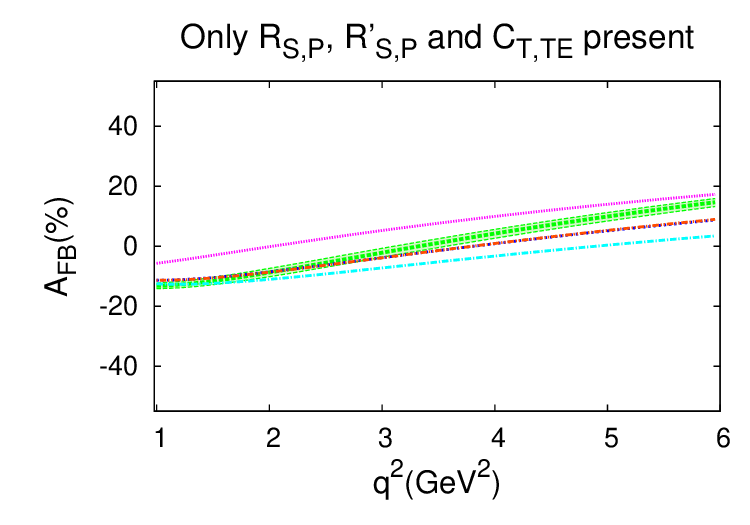}
\includegraphics[width=0.4\linewidth]{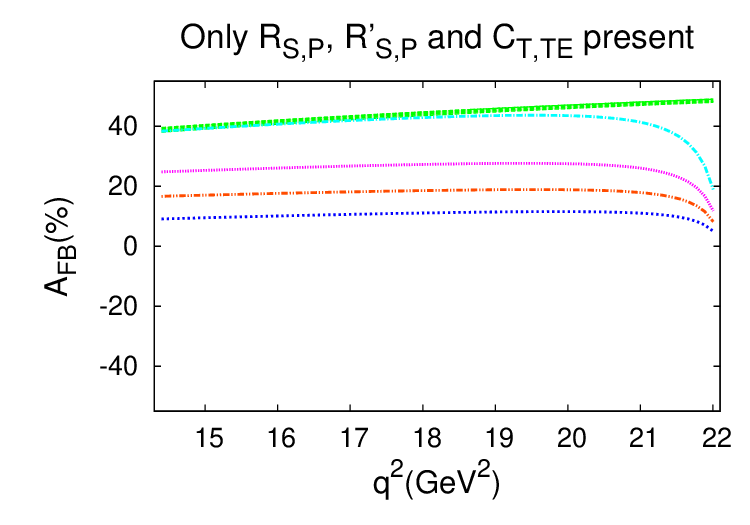} \\
  \includegraphics[width=0.4\linewidth]{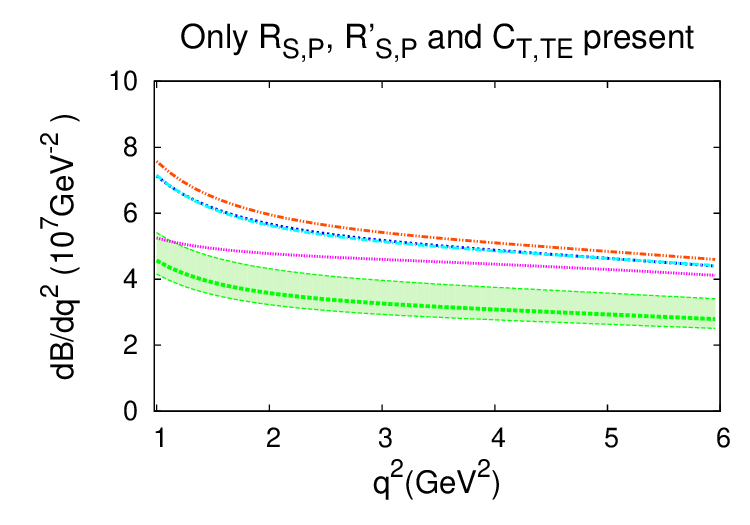}
 \includegraphics[width=0.4\linewidth]{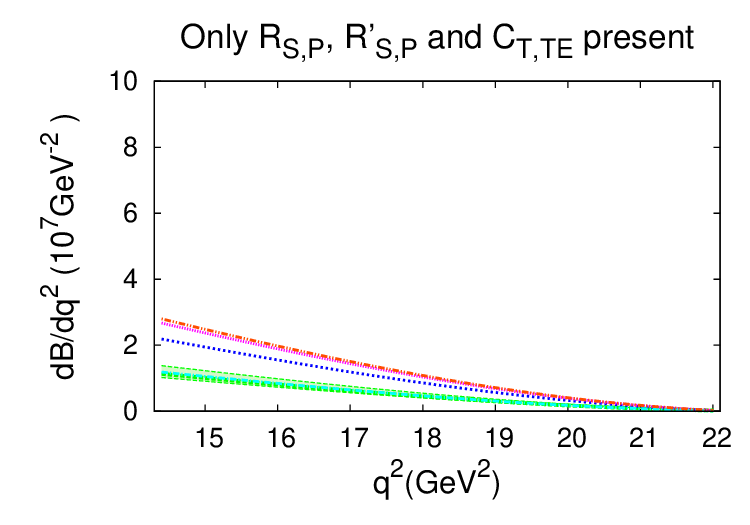}
\caption{The left (right) panels of the figure show $\AFB$ and DBR for
  $\BXsmumu$ in the low-$q^2$ (high-$q^2$) region, in the scenario
  where both SP and T terms are present.  The band corresponds to the
  SM prediction and its uncertainties; the lines show predictions for
  some representative values of NP parameters $(R_S, R_P, R'_S, R'_P ,
  C_T, C_{TE})$.  For example, the magenta curves in the low-$q^2$ and
  high-$q^2$ regions correspond to
  $(-1.23,-1.79,-0.86,-1.85,0.27,-0.36)$ and
  $(-1.23,-0.23,-1.35,0.08,1.37,0.01)$, respectively.
\label{fig:incl-spsppt}}
}

\subsection{Polarization fractions $f_L$ and $f_T$}

In Ref.~\cite{Lee:2006gs} it was pointed out that, besides the
dilepton invariant mass spectrum and the forward-backward asymmetry, a
third observable can be obtained from $\BXsmumu$, namely the double
differential decay width:
 \begin{equation}
 \frac{d^2B}{dz\,d\cos\theta_\mu}=
\frac{3}{8}\Big[ (1 + \cos^2\theta_\mu)H_T(z) 
+ 2\cos\theta_\mu H_A(z) + 2(1-\cos^2\theta_\mu)H_L(z)\Big]\;.
 \end{equation}
The functions $H_i(z)$ do not depend on $\cos \theta_\mu$.  The sum
$H_L(z)+H_T(z)$ gives the differential branching ratio $dB/dz$, while
the forward-backward asymmetry is given by $3\,H_A/4(H_L+H_T)$.
Splitting $dB/dz$ into longitudinal and transverse parts separates the
contributions with different $q^2$ dependences, providing a third
independent observable. This does not require measuring any additional
kinematical variable -- $q^2$ and $\cos \theta_\mu$ are sufficient.
Including all the NP interactions, and neglecting terms suppressed by
$m_\mu/m_b$ and $m_s/m_b$, $H_L(z)$ and $H_T(z)$ are given by
\begin{equation}
H_L(z)=H^{SM}_L(z)+H^{SM-VA}_L(z)+H^{VA}_L(z)+H^{SP}_L(z)+H^{T}_L(z)\;,
\end{equation}
\begin{equation}
H_T(z)=H^{SM}_T(z)+H^{SM-VA}_T(z)+H^{VA}_T(z)+H^{SP}_T(z)+H^{T}_T(z)\;,
\end{equation}
where the $H$ functions are given in Appendix~\ref{app-bxsmumu}.
The superscripts indicate the Lorentz structures contributing to
the term. 
The polarization fractions  $f_L$ and $f_T$ can be defined as
\be
f_L=\frac{H_L(z)}{H_L(z)+H_T(z)} ~, \qquad \qquad f_T=\frac{H_T(z)}{H_L(z)+H_T(z)} ~.
\ee
In the SM, $f_L$ can be as large as 0.9 at low $q^2$, and it decreases
to about 0.3 at high $q^2$.

Fig.~\ref{fig:incl-fl-vap} shows that when only $R_{V,A}$ couplings
are present, in the low-$q^2$ region $f_L$ can be suppressed
substantially, or even enhanced up to 1. A similar effect -- small
enhancement or a factor of two suppression -- is possible at high
$q^2$.  The suppression at low-$q^2$ is typically correlated with an
enhancement at high-$q^2$.  The effect of $R'_{V,A}$ couplings is
similar, but much milder, as expected.  SP and T operators,
individually or together, can only have an marginal effect on $f_L$.

\FIGURE[t]{
\includegraphics[width=0.4\linewidth]{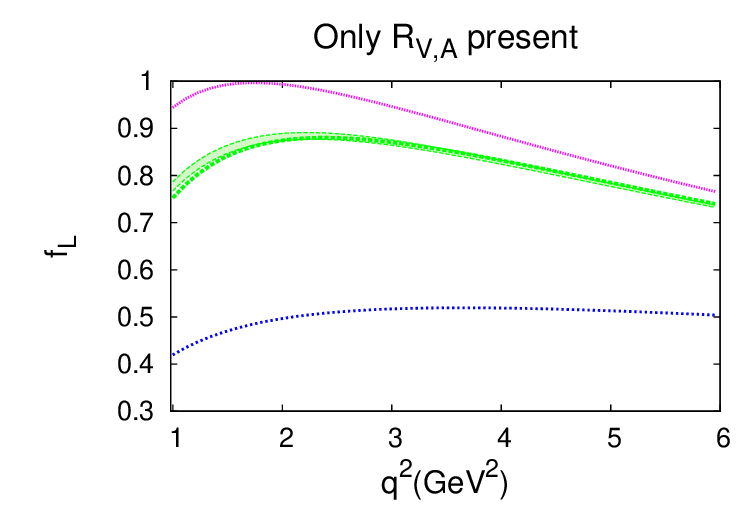}
\includegraphics[width=0.4\linewidth]{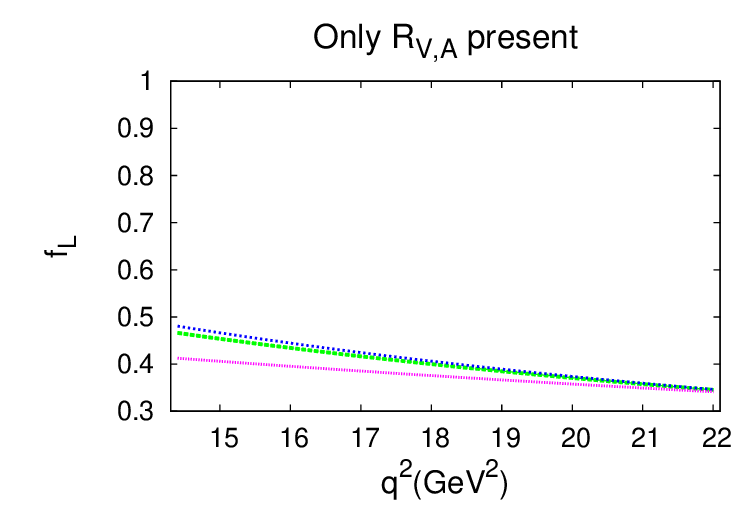}
\caption{The left (right) panels of the figure show $f_L$ for
  $\BXsmumu$ in the low-$q^2$ (high-$q^2$) region, in the scenario
  where only $(R_V, R_A)$ terms are present.  The band corresponds to
  the SM prediction and its uncertainties; the lines show predictions
  for some representative values of NP parameters $(R_V, R_A)$. For
  example, the blue curves in the low-$q^2$ and high-$q^2$ regions
  correspond to $(-8.14,5.75)$ and $(1.87,4.85)$, respectively.
\label{fig:incl-fl-vap}}
}

\section{\boldmath $\Bsmumugamma$
\label{Bsmumugamma}}

In this section we examine the NP contributions to the radiative
leptonic decay $\Bsmumugamma$.  This decay has not been detected as
yet.  The SM prediction for the BR in the range $q^2 \le 9.5$ GeV$^2$
and $q^2 \ge 15.9$ GeV$^2$ is $\approx 18.9 \times 10^{-9}$
\cite{Melikhov:2004mk}.  Although this decay needs the emission of an
additional photon as compared to $\Bsmumu$, which would suppress the
BR by a factor of $\alpha_{em}$, the photon emission also frees it
from helicity suppression, making its BR much larger than $\Bsmumu$.

This decay has contributions from many channels
\cite{Eilam:1996vg,Aliev:1996ud,Geng:2000fs,Dincer:2001hu,Melikhov:2004mk,Melikhov:2004fs}:
(i) direct emission of real or virtual photons from valence quarks of
the ${\bar B}^0_s$, (ii) real photon emitted from an internal line of
the $b \to s$ loop, (iii) weak annihilation due to the axial anomaly,
and (iv) bremsstrahlung from leptons in the final state.  The photon
emission from the $b \to s$ loop is suppressed by $m_b^2/m_W^2$
\cite{Aliev:1996ud}, and the weak annihilation is further suppressed
by $\Lambda_{QCD}/m_b$ \cite{Melikhov:2004mk}.  These two
contributions can then be neglected.  The bremsstrahlung contribution
is suppressed by $m_\mu/m_b$, and dominates only at extremely low
photon energies due to the infrared divergence.  The virtual photon
emission dominates in the low-$q^2$ region around the $\phi$
resonance.  If we choose the regions $2$ GeV$^2 \le q^2 \le 6$ GeV$^2$
and $14.4$ GeV$^2 \le q^2 \le 25$ GeV$^2$ as the low-$q^2$ and
high-$q^2$ regions, respectively, then the dominating contribution
comes from the diagrams in which the final-state photon is emitted
either from the $b$ or the $s$ quark. Then the $\Bsmumugamma$ decay is
governed by the effective Hamiltonian describing the $\bsmumu$
transition, as given in Eq.~(\ref{HSM}), and our formalism is
applicable.  Here we consider the the DBR and $\AFB$ in
$\Bsmumugamma$.

\subsection{Differential branching ratio and forward-backward asymmetry}

We begin with the differential branching ratio. The SP operators do
not contribute to the amplitude of $\Bsmumugamma$ and hence do not
play any role in the decay.  

 In terms of the dimensionless parameter $x_\gamma=2 E_\gamma/m_{B_s}$,
 where $E_\gamma$ is the photon energy in the $\bsbar$ rest frame,
 one can calculate the double differential decay rate to be
 \begin{eqnarray}
 \frac{\text{d}^2\Gamma}{\text{d}x_{\gamma}\text{d}(\cos\theta_\mu)} =
 \frac{1}{2 m_{B_s}} \dfrac{2 v\, m_{B_s}^2 x_{\gamma}}{(8\pi)^3}
 {\cal M}^{\dagger}{\cal M} \; ,
 \label{ddbr-mumugamma1}
 \end{eqnarray}
 where
 $v \equiv \sqrt{1- 4 m_{\mu}^2/[m_{B_s}^2(1-x_\gamma)]}$.
 From Eq.~(\ref{ddbr-mumugamma1}) we get the DBR to be
 \begin{eqnarray}
 \frac{\text{d}B}{\text{d} x_{\gamma}} &=&
 \tau_{B_s}\int_{-1}^{1} \frac{\text{d}^2\Gamma}
 {\text{d} x_{\gamma}\text{d}(\cos\theta_\mu)}\,
 \text{d}\cos\theta_\mu  \nn \\
 &=&\tau_{B_s}\Bigg[\frac{1}{2 m_{B_s}}
 \dfrac{2 v m_{B_s}^2}{(8\pi)^3}\Bigg]
 \Bigg[\frac{1}{4} ~ \frac{16 G_F^2}{2}
 \frac{\alpha_{em}^2}{16 \pi^2} |V_{tb}V_{ts}^*|^2 e^2\Bigg] \Theta \; .
 \label{dbr:bsg:main-1}
 \end{eqnarray}
 Here the quantity $\Theta$ has the form
 \beq
 \Theta = \frac{2}{3}~m_{B_s}^4~x^3_{\gamma}
 \Big[X_{VA}+X_{T}+X_{VA{\hbox{-}}T}\Big] \; ,
 \label{mumugamma-theta-expansion-1}
 \eeq
 where the $X$ terms are given in Appendix~\ref{app-Bsmumugamma}.
The subscripts of the $X$ terms denote the Lorentz structure(s)
contributing to that term.
For the sake of brevity, we have included the SM contributions in
$X_{VA}$. 

 The normalized forward-backward asymmetry of muons in
 $\Bsmumugamma$ is defined as
 \beq
 \AFB(q^2) = \frac {\displaystyle \int_{0}^{1} d\cos\theta_\mu
 \frac{d^2B}{dq^2 d\cos\theta_\mu} - \int_{-1}^{0} d\cos\theta_\mu
 \frac{d^2B}{dq^2 d\cos\theta_\mu} }{\displaystyle \int_{0}^{1}
 d\cos\theta_\mu \frac{d^2B}{dq^2 d\cos\theta_\mu} + \int_{-1}^{0}
 d\cos\theta_\mu \frac{d^2B}{dq^2 d\cos\theta_\mu} } ~,
 \eeq
 where $\theta_\mu$ is the angle between the three-momentum vectors of the
 $\bsbar$ and the $\mu^+$ in the dimuon center-of-mass frame. The
 calculation of $\AFB$ gives
 \begin{eqnarray}
 A_{FB}(q^2) &=& \frac{1}{\Theta}~
\Bigg(2 m^4_{B_s} v ~ x_{\gamma}^3\Bigg)\Bigg[ 
Y_{VA} + Y_{VA{\hbox{-}}T} \Bigg] \; ,
\label{afb-y} 
\end{eqnarray}
 with the $Y$ terms are defined in Appendix~\ref{app-Bsmumugamma}.

The details of the calculation are given in Appendix~\ref{app-Bsmumugamma}.
For the numerical calculations, we use the matrix elements given in
Ref.~\cite{Kruger:2002gf}. 
The parameters involved in the form factor calculations are chosen
in such a way that the LEET relations
between form factors are satisfied to a 10\% accuracy \cite{Kruger:2002gf}.
In our numerical analysis we take the errors in these form factors 
to be $\pm 10\%$.

Within the SM, $A_{FB}(q^2)$ is predicted to vanish around $q^2
\approx 4.3$ GeV$^2$ (i.e. $x_\gamma \approx 0.85$)
\cite{Kruger:2002gf}, and the crossing is predicted to be negative.
It is therefore interesting to see the effects of various NP operators
and their combinations on $A_{FB}$.  In the extreme LEET limit, using
the form-factor relations given in Ref.~\cite{Kruger:2002gf}, one can
easily see that the $A_{FB}$ is independent of the form factors.  In
Fig.~\ref{fig:bsg-va} we see large bands in the SM predictions of
$A_{FB}$ in the low $q^2$ region.  One may tend to interpret these as
large corrections to the LEET limit, however this would be somewhat
misleading, as we take the errors in the form factors, due to
corrections from the LEET limit, to be uncorrelated.  In realistic
models, LEET corrections to the form factors will be correlated,
leading to a smaller uncertainty band for $A_{FB}$ in the SM.

\FIGURE[t]{
\includegraphics[width=0.4\linewidth]{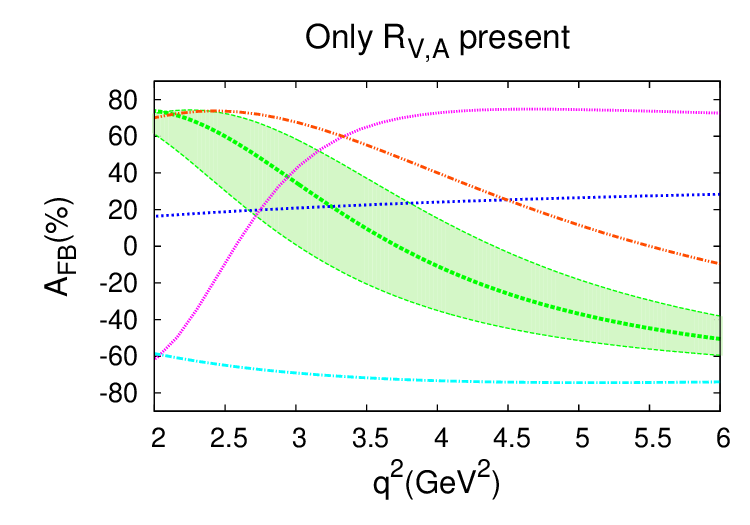}
\includegraphics[width=0.4\linewidth]{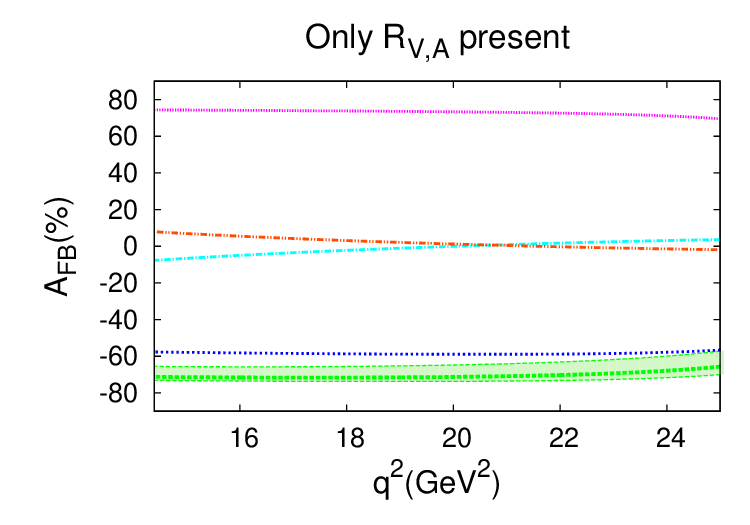} \\
  \includegraphics[width=0.4\linewidth]{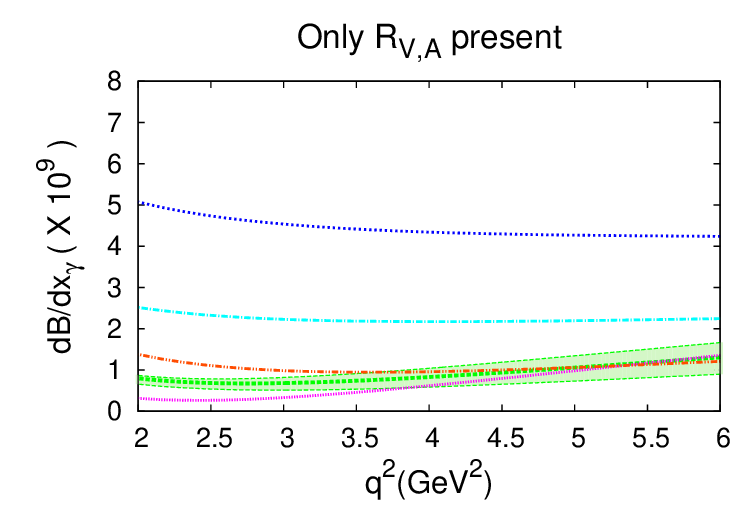}
 \includegraphics[width=0.4\linewidth]{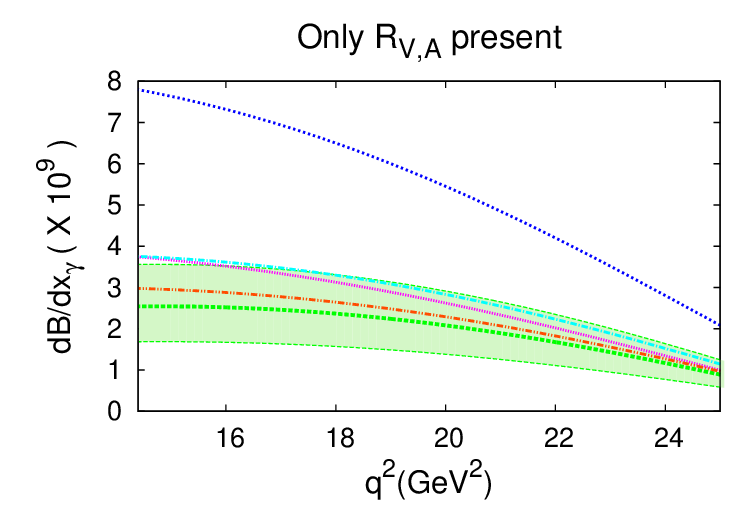}
\caption{The left (right) panels of the figure show $\AFB$ and DBR for
  $\Bsmumugamma$ in the low-$q^2$ (high-$q^2$) region, in the scenario
  where only $(R_V, R_A)$ terms are present.  Note that here $q^2 =
  m_B^2 (1-x_\gamma)$.  The band corresponds to the SM prediction and
  its uncertainties; the lines show predictions for some
  representative values of NP parameters $(R_V, R_A)$.  For example,
  the magenta curves in the low-$q^2$ and high-$q^2$ regions
  correspond to $(2.47,7.08)$ and $(-7.14,-0.42)$, respectively.
\label{fig:bsg-va}}
}

Fig.~\ref{fig:bsg-va} also shows $\AFB$ and DBR in the presence of NP
in the form of $R_{V,A}$ couplings.  With the large allowed values of
$|R_{V,A}|$ and the absence of any helicity suppression, we expect VA
operators to have a significant impact on the observables.  As can be
seen from the figure, the maximum allowed value of DBR can be 2-3
times larger than the SM prediction. The BR can also be suppressed
below the SM prediction due to destructive interference.  In the
low-$q^2$ region, the suppression can be large.  The features of the
zero-crossing predicted by the SM can be affected: it can be positive
or negative, can take place at any value of $q^2$, and can disappear
altogether. As expected, the impact of $R'_{V,A}$ couplings is much
milder.  In particular, the zero-crossing is always positive and in
the low-$q^2$ region.

With new tensor couplings, an enhancement of the DBR by up to a factor
of 3 in comparison to the SM prediction is possible.  Moreover, in the
limit of neglecting the muon mass, T operators do not contribute to
the $Y$-terms in Eq.~(\ref{afb-y}); their contribution is only to
$\Theta$.  As a result, they can only suppress $A_{FB}$ from its SM
value.

When all NP operators are allowed, we find that $B(\Bsmumugamma)$ can
be enhanced by a factor of 4, or it can be suppressed significantly.
The shape of $\AFB(q^2)$ is determined by the new VA couplings, while
its magnitude can be suppressed if the T couplings are significant.

\section{\boldmath $\BKmumu$
\label{BKmumu}}

The decay mode $\BKmumu$ is interesting primarily because the
forward-backward asymmetry of muons is predicted to vanish in the SM.
This is due to the fact that the hadronic matrix element for the
$\bdbar \to \bar{K}$ transition does not have any axial-vector
contribution.  $\AFB$ can have a nonzero value only if it receives a
contribution from new physics in the form of SP or T operators. Thus, the
information from this decay is complementary to that from the other
decays considered earlier, which were more sensitive to new physics VA
operators.

The total branching ratio of $\BKmumu$ has been measured to be
\cite{Barberio:2008fa}
\beq
B(\BKmumu) = \left(4.5 ^{+1.2} _{-1.0} \right) \times 10^{-7} ~,
\label{BR-BKmumu}
\eeq
which is consistent with the SM prediction \cite{Ali:2002jg}
\beq
B(\BKmumu)_{\rm SM} = (3.5 \pm 1.2) \times 10^{-7} \; .
\eeq
The integrated asymmetry, $\langle A_{FB} \rangle$, has been measured
by BaBar \cite{babar-06} and Belle \cite{Belle-oldKstar,ikado-06} to be
\begin{equation}
\left\langle A_{FB}\right\rangle  =  (0.15_{-0.23}^{+0.21} \pm 0.08)\,
\, \, \, \, \,
 {\rm (BaBar)} \, ,
\end{equation}
\begin{equation}
\left\langle A_{FB}\right\rangle  = (0.10 \pm 0.14 \pm 0.01) \, \,
\, \, {\rm (Belle)}. \label{fb_exp}
\end{equation}
These measurements are consistent with zero.  However, within
$2\sigma$ they can be as large as $\sim 40\%$.  Experiments such as
the LHC or a future Super-$B$ factory will increase the statistics by
more than two orders of magnitude.  For example, at ATLAS at the LHC,
after analysis cuts the number of $\BKmumu$ events is expected to be
$\sim 4000$ with $30$ fb$^{-1}$ of data \cite{Adorisio}.
Thus, $\langle A_{FB} \rangle$ can soon be probed to values as low as $5\%$.
With higher statistics, one will even be able to measure $\AFB$ as
a function of the invariant dimuon mass squared $q^2$.
This can provide a stronger handle on this quantity than just its
average value $\langle A_{FB} \rangle$.

The effect of NP on $\langle \AFB \rangle$ and the $\AFB(q^2)$
distribution in $\BKmumu$ was studied in Refs.~\cite{Bobeth:2007dw}
and \cite{Alok:2008wp} respectively. In the latter, it was shown that
simultaneous new-physics SP and T operators can lead to a large
enhancement of $\AFB(q^2)$ in the high-$q^2$ region. However, NP
effects due to other operators were not studied. Here we present a
complete analysis of the effect of NP on the $\AFB(q^2)$ distribution
in $\BKmumu$ by taking into account all possible NP operators and
their combinations. In addition, we study the possible zero crossing
of $\AFB(q^2)$ and the correlations between the DBR and $\AFB$
features.

 \subsection{Differential branching ratio and forward-backward asymmetry}

The differential branching ratio for this mode is given by
 \begin{eqnarray}
 \frac{dB}{dz} & = & B'_0\, \phi^{1/2}\,\beta_\mu
 \Bigg[X'_{VA} + X'_{SP} + X'_T + X'_{VA{\hbox{-}}SP} + X'_{VA{\hbox{-}}T} \Bigg] \; ,
 \end{eqnarray}
where the normalization factor $B_0'$, the phase factor $\phi$ and the
$X'$ terms are given in Appendix~\ref{app-bkmumu}.  The subscripts for
the $X'$ terms denote the Lorentz structure(s) contributing to that
term.

 The normalized forward-backward asymmetry for the muons in $\BKmumu$
 is defined as
 \beq
 \AFB(q^2) = \frac {\displaystyle \int_{0}^{1} d\cos\theta_\mu
 \frac{d^2B}{dq^2 d\cos\theta_\mu} - \int_{-1}^{0} d\cos\theta_\mu
 \frac{d^2B}{dq^2 d\cos\theta_\mu} }{\displaystyle \int_{0}^{1}
 d\cos\theta_\mu \frac{d^2B}{dq^2 d\cos\theta_\mu} + \int_{-1}^{0}
 d\cos\theta_\mu \frac{d^2B}{dq^2 d\cos\theta_\mu} } ~,
 \eeq
 where $\theta_\mu$ is the angle between the three-momenta of the $\bdbar$
 and the $\mu^+$ in the dimuon center-of-mass frame.
 The calculation of $\AFB(q^2)$ gives
 \begin{equation}
 A_{FB}(q^2)=\frac{2B'_0 \, \beta_\mu \, \phi}{dB/dz}
 \Bigg[Y'_{VA{\hbox{-}}SP} + Y'_{VA{\hbox{-}}T} + Y'_{SP{\hbox{-}}T} \Bigg]
 \label{afb-Kmumu-main-1}
 \end{equation}
 where the Y terms are given in Appendix~\ref{app-bkmumu}.

The largest source of uncertainty in the calculations are the $\bar{B}
\to \bar{K}$ form factors. As these cannot be calculated from first
principles within QCD, one has to rely on models.  In the numerical
calculations, we use the form factors as calculated in
Ref.~\cite{ali-00} in the framework of QCD light-cone sum rules; the
details are given in Appendix~\ref{app-bkmumu}. There are, however,
certain limits in which relations between form factors can be
rigorously obtained. In the large energy (LEET) limit, these relations
are valid up to $\alpha_s$, $1/E_K$ and $1/m_b$ corrections
\cite{Charles, Beneke}.

In the LEET limit, using the form-factor relations in
Eq.~(\ref{leet_rel}), one can verify that $A_{FB}$ is independent of
the form factors. This is quite useful as it implies that the
measurement of $A_{FB}$ can be used to extract the parameters of the
new-physics operators without form-factor uncertainties in this limit.

In the low-energy, large $q^2$, region one can also derive relations
between form factors in the heavy-quark limit \cite{grin1,
  grin2}. However, these relations do not completely eliminate the
form-factor dependence of the calculated quantities, and hence we do
not consider these relations. An analysis where these relations have
been used in the context of $\bsmumu$ can be found in
Refs.~\cite{hill1,hill2}.

{}From Eq.~(\ref{afb-Kmumu-main-1}), clearly new VA couplings alone
cannot give rise to $\AFB$, which vanishes in the SM in any case. Note
that this is one of the few cases where the VA couplings fail to
influence an asymmetry significantly, in spite of the large allowed
values of the couplings. This is because the argument about the
hadronic matrix element $\bar{B}^0_d \to \bar{K}$ not having any
axial-vector contribution stays valid even in the presence of NP.  The
DBR can, however, be enhanced by up to a factor of 2, or marginally
suppressed.

\FIGURE[t]{
\includegraphics[width=0.4\linewidth]{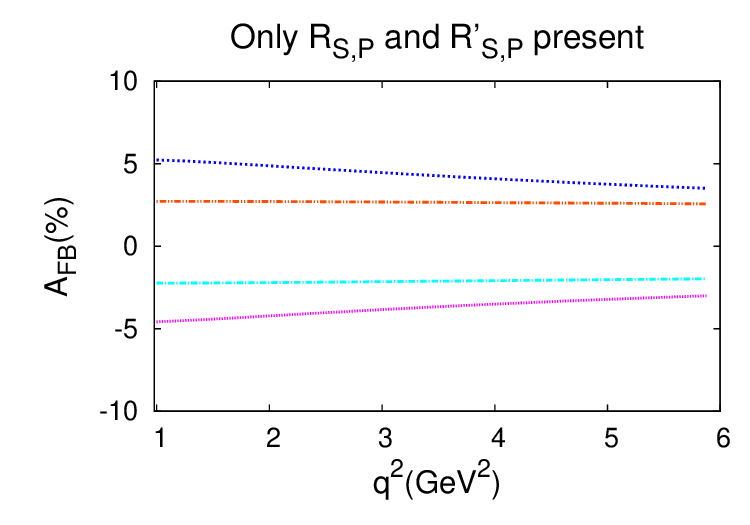}
\includegraphics[width=0.4\linewidth]{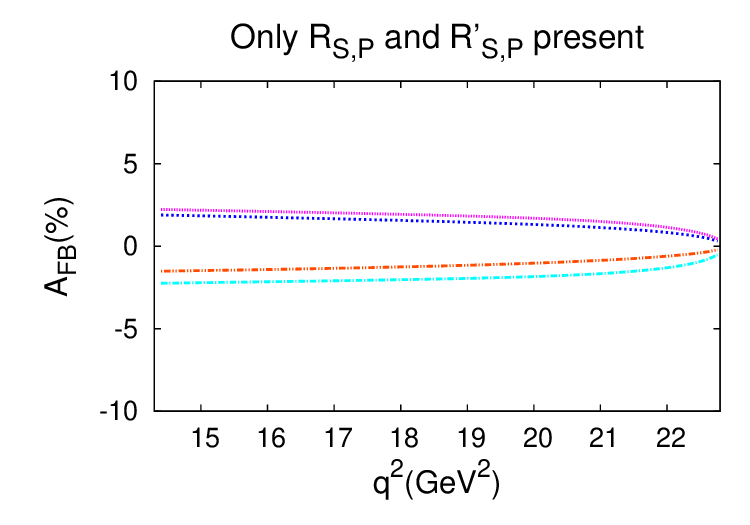} \\
  \includegraphics[width=0.4\linewidth]{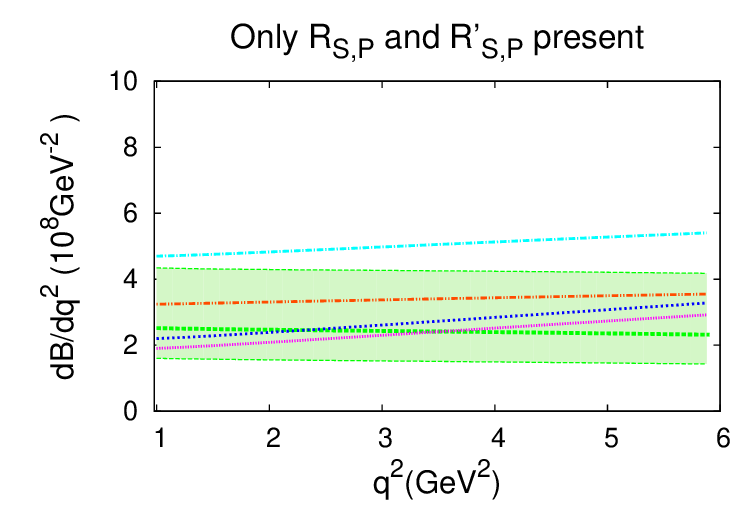}
 \includegraphics[width=0.4\linewidth]{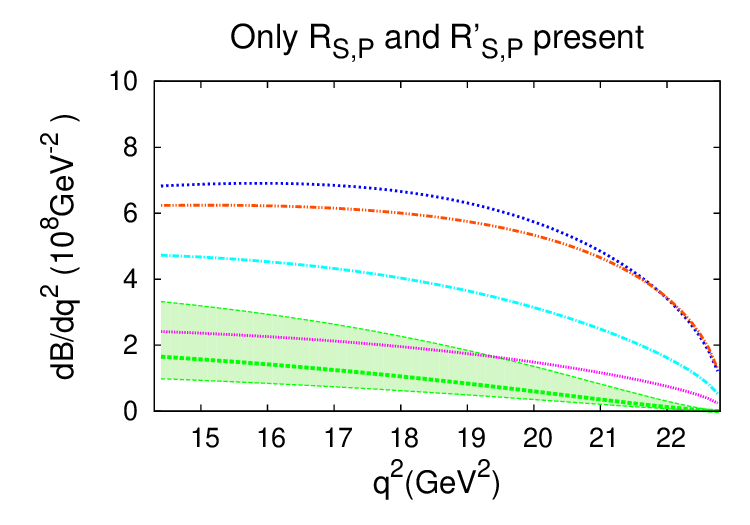}
\caption{The left (right) panels of the figure show $\AFB$ and DBR for
  $\BKmumu$ in the low-$q^2$ (high-$q^2$) region, in the scenario
  where all NP SP couplings are present.  The band corresponds to the
  SM prediction and its uncertainties; the lines show predictions for
  some representative values of NP parameters $(R_S, R_P, R'_S, R'_P)$
  .  For example, the blue curves in the low-$q^2$ and high-$q^2$
  regions correspond to $(-2.50,6.18,-2.84,-5.64)$ and
  $(-2.41,1.86,-2.07,1.42)$, respectively.
\label{fig:bkll-spspp}}
}

The contribution of SP operators through the $Y'_{VA{\hbox{-}}SP}$
terms can give rise to $\AFB$, where the VA contribution comes from
the SM operators. The effect is rather small when only $R_{S,P}$ or
only $R'_{S,P}$ couplings are present, due to the strong constraints
on their values. The peak value of $A_{FB}$ in the low-$q^2$ region
stays below the percent level, while in the the high-$q^2$ region it
can be enhanced up to $2\%$ at the extreme end point ($q^2 \gsim 22$
GeV$^2$), which is virtually impossible to observe.  However if both
the primed and unprimed SP couplings are present simultaneously, the
constraints on them are weakened. In such a situation, the peak value
of $A_{FB}$ in the low-$q^2$ (high-$q^2$) can become $\sim 5\%$ ($\sim
3\%$).  This may be seen in Fig.~\ref{fig:bkll-spspp}. It is also
observed that $\AFB$ is always positive or always negative,
i.e.\ there is no zero crossing.  The DBR also is significantly
affected only if both the primed and unprimed SP couplings are
present: it can be enhanced by up to a factor of 3.

\FIGURE[]{
\includegraphics[width=0.4\linewidth]{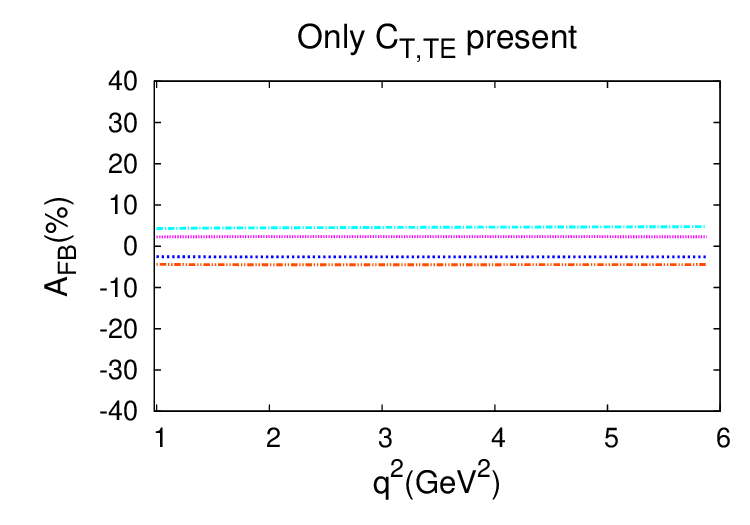}
\includegraphics[width=0.4\linewidth]{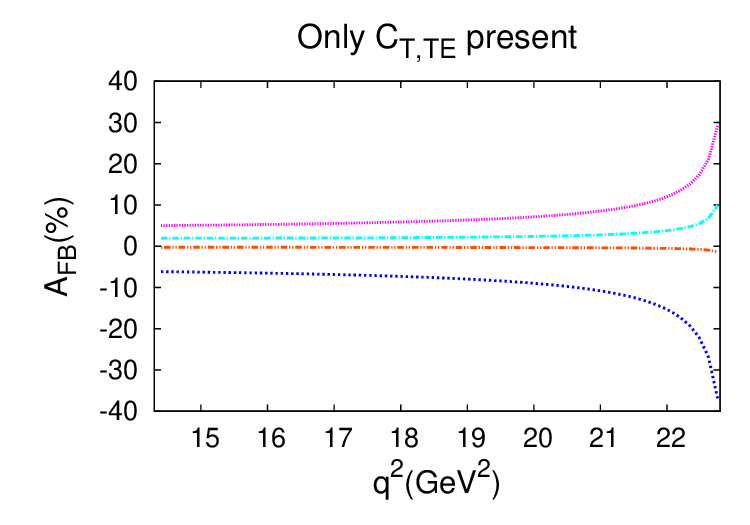} \\
  \includegraphics[width=0.4\linewidth]{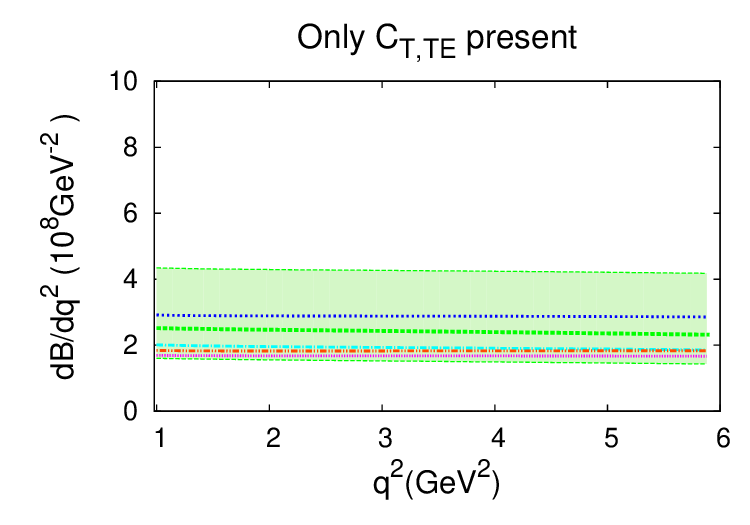}
 \includegraphics[width=0.4\linewidth]{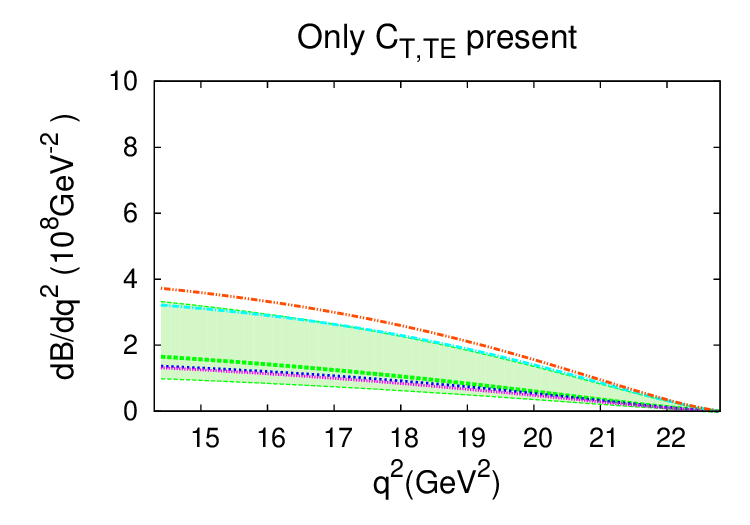}
\caption{The left (right) panels of the figure show $\AFB$ and DBR for
  $\BKmumu$ in the low-$q^2$ (high-$q^2$) region, in the scenario
  where only T terms are present.  The band corresponds to the SM
  prediction and its uncertainties; the lines show predictions for
  some representative values of NP parameters $(C_T, C_{TE})$.  For
  example, the blue curves in the low-$q^2$ and high-$q^2$ regions
  correspond to $(0.30,0.37)$ and $(0.49,0.57)$, respectively.
\label{fig:Kmumu-T}}
}

New T couplings are also expected to give rise to $\AFB$ through the
$Y'_{VA{\hbox{-}}T}$ terms in Eq.~(\ref{afb-Kmumu-main}).  It is
observed from Fig.~\ref{fig:Kmumu-T} that $A_{FB}(q^2)$ can be
enhanced up to 5-6\% in almost the entire $q^2$ region.  Moreover, at
$q^2 \gtrsim 21$ GeV$^2$, the peak value of $A_{FB}(q^2)$ reaches a
larger value ( $\sim 30\%$). The value of $A_{FB}(q^2)$ is always
positive or always negative, i.e.\ there is no zero crossing point.
The DBR values do not go significantly outside the SM-allowed range.

\FIGURE[]{
\includegraphics[width=0.4\linewidth]{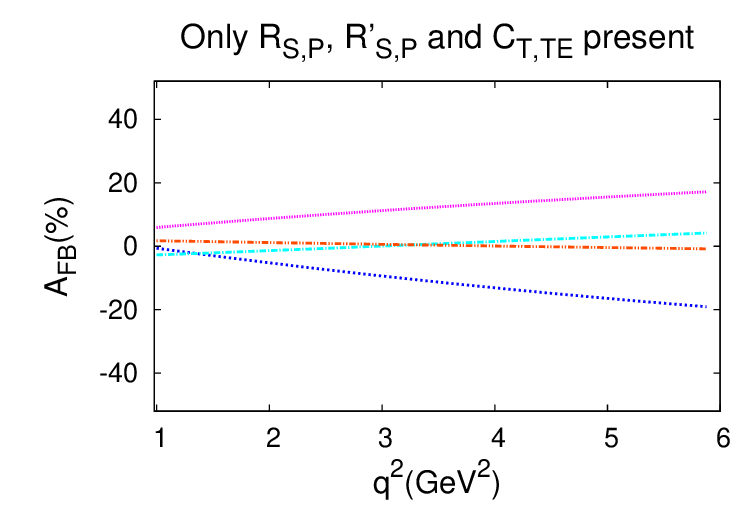}
\includegraphics[width=0.4\linewidth]{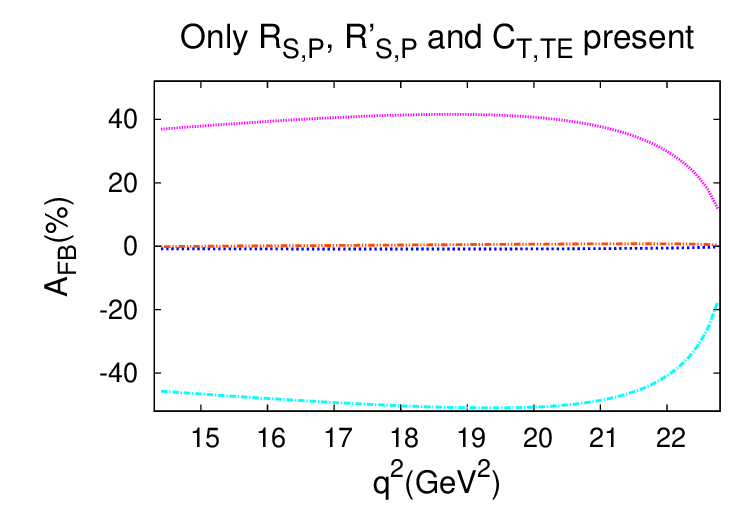} \\
  \includegraphics[width=0.4\linewidth]{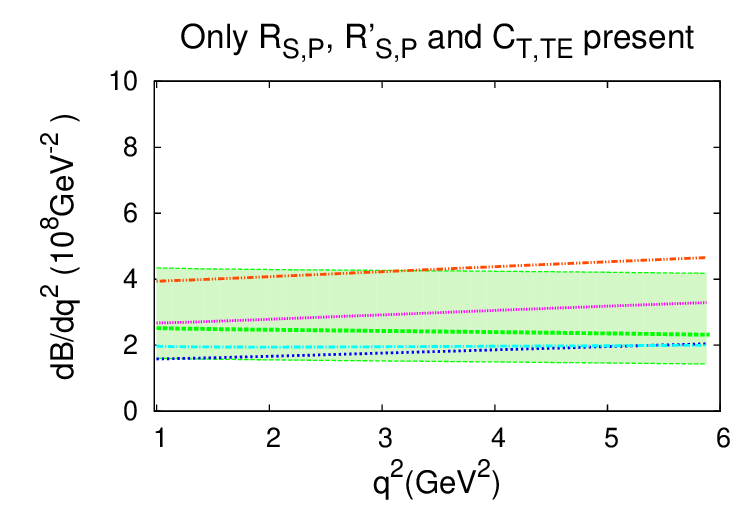}
 \includegraphics[width=0.4\linewidth]{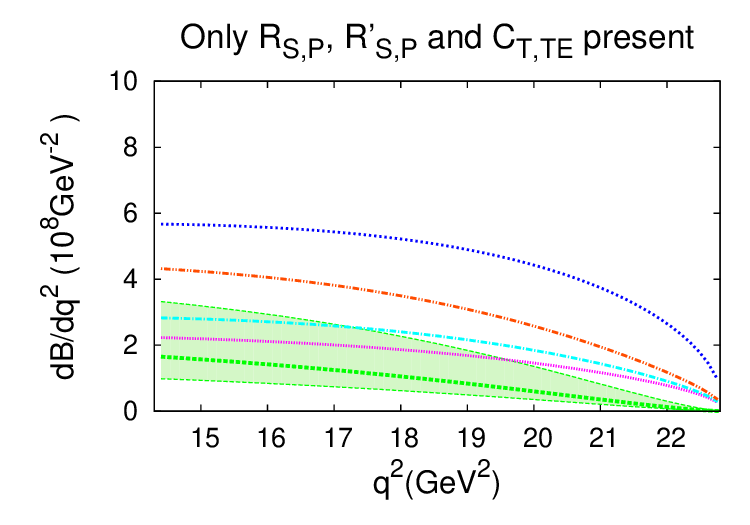}
\caption{The left (right) panels of the figure show $\AFB$ and DBR for
  $\BKmumu$ in the low-$q^2$ (high-$q^2$) region, in the scenario
  where both SP and T terms are present.  The band corresponds to the
  SM prediction and its uncertainties; the lines show predictions for
  some representative values of NP parameters $(R_S, R_P, R'_S, R'_P ,
  C_T, C_{TE})$.  For example, the magenta curves in the low-$q^2$ and
  high-$q^2$ regions correspond to
  $(-0.09,-2.24,0.16,-2.14,-0.33,-0.40)$ and
  $(-0.40,1.87,-0.59,1.88,-0.34,0.66)$, respectively.
\label{fig:bkll-spsppt}}
}

When VA and T couplings are present simultaneously, a DBR enhancement
of up to a factor of 2 is possible, while $\AFB$ can be large only at
extremely high $q^2$.  On the other hand, when SP and T couplings are
present simultaneously, their interference can have a large impact on
$A_{FB}$.  The interference term $Y'_{SP{\hbox{-}}T}$ that contributes
to $\AFB$ is not suppressed by $m_\mu/m_b$, and therefore a large
$\AFB$ is possible, as can be seen from Fig.~\ref{fig:bkll-spsppt}.
This is also the only combination of NP couplings where a zero
crossing may occur.  Among the asymmetries considered in this paper,
this is the one where the SP and T operators can have the largest
impact.  The DBR can also be enhanced by up to a factor of 2-3 at
large $q^2$ due to the simultaneous presence of primed and unprimed SP
operators.

\section{\boldmath $\BKstarmumu$
\label{BKstarmumu}}

The measurement of the forward-backward asymmetry in $\BKstarmumu$ by
the Belle collaboration \cite{Belle-oldKstar,Belle-newKstar}, which
showed a deviation from the SM prediction, indicates the possibility
of the presence of new physics.  According to the SM, $\AFB$ is $\leq
20\%$ and negative at low $q^2$, has a zero crossing at $q^2 \approx
4$ GeV$^2$, and is positive but $\leq 40\%$ for larger $q^2$
values. The experiment showed the asymmetry to be positive throughout
the range of $q^2$ -- consequently no zero crossing -- and $\AFB
\approx 60\%$ at large $q^2$ values. This has generated a special
interest in this decay.

There have already been a number of theoretical studies, both within
the SM \cite{Kruger:2000zg,Beneke:2001at,Egede:2008uy} and in specific
NP scenarios
\cite{kruger-matias,Lunghi:2006hc,Altmannshofer:2008dz,AFBNP},
focusing on the branching fraction and $\AFB$ of $\BKstarmumu$.  For
example, Ref.~\cite{HHM} has pointed out that $\AFB(q^2)$ is a
sensitive probe of NP that affects the SM Wilson coefficients. Other
observables based on the $K^*$ spin amplitudes of this decay are at
present under active theoretical and experimental analysis
\cite{kruger-matias,Lunghi:2006hc,Egede:2008uy}.  Finally, more
challenging observables, such as the polarized lepton forward-backward
asymmetry \cite{Aliev:2001pq, Bensalem:2002ni,
  Aliev:2003fy,Aliev:2004hi}, have also been considered, though the
measurement of this quantity is still lacking.

In the coming years, the LHCb experiment will collect around 3000
events of $\BKstarmumu$ per fb$^{-1}$ in the full range of $q^2$.
An integrated luminosity of 2 fb$^{-1}$ already would allow
the extraction of the SM zero of $\AFB$ (if it is there)
with a precision of $\pm 0.5$ GeV$^2$ \cite{lhc-roadmap}.  Indeed,
a dataset of 100 pb$^{-1}$ would already improve the world precision
obtained by Babar, Belle and CDF.  These measurements would also
permit many of the additional tests for NP mentioned above.

The decay $\BKstarmumu$, with $\bar{K^*}$ decaying to $\bar{K}\pi$,
has four particles in the final state. This implies that there are
three physical angles that can specify the relative directions of
these four final-state particles. The differential decay rate as a
function of these three angles has much more information than just the
forward-backward asymmetry. Indeed, $\AFB$ is just one of the
observables that can be derived from the complete angular analysis of
this decay. In this section we also consider other CP-conserving
observables.

\subsection{Angular analysis}

The complete angular distribution in $\BKstarmumu$ has been calculated
in Refs.~\cite{Chen:2002bq,Chen:2002zk} within the SM. In this
section, we calculate the angular distribution in the presence of NP,
which is a new result.  The full transition amplitude for
$\bar{B}(p_B)\rightarrow \bar{K}^*(p_{K^*},\epsilon^*) \mu^+(p_\mu^+)
\mu^-(p_\mu^-)$ is
 \bea
\label{TABKstmumu}
i{\cal M}(\BKstarmumu) & = & (-i)\frac{1}{2}~\Bigg[\frac{4~G_F}{\sqrt{2}}
\frac{\alpha_{em}}{4 \pi} (V_{ts}^* V_{tb})\Bigg] \times
\nn \\
&& \hspace{-3.2cm}
[ M_{V\mu}  L^\mu+ M_{A\mu}  L^{\mu5}+M_S  L+M_P  L^5 +M_{T\mu \nu}  L^{\mu\nu}
+i M_{E\mu \nu}  L_{\alpha \beta} \epsilon^{\mu\nu\alpha\beta}] \; ,
\eea
where the $L$'s are defined in Eq.~(\ref{Ldefs}). The $M$'s are given in Appendix~\ref{app-bkstarmumu}.

The complete three-angle distribution for the decay
$\bar{B}\rightarrow \bar{K}^*  (\rightarrow \bar{K}\pi)\mu^+\mu^-$
can be expressed in terms of $q^2$,
two polar angles $\theta_\mu$, $\theta_{K}$, and the angle between
the planes of the dimuon and $K \pi$ decays, $\phi$.
These angles are described in Fig.~\ref{KstmumuAD}.
We choose the momentum and polarization four-vectors of the $K^*$ meson
in the dimuon rest frame as
\bea
\label{KstKin1}
& & p_{K^{*}} =(E_{K^{*}},0,0, |\vec{p}_{K^{*}}|) \; , \nn \\
& & \varepsilon(0) =\frac{1}{m_{K^{*}}}(|\vec{p}_{K^{*}}|,0,0,E_{K^{*}}) \; ,
\quad \varepsilon(\lambda=\pm 1)= \mp \frac{1}{\sqrt{2}} (0,1,\pm i,0) \; ,
\eea
with
\bea
\label{KstKin2}
E_{K^{*}} &=& \frac{m^2_B-m^2_{K^*}-q^2}{2 \sqrt{q^2}},
~~|\vec{p}_{K^{*}}|=\sqrt{E^2_{K^{*}}-m^2_{K^*}} \; .
\eea

\FIGURE[t]{
\centerline{
\includegraphics[width=9.5cm]{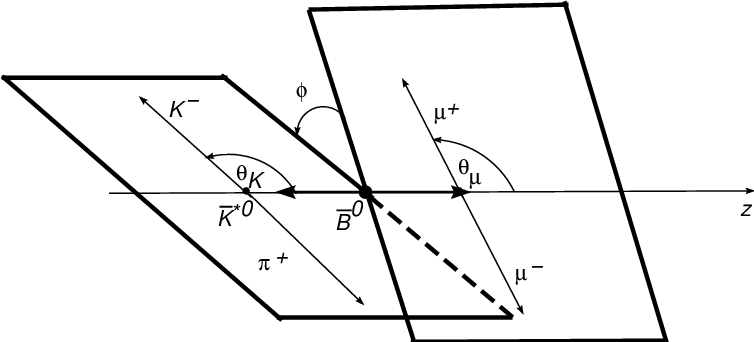}}
      \caption{The description of the angles $\theta_{\mu,K}$ and $\phi$ in the angular distribution of $\bar{B}\rightarrow \bar{K}^*  (\rightarrow\bar{K}\pi)\mu^+\mu^-$ decay.}
\label{KstmumuAD}
}

The three-angle distribution can be obtained using the
helicity formalism:
\bea
\label{ADKst}
&& \frac{d^4\Gamma}{dq^2d\cos{\theta_\mu} d\cos{\theta_{K}} d\phi } =N_F
\times \nl
\Bigg\lbrace &&  \cos^2{\theta_{K}} \Big(I^0_1 + I^0_2 \cos{2 \theta_\mu} + I^0_3 \cos{ \theta_\mu} \Big) + \sin^2{\theta_{K}} \Big(I^T_1   + I^T_2  \cos{2 \theta_\mu}   + I^T_3  \cos{ \theta_\mu} \nl &&+ I^T_4  \sin^2{\theta_\mu} \cos{2 \phi}+ I^T_5  \sin^2{\theta_\mu} \sin{2 \phi} \Big) + \sin{2\theta_{K}}\Big( I^{LT}_1  \sin{2\theta_\mu}\cos{ \phi} \nl &&  + I^{LT}_2  \sin{2\theta_\mu}\sin{ \phi}  + I^{LT}_3  \sin{\theta_\mu}\cos{ \phi} + I^{LT}_4  \sin{\theta_\mu}\sin{ \phi}\Big)  \Bigg\rbrace \; ,
\eea
where the normalization factor $N_F$ is
\bea
\label{NF}
N_F &=& \frac{3 \alpha^2_{em}G^2_F|V^*_{ts} V_{tb}|^2 |\vec{p}^B_{K^*}|
\beta_\mu}{2^{14}\pi^6 m^2_B}Br(K^*\rightarrow K\pi) \; .
\eea
Here $\beta_\mu =\sqrt{1-4 m^2_\mu/q^2}$, and $|\vec{p}^B_{K^*}|$ is
the magnitude of the $K^*$ momentum in the $B$-meson rest frame:
\bea
\label{KstmominB}
|\vec{p}^B_{K^*}| &=& \frac{1}{2 m_B}\sqrt{m^4_B+m^4_{K^*}+q^4-2
[q^2 m^2_B+m^2_{K^*}(m^2_B+q^2)]}\; .
\eea
The twelve angular coefficients $I$ depend on the couplings, kinematic
variables and form factors, and are given in
Appendix~\ref{app-bkstarmumu}.  In this paper we concentrate on the
CP-conserving observables: the DBR, the forward-backward asymmetry
$\AFB$, the polarization fraction $f_L$, and the asymmetries
$A_T^{(2)}$ and $A_{LT}$.

The theoretical predictions for the relevant $B \to K^*$ form factors
are rather uncertain in the region ($7$~GeV$^2 \le q^2 \le
12$~GeV$^2$) due to nearby charmed resonances.  The predictions are
relatively more robust in the lower and higher $q^2$ regions.  We
therefore concentrate on calculating the angular distribution in the
low-$q^2$ ($1~{\rm GeV^2} \le q^2 \le 6~{\rm GeV^2}$) and the
high-$q^2$ ($q^2 \ge 14.4~{\rm GeV^2}$) regions.  For numerical
calculations, we follow Ref.~\cite{AFBNP} for the form factors: in the
low-$q^2$ region, we use the form factors obtained using QCD
factorization, while in the high-$q^2$ region, we use the form factors
calculated in the light-cone sum-rule approach.

\subsection{Differential branching ratio and forward-backward asymmetry}

The forward-backward asymmetry for the muons is defined by
\bea
\label{FBA}
A_{FB}(q^2) &=&\frac{\int^1_0 d\cos{\theta_\mu}\frac{d^2\Gamma}{dq^2d\cos{\theta_\mu}  }-\int^0_{-1} d\cos{\theta_\mu}\frac{d^2\Gamma}{dq^2d\cos{\theta_\mu}  }}{\int^1_0 d\cos{\theta_\mu} \frac{d^2\Gamma}{dq^2d\cos{\theta_\mu}  }+\int^0_{-1} d\cos{\theta_\mu}\frac{d^2\Gamma}{dq^2d\cos{\theta_\mu}  }} \; .
\eea
It can be obtained by integrating over the two angles  $\theta_{K}$
and $\phi$ in Eq.~(\ref{ADKst}).
We obtain the  double differential decay rate  as
\bea
\label{doubDR2}
\frac{d^2\Gamma}{dq^2d\cos{\theta_\mu}  } &=&\frac{8 \pi N_F}{3}  \Big[\frac{1}{2} \Big(
I^0_1 + I^0_2 \cos{2  \theta_\mu} + I^0_3  \cos{\theta_\mu} \Big)+\Big (
I^T_1  + I^T_2 \cos{2  \theta_\mu} \nl && +  I^T_3  \cos{\theta_\mu} \Big)\Big]
 \; .
\eea

Further integration over the angle $\theta_\mu$ gives the differential
decay rate.
The contribution of the NP operators to the differential branching
ratio and forward-backward asymmetry of $\BKstarmumu$ was examined in detail
in Ref.~\cite{AFBNP}.
We do not reproduce the analysis here, but only give
the results below.

If only $R_{V,A}$ couplings are present, $\AFB$ can be enhanced at low
$q^2$, while keeping it positive, so that there is no zero crossing as
indicated by the recent data
\cite{Belle-oldKstar,Belle-newKstar,BaBar-Kmumu,BaBar-Kstarmumu}.
However, an enhancement at high $q^2$, also indicated by the same
data, is not possible.  On the other hand, if only $R'_{V,A}$
couplings are present, $\AFB$ can become large and positive at high
$q^2$, but then it has to be large and negative at low $q^2$. These
couplings are therefore unable to explain the positive values of
$\AFB$ at low $q^2$.  Thus, in order to reproduce the current
$\BKstarmumu$ experimental data, one needs both unprimed and primed NP
VA operators.  The NP coupling values that come closest to the data
typically correspond to suppressed DBR at low $q^2$.  (See
Fig.~\ref{fig:K*mumu-Afb}.)  But it is also possible to have a large
$A_{FB}$ (up to 60\%) in the entire $q^2$ region while being
consistent with the SM prediction for the DBR.  At present, the errors
on the measurements are quite large. However, if future experiments
reproduce the current central values with greater precision, this will
put important constraints on any NP proposed to explain the data.

\FIGURE[t]{
\includegraphics[width=0.4\linewidth]{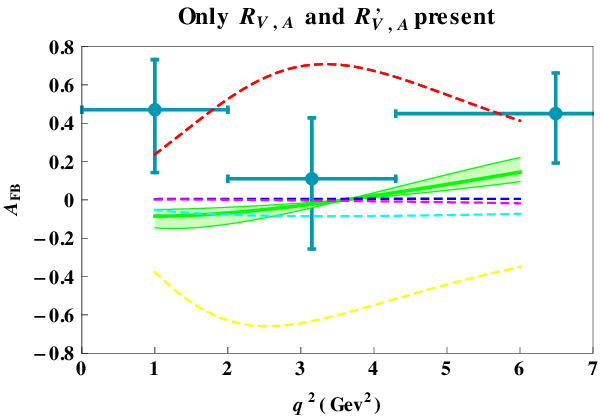}
\includegraphics[width=0.4\linewidth]{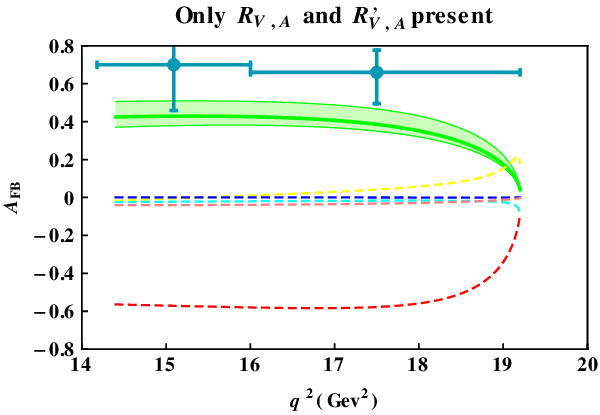} \\
  \includegraphics[width=0.4\linewidth]{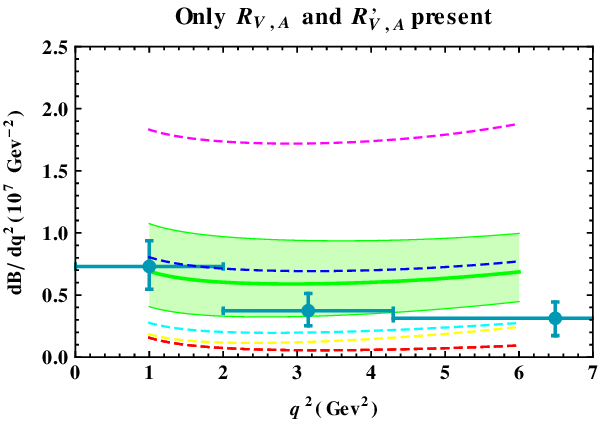}
 \includegraphics[width=0.4\linewidth]{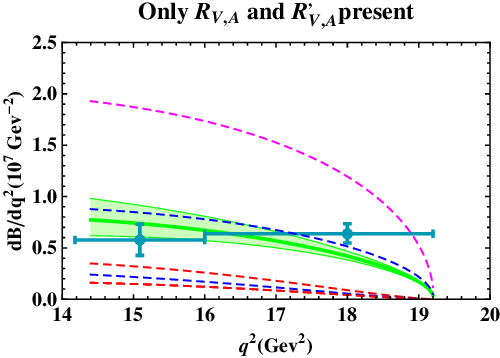}
\caption{The left (right) panels of the figure show $\AFB$ and DBR for
  $\BKstarmumu$ in the low-$q^2$ (high-$q^2$) region, in the scenario
  where both $(R_V, R_A)$ and $(R'_V, R'_A)$ terms are present.  The
  band corresponds to the SM prediction and its uncertainties; the
  lines show predictions for some representative values of NP
  parameters $(R_V, R_A, R'_V, R'_A)$.  For example, the red curves
  for $\AFB$ in the low and high $q^2$ regions correspond to $(-1.55,
  1.75, 6.16, 1.73)$ and $(-5.79, 1.10, 0.47,-3.34)$,
  respectively. The pink curves for DBR in the low-$q^2$ and high-$q^2$
  regions correspond to $(1.96, -4.09, 4.61, 0.13)$.  For comparison,
  the experimental data are also displayed in blue cross lines.
\label{fig:K*mumu-Afb}}
}


New SP couplings by themselves cannot significantly affect either the
DBR or the $\AFB$ predictions of the SM.  New T couplings in general
tend to enhance DBR significantly, by up to a factor of 2, while not
contributing any additional terms to the asymmetry.  As a result, the
magnitude of $\AFB$ is suppressed.  The zero crossing can be anywhere
in the entire $q^2$ range, or it may disappear altogether. However,
whenever it is present, it is always a SM-like (positive) crossing.
When SP and T couplings are present simultaneously, additional
contributions to $\AFB$ that are not suppressed by $m_\mu/m_B$ are
possible.  As a result, $\AFB$ obtained with this combination can be
marginally enhanced as compared to the case with only T operators.  It
is then possible to have no zero crossing. However, the magnitude of
$\AFB$ cannot be large in the high-$q^2$ region.

\subsection{Polarization fraction $f_L$}

The differential decay rate and $K^*$ polarization fractions can be
found by integrating over the three angles in Eq.~(\ref{ADKst}) to get
\bea
\frac{d\Gamma}{dq^2 } &=& \frac{8 \pi N_F}{3} (A_{L}+A_{T}) ~,
\eea
where the longitudinal and transverse polarization amplitudes
$A_L$ and $A_T$ are obtained from Eq.~(\ref{doubDR2}):
\bea
\label{HL}
A_L &=& \Big(I^0_1  - \frac{1}{3} I^0_2 \Big),\quad A_T  = 2 \Big(I^T_1  - \frac{1}{3} I^T_2 \Big) ~.
\eea
It can be seen from the expressions for the $I$'s in Appendix~\ref{app-bkstarmumu}
[see Eq.~(\ref{eq:IT})] that SP couplings cannot affect $A_T$.
The longitudinal and transverse polarization fractions, $f_L$ and $f_T$,
respectively, are defined as
\bea
\label{flft}
f_L &=& \frac{A_L}{A_L+A_T} ~~,~~~~
f_T = \frac{A_T}{A_L+A_T} ~.
\eea
In the SM, $f_L$ can be as large as 0.9 at low $q^2$, and it decreases
to about 0.3 at high $q^2$.  As can be seen from Fig.~\ref{fig:fL-VA},
new VA couplings can suppress $f_L$ substantially: it can almost
vanish in some allowed parameter range.

\FIGURE[t]{
\includegraphics[width=0.4\linewidth]{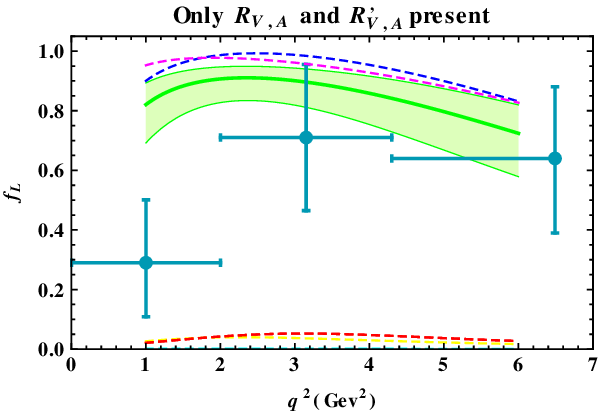}
\includegraphics[width=0.4\linewidth]{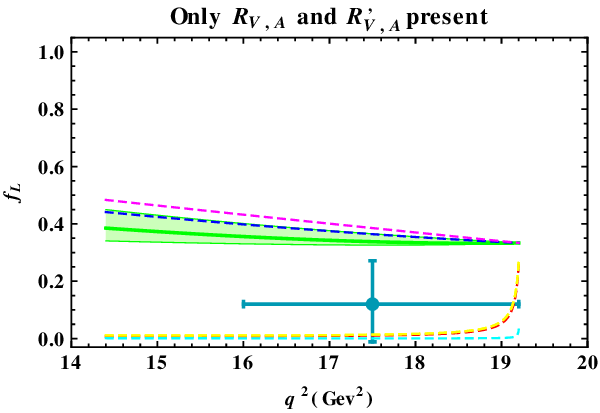} \\
\caption{The left (right) panel of the figure shows $f_L$ for
  $\BKstarmumu$ in the low-$q^2$ (high-$q^2$) region, in the scenario
  where both $(R_V, R_A)$ and $(R'_V, R'_A)$ terms are present.  For
  example, the blue curves in the low-$q^2$ and high-$q^2$ regions
  correspond to $(1.64, -0.90, 4.27, -0.91)$ and $(1.96, -4.09, 4.61,
  0.13)$, respectively.  For comparison, the experimental data are
  also displayed in blue cross lines.
\label{fig:fL-VA}}
}

New SP couplings cannot change the value of $f_L$ outside the range
allowed by the SM.  This may be attributed to the strong constraints
on the values of these couplings.  New T couplings tend to suppress
$f_L$, except at $q^2 \approx 1$-$2$ GeV$^2$, where the value of
$f_L$ cannot be less than 0.5 as may be seen from Fig.~\ref{fig:fL-T}.
Since both VA and T couplings tend to suppress $f_L$, their combined
effect results in a similar behavior.

\FIGURE[t]{
\includegraphics[width=0.4\linewidth]{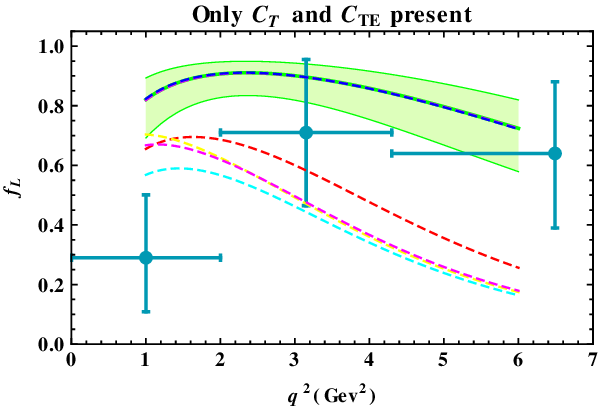}
\includegraphics[width=0.4\linewidth]{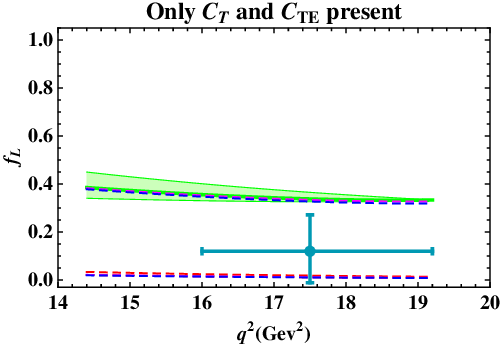} \\
\caption{The left (right) panel of the figure shows $f_L$ for
  $\BKstarmumu$ in the low-$q^2$ (high-$q^2$) region, in the scenario
  where only new T couplings are present.  The band corresponds to the
  SM prediction and its uncertainties; the lines show predictions for
  some representative values of NP parameters $(C_T,C_{TE})$.  For
  example, the red curves in the low-$q^2$ and high-$q^2$ regions correspond
  to $(0.66, -0.14)$ and $(0.3, -0.46)$, respectively.
\label{fig:fL-T}}
}

\subsection{Angular asymmetries $A_T^{(2)}$ and $A_{LT}$}

In this subsection we consider the two angular asymmetries $A_T^{(2)}$
and $A_{LT}$. The first quantity was discussed before in
Ref.~\cite{kruger-matias}, while $A_{LT}$ is introduced here for the
first time.

The CP-conserving transverse asymmetry $A_T^{(2)}$
can be  defined through the double differential decay rate
\bea
\label{doubDR3}
\frac{d^2\Gamma}{dq^2 d\phi } &=&\frac{1}{2\pi} \frac{d\Gamma}{dq^2  }
\Big[ 1+ f_T \left(A^{(2)}_T \cos{2 \phi} + A^{(im)}_T \sin{2 \phi}\right)
\Big] \; .
\eea
Here $ A^{(im)}_T$ depends on the imaginary part of a certain combination
of amplitudes and can be used to construct CP-violating observables.
We will not consider it any further in this work. The asymmetry
$ A^{(2)}_T $ can be obtained by integrating over the two polar
angles $\theta_{\mu}$ and $\theta_{K}$ in Eq.~(\ref{ADKst}).
It can be expressed as
\bea
 \label{AT2}
A^{(2)}_T &=& \frac{4 I^T_4}{3 A_T}.
\eea
We observe that $A_T^{(2)}$ cannot be affected by SP couplings.

In the SM,
\bea
A^{(2)}_T & \approx& \frac{4 \beta^2_\mu  \Big(|A^V_\perp|^2-|A^V_\parallel|^2 + |A^A_\perp|^2-|A^A_\parallel|^2\Big)}{3 A_T} \; .
\eea
The transversity amplitudes $A_{\parallel, \perp}$ are defined through
Eqs.~(\ref{Ksthelamp}) and (\ref{Kstrasamp1}) given in
Appendix~\ref{app-bkstarmumu}. At leading order in
$\Lambda_{QCD}/E_{K^{*}}$, $\Lambda_{QCD}/m_b$ and $\alpha_s$ (the
LEET limit), one can use the form-factor relations of
Refs.~\cite{Charles, Beneke} and neglect terms of ${\cal
  O}(m^2_{K^*}/m^2_B)$ to obtain
\beq
A^{+}_{V} \approx  0 ~~,~~~~ A^{+}_{A} \approx  0 \; .
\label{leet_kstar}
\eeq
Thus, in the low-$q^2$ region, 
\beq
A^i_\parallel \approx  \frac{A^{-}_i}{\sqrt{2}} ~~,~~~~ A^i_\perp \approx  -\frac{A^{-}_i}{\sqrt{2}}  \quad \mathrm{for}\quad i= V, A  \; ,
\label{leet_kstar2}
\eeq
which corresponds to the LEET limit.  $A_{T}^{(2)} \approx 0$ in the
SM and is independent of form factors up to corrections of order
$\Lambda_{QCD}/E_{K^{*}}$, $\Lambda_{QCD}/m_b$ and $\alpha_s$,
i.e.\ the hadronic uncertainty is small. This can be seen in
Figs.~\ref{fig:AT2-VA} and \ref{fig:AT2-T}. This indicates that
corrections to the LEET limit are small, and makes $A_{T}^{(2)}$ an
excellent observable to look for new-physics effects
\cite{kruger-matias}.

We now examine the longitudinal-transverse asymmetry $A_{LT}$, defined by
 \bea
 \label{ALT1-def}
 A_{LT} &=& \frac{\int^{\pi/2}_{-\pi/2}d\phi(\int^1_0 d\cos {\theta_{K}} \frac{d^3\Gamma}{dq^2d\phi d\cos {\theta_{K}}}-\int^0_{-1} d\cos {\theta_{K}} \frac{d^3\Gamma}{dq^2d\phi d\cos {\theta_{K}}})}{\int^{\pi/2}_{-\pi/2}d\phi(\int^1_0 d\cos {\theta_{K}} \frac{d^3\Gamma}{dq^2d\phi d\cos {\theta_{K}}}+\int^0_{-1} d\cos {\theta_{K}} \frac{d^3\Gamma}{dq^2d\phi d\cos {\theta_{K}}})} \; .
 \eea
One can compare $A_{LT}$ to $A_{FB}$. In $A_{FB}$ the angle $\phi$ is
integrated over its entire range, while in $A_{LT}$ $\phi$ is only
integrated over the range $(- \pi/2,\pi/2)$. This choice of
integration range eliminates all terms which depend on the imaginary
part of combinations of amplitudes in the angular distribution. (These
eliminated terms can be used to construct CP-violating observables and
will not be discussed here.)  In $A_{LT}$ only the CP-conserving parts
of the angular distribution survive. Note that, in the CP-conserving
limit, $A_{LT}$ is the same as the observable $S_5$ defined in
Ref.~\cite{Altmannshofer:2008dz}, apart from a normalization constant.
The quantity $A_{LT}$ can also be expressed in terms of the
observables $A_{T}^{(3)}$ and $A_{T}^{(4)}$ defined in
Ref.~\cite{Egede:2008uy}. However, $A_{LT}$ is easily extracted from
the angular distribution and has different properties in the LEET
limit than $A_{T}^{(3)}$ and $A_{T}^{(4)}$.

Using Eq.~(\ref{ADKst}), the asymmetry $A_{LT}$ can be expressed as
\bea
\label{ALT1-expr}
 A_{LT} &=& \frac{I^{LT}_3}{2 (A_L + A_T) }.
 \eea
We observe from Eq.~(\ref{eq:ILT}) that $A_{LT}$ depends on the VA couplings, as
well as on V-S, S-TE, P-T, and V-T interference terms.
In the SM,
 \bea
 \label{ALT1-exprSM}
 A_{LT} &=& \frac{\beta_\mu {\rm Re}[ A^{L}_{0,VA}(A^{V*}_\perp - A^{A*}_\perp)-A^{R}_{0,VA}(A^{V*}_\perp + A^{A*}_\perp)]}{\sqrt{2}  (A_L + A_T) }\; .
 \eea

Now, in the LEET limit, $A^{+}_{V,A} \approx 0$. Hence, in this limit,
\bea
A_{LT}^{LEET} &\propto& \frac{ {\rm Re}[A^0_V A^{-*}_A + A^0_A A^{-*}_V]}{A_L + A_T} \; .
 \eea
{}From this it can be shown that the SM predicts ${A}_{LT} = 0$ at
\bea
q^2 \approx -\frac{C^{eff}_7 m_b m^2_B}{C^{eff}_7 m_b +C^{eff}_9 m_B }
\approx 1.96 ~{\rm GeV}^2 \; .
\eea
Thus, just like $A_{FB}$, the quantity $A_{LT}$ also has a zero
crossing which is independent of form factors in the LEET limit.  Note
that the zero crossing of $A_{LT}$ is different from that of $A_{FB}$.
Figs.~\ref{fig:AT2-VA} and \ref{fig:AT2-T} also demonstrate that the
zero crossing of $A_{LT}$ has a very small hadronic uncertainty. This
indicates small corrections to the LEET limit, making the position of
the zero crossing of $A_{LT}$ a robust prediction of the SM. This quantity 
would therefore be very useful in searching for new-physics effects. 

New VA couplings can affect $A_T^{(2)}$ significantly: they can
enhance its magnitude by a large amount, change its sign, and change
its $q^2$-dependence.  The zero-crossing point may be at a value of
$q^2$ different from that predicted by the SM.

Since $A_{LT}$ here is identical to the observable $S_5$ in
Refs.~\cite{ Altmannshofer:2008dz,aoife} in the CP-conserving limit
(apart from a normalization factor), the zero-crossing in both of
these observables is expected to take place at the same $q^2$.
Indeed, the results agree at LO, while the NLO corrections can shift
the $q^2$ at the zero-crossing to $q^2= 2.24^{+0.06}_{-0.08}$
\cite{Altmannshofer:2008dz}.  Note that the deviation due to new VA
couplings can be much larger than the effects due to NLO corrections.

Except at very low $q^2$, the magnitude of $A_{LT}$ is generally
suppressed by new VA couplings. The primed VA couplings can be
constrained by $A_{LT}$ better than the unprimed VA couplings. In both
cases, the value of $A_{LT}$ can be anywhere in the $q^2$ range, and
can be positive or negative. In particular, there may or may not be a
zero crossing, and if there is, its position can be different from
that of the SM.

\FIGURE[t]{
\includegraphics[width=0.4\linewidth]{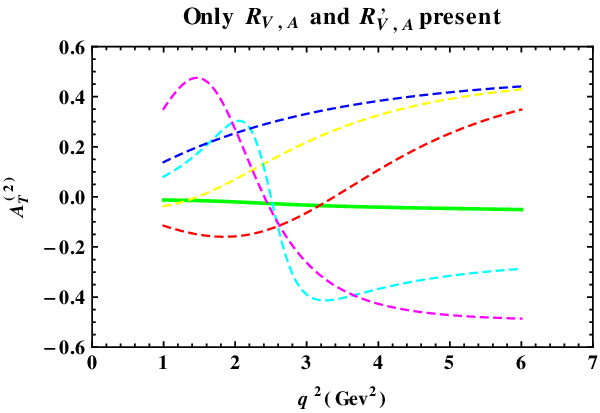}
\includegraphics[width=0.4\linewidth]{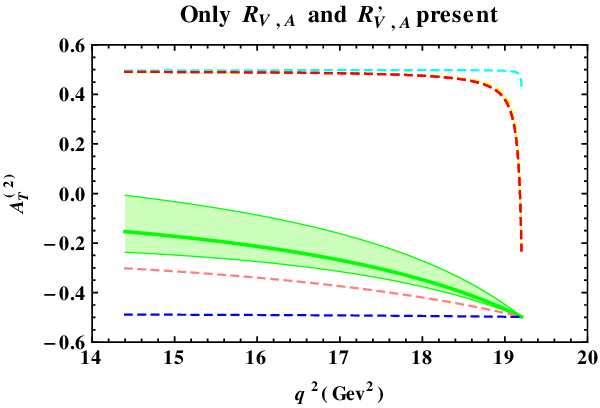} \\
  \includegraphics[width=0.4\linewidth]{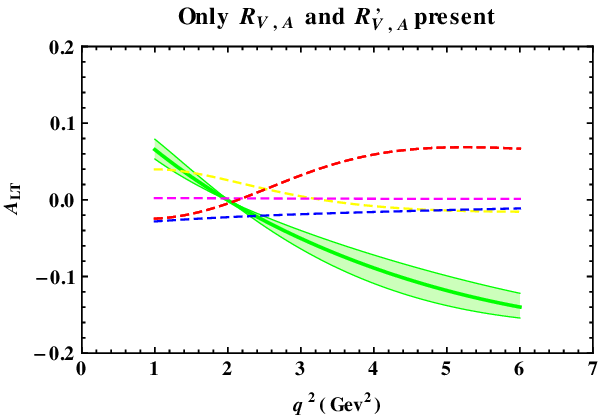}
 \includegraphics[width=0.4\linewidth]{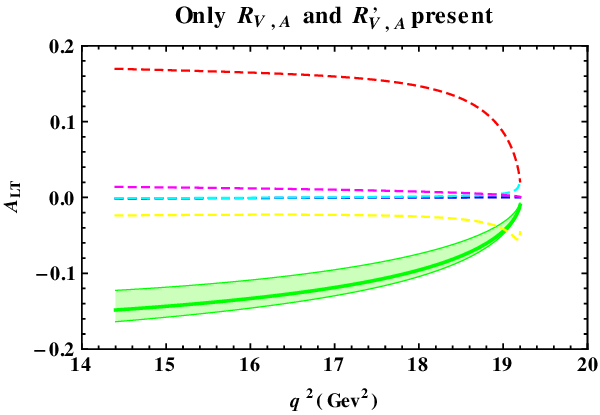}
\caption{The left (right) panels of the figure show $A_T^{(2)}$ and
  $A_{LT}$ for $\BKstarmumu$ in the low-$q^2$ (high-$q^2$) region, in
  the scenario where both $(R_V, R_A)$ and $(R'_V, R'_A)$ terms are
  present.  The band corresponds to the SM prediction and its
  uncertainties; the lines show predictions for some representative
  values of NP parameters $(R_V, R_A, R'_V, R'_A)$.  For example, the
  pink curves for $A_T^{(2)}$ in the low-$q^2$ and high-$q^2$ regions
  correspond to $(1.96, -4.09, 4.61,0.13)$ and $(1.64, -0.90, 4.27,
  -0.91)$, respectively. The red curves for $A_{LT}$ in the low-$q^2$ and
  high-$q^2$ regions correspond to $(-1.55, 1.75, 6.16, 1.73)$ and
  $(-5.79, 1.10, 0.47, -3.33)$, respectively.
\label{fig:AT2-VA}}
}

New SP couplings do not affect $A_T^{(2)}$, and $A_{LT}$ qualitatively
behaves similarly to the SM.  New T couplings in general tend to
suppress the magnitudes of both asymmetries (see
Fig.~\ref{fig:AT2-T}).

\FIGURE[t]{
\includegraphics[width=0.4\linewidth]{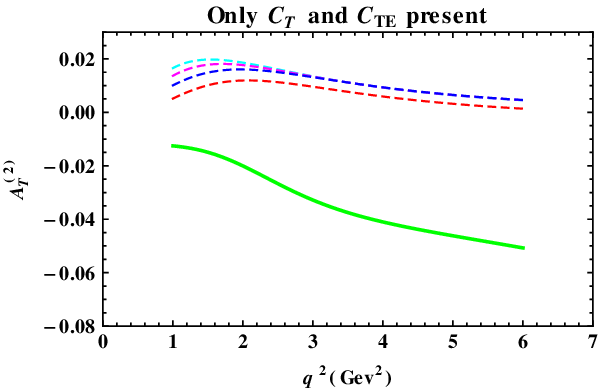}
\includegraphics[width=0.4\linewidth]{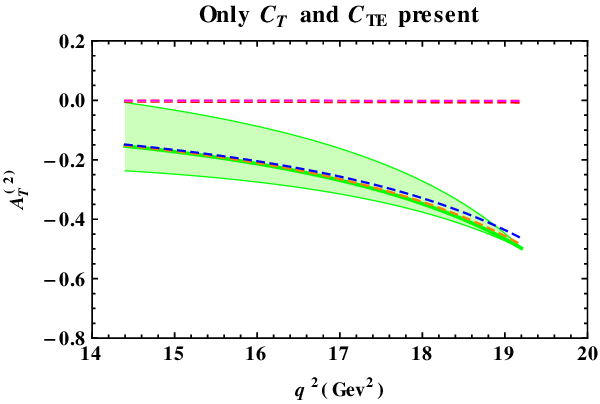} \\
  \includegraphics[width=0.4\linewidth]{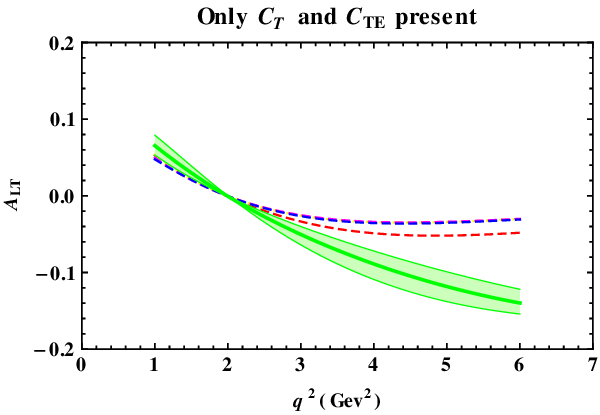}
 \includegraphics[width=0.4\linewidth]{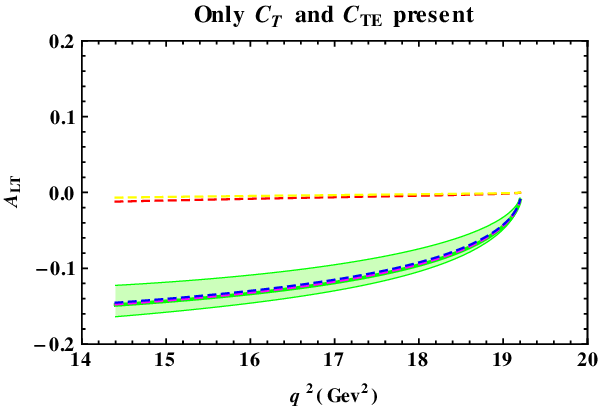}
\caption{The left (right) panels of the figure show $A_T^{(2)}$ and
  $A_{LT}$ for $\BKstarmumu$ in the low-$q^2$ (high-$q^2$) region, in
  the scenario where only new T couplings are present. The band
  corresponds to the SM prediction and its uncertainties; the lines
  show predictions for some representative values of NP parameters
  $(C_T, C_{TE})$.  For example, the blue curves for $A_T^{(2)}$ in
  the low-$q^2$ and high-$q^2$ regions correspond to $(0.3, -0.46)$ and
  $(-0.005, 0.014)$, respectively. The red curves for $A_{LT}$ in the
  low-$q^2$ and high-$q^2$ regions correspond to $(0.3, -0.46)$ and $(0.66,
  -0.14)$, respectively.
\label{fig:AT2-T}}
}

\section{Discussion and Summary}
\label{summary}

Flavor-changing neutral current (FCNC) processes are expected to be
incisive probes of new physics.  In the SM, they occur only at loop
level, and hence are suppressed.  This may allow the new-physics (NP)
effects to be identifiable.  Of course, since we have no clue about
what form the NP takes, the observations from a variety of processes
are necessary.  In this paper, we have focussed on the processes that
involve the effective transition $\bsmumu$.

The transition $\bsmumu$ is responsible for many decay modes such as
$\Bsmumu$, $\BXsmumu$, $\Bsmumugamma$, $\BKmumu$, $\BKstarmumu$.
While some of these processes (e.g.\ $\Bsmumu$) have not yet been
observed, the upper bounds on their branching ratios have already
yielded strong constraints on NP.  Some of these processes have been
observed and the measurements of their branching fractions, as well as
of additional observables such as the forward-backward asymmetries,
are available.  Indeed, the recently-observed muon forward-backward
asymmetry in $\BKstarmumu$ has been found to deviate slightly from the
SM predictions.  If this is in fact due to the presence of NP, such NP
should contribute to all the other decays involving the effective
transition $\bsmumu$.  The effects of this NP on these decay modes
would be correlated, and hence a combined analysis of all these decay
modes would be invaluable in discerning the type of NP present.

While specific models of NP may be used and their effect on the
relevant observables studied, we have chosen to explore the NP in a
model-independent way, in terms of the Lorentz structures of the NP
operators that contribute to the effective $\bsmumu$ Hamiltonian.  We
have performed a general analysis that includes NP vector-axial vector
(VA), scalar-pseudoscalar (SP), and/or tensor (T) operators.  We have
computed the effects of such NP operators, individually and in all
combinations, on these decays.  We have taken the couplings to be real
and have considered the CP-conserving observables in this paper; the
CP-violating observables are discussed in Ref.~\cite{CPviol}.  The aim
is to find NP signals, and using them, to identify the Lorentz
structure of the NP.  As the first step towards this goal, we
calculate the constraints on the NP couplings, and, keeping the
couplings within these bounds, we look for the observables where the
NP signal can potentially stand out above the SM background.

It is crucial to understand this SM background, which makes it
imperative to use observables whose values are predicted reasonably
accurately within the SM.  The main source of the SM uncertainties is
the hadronic matrix elements, whose theoretical calculations often
have errors of the order of tens of percent.  We have handled this on
many levels.  First, we have tried to identify observables that will
not be very sensitive to the hadronic uncertainties.  For example in
$\BKmumu$, the SM prediction for the forward-backward asymmetry is
simply zero, independent of any hadronic elements.  Also, while the
differential branching ratios may be strongly dependent on the
hadronic matrix elements, the forward-backward asymmetries are less
so.  Furthermore, the large-energy effective theory (LEET) limits can
be used to control the uncertainties in the low-$q^2$ region for
observables like $A_{FB}$ and $A_T^{(2)}$.  For example, certain
observables, such as the zero-crossing of $A_{FB}$ in $\BKstarmumu$,
can be shown to be robust under form-factor uncertainties in the LEET
limit.  The longitudinal-transverse asymmetry $A_{LT}$ in $\BKstarmumu$ 
also has a zero crossing in the SM with small
hadronic uncertainties.  These measurements can even be used to
extract the parameters of the NP operators, to a very good
approximation.

Also, we focus only on the situations where the NP contribution can be
so significant that it will stand out even if the SM errors were
magnified.  Our figures show bands for SM predictions that include the
form-factor uncertainties as quoted in the form-factor calculations,
and these are overlaid with some examples of the allowed values of
these observables when NP contributions are included.  This allows the
scaling of these uncertainties to be easily visualized.  We identify
and emphasize only those situations where the results with the NP can
be significantly different from those without the NP, even if the
hadronic uncertainties were actually much larger.  Note that further
inclusion of the NLO QCD corrections would affect the central values
of the SM predictions to a small extent, while also decreasing the
renormalization scale uncertainty. However, since our primary interest
is looking for observables where the NP effects are large, a LO
analysis is sufficient.

\afterpage{\clearpage}

\TABLE[!htb]{
{\footnotesize 
\begin{tabular}{p{2.6cm}|p{2.5cm}|p{2.7cm}|p{2.6cm}|p{2.6cm}}
\hline
Observable & SM & {Only new VA} & {Only new SP} & {Only new T} \\
\hline
$\Bsmumu$ & & & & \\
\hfill BR & $(3.35 \pm 0.32) \times 10^{-9}$ &
$\bullet$ Marginal E \newline $\bullet$ Significant S
& $\bullet$ Large E \newline
$\bullet$ Maximal S
& No effect \\
\hline
$\BXsmumu$ & & & & \\
\hfill DBR & & $\bullet$ E ($\times 2$) \newline
$\bullet$ S ($\div 2$)
& $\bullet$ Marginal E  & $\bullet$ E ($\times 2$) \\
 & & & & \\
\hfill $\AFB$ & ZC$\approx 3.5$ GeV$^2$ &
$\bullet$ E(30\%) low $q^2$ \newline
$\bullet$ ZC shift / \newline disappearence
& $\bullet$ Marginal S & $\bullet$ Marginal S \\
 & & & & \\
\hfill $f_L$ & $\bullet$ $0.9 \to 0.3$ \newline
(low$\to$high $q^2$)
& $\bullet$ Large S at low $q^2$
& $\bullet$ Marginal S  & $\bullet$ Marginal E \\
\hline
$\Bsmumugamma$ & & & & \\
\hfill DBR &
& $\bullet$ E ($\times 2- \times 3$) \newline $\bullet$ S (low $q^2$)
& No effect & $\bullet$ E ($\times 3$) \\
 & & & & \\
\hfill $\AFB$ & ZC$\approx 4.3$ GeV$^2$
& $\bullet$ ZC shift / \newline disappearence
& No effect & $\bullet$ Large S \\
\hline
$\BKmumu$ & & & & \\
\hfill DBR &
& $\bullet$ E ($\times 2$) \newline $\bullet$ Marginal S
& $\bullet$ E at high $q^2$ & $\bullet$ Small effect \\
 & & & & \\
\hfill $\AFB$ & Vanishes
& $\bullet$ No effect
& $\bullet$ E at low $q^2$ \newline
$\bullet$ No ZC  & $\bullet$ E at high $q^2$ \newline
$\bullet$ No ZC \\
\hline
$\BKstarmumu$ & & & & \\
\hfill DBR &
& $\bullet$ E ($\times 2$) \newline $\bullet$ S ($\div 2$)
& No effect  & $\bullet$ E ($\times 2$) \\
 & & & & \\
\hfill $\AFB$ & ZC$\approx 3.9$ GeV$^2$
& $\bullet$ E at low $q^2$ \newline
$\bullet$ ZC shift / \newline disappearence
& No effect & $\bullet$ Significant S \newline
$\bullet$ ZC shift \\
 & & & & \\
\hfill $f_L$ & $\bullet$ $0.9 \to 0.3$ \newline
(low$\to$high $q^2$)
& $\bullet$ Large S
& No effect  & $\bullet$ Significant S \\
 & & & & \\
\hfill $A_T^{(2)}$ & $\bullet$ $\uparrow$ with $q^2$ \newline
$\bullet$ No ZC
& $\bullet$ E ($\times 2$) \newline $\bullet$ ZC possible
& No effect  & $\bullet$ Significant S \\
 & & & & \\
\hfill $A_{LT}$ & $\bullet$ ZC at low $q^2$ \newline
$\bullet$ more -ve \newline at large $q^2$
& $\bullet$ Significant S \newline $\bullet$
 ZC shift / \newline disappearence
& No effect & $\bullet$ Significant S \\
\hline
\end{tabular}
}
\caption{The effect of NP couplings on observables.
E($\times n$): enhancement by up to a factor of $n$,
S($\div n$): suppression by up to a factor of $n$,
ZC: zero crossing.
\label{tab:summary}}
}

Our results are summarized in Table~\ref{tab:summary}, for the cases
where the NP has only one type of Lorentz structure: VA, SP or T.  We
note certain generic features of the influence of different NP Lorentz
structures.

New VA operators are the ones that influence the observables strongly
in most cases. They typically can interfere with the SM terms
constructively or destructively, thus enhancing or suppressing the
differential branching ratios by up to factors of 2 or 3.  They also
are able to enhance almost all the asymmetries, the notable exception
being $A_{FB}$ in $\BKmumu$, where the VA operators cannot
contribute. But for most other observables, this kind of NP can
potentially be observed. This can be traced to the large magnitudes of
the NP couplings still allowed by data, which in turn can be traced to
the possibility of interference between the new VA operators with the
SM operators that allows more freedom for the new VA couplings.
Typically, the $R_{V,A}$ couplings are constrained more weakly than
the $R'_{V,A}$ couplings, since the corresponding operators have the
same structure as those of the SM, allowing strong destructive
interferences.  Consequently, the operators with $R_{V,A}$ couplings
are more likely to show themselves over and above the SM
background. We point out that the exception to this rule is the
$A_{FB}$ in $\BKstarmumu$ at large $q^2$, where the $R'_{V,A}$
couplings can cause a larger enhancement.

The SP operators, on the other hand, are handicapped by the stringent
constraints from the upper bound on $B(\Bsmumu)$. If only $R_{S,P}$ or
$R'_{S,P}$ couplings are present, the constraints become even more
severe. It is for this reason that, even when the SP contributions are
unsuppressed by $m_\mu/m_b$, they are not often large enough to stand
apart from the SM background.

The couplings of the T operators, viz.\ $C_T$ and $C_{TE}$, are not as
suppressed as those of the SP operators. Therefore, they typically
contribute significantly to the DBRs. However, the interference terms
of these operators with the SM operators often suffer from the
$m_\mu/m_b$ helicity suppression, and hence they tend to suppress the
magnitudes of the asymmetries.

The combination of multiple Lorentz structures in general gives rise
to the combination of features of the individual Lorentz structures
involved.  In particular, if the VA operators appear in conjunction
with another Lorentz structure, the effects of the VA operators
typically dominate.  The T operators can interfere with the SP
operators without the $m_\mu/m_b$ helicity suppression, but the strong
constraints on the SP operators hold them back.  A remarkable
exception is the combination of SP and T operators in the
forward-backward asymmetry in $\BKmumu$. This asymmetry, which
vanishes in the SM, can be enhanced to $\sim 5 \%$ at low $q^2$ with
only SP operators, and can be enhanced to $\sim 30\%$ with T operators
but only at $q^2 \approx m_B^2$. However, the presence of both SP and
T operators allows the asymmetry to be $\sim 40\%$ in the whole
high-$q^2$ region.  A similar feature, though to a less-spectacular
extent, is observed in $\AFB$ of $\BKstarmumu$ \cite{AFBNP}.

With the large amount of data expected from the LHC experiments and
$B$-factories in the coming years, we may be able to detect confirmed
NP signals in the above processes.  In that case, a combined analysis
of all these decay modes, as carried out in this paper, would enable
us to identify the Lorentz structure of the NP operators.  This will
be important in establishing precisely what type of NP is present.

\bigskip
\noindent
{\bf Acknowledgments}: We thank Gagan Mohanty and Zoltan Ligeti for
useful comments, and S. Uma Sankar and Alejandro Szynkman for helpful
collaboration on several parts of this analysis. M.D. would like to
thank Wolfgang Altmannshofer for useful discussions. This work was
financially supported by NSERC of Canada (AKA, DL).

\bigskip
\noindent
{\bf Notes added}: After this paper was submitted, the CDF
Collaboration reported \cite{CDFmeas} the measurement of
\beq
B(\Bsmumu) = (1.8^{+1.1}_{-0.9}) \times 10^{-8} ~.
\eeq
On the other hand, the recent LHCb update does not confirm this result
\cite{LHCbupdate}. They improve the present upper bound on
$B(\Bsmumu)$ to
\beq
B(\Bsmumu) \le 1.3 \times 10^{-8} ~~~{\hbox{(90\% C.L.)}}
\eeq
In addition, LHCb has measured various observables in $\BKstarmumu$
\cite{LHCbupdate2}. Their measurement of the $\AFB$ distribution is
consistent with the SM prediction, except in the high-$q^2$ region,
where we now see a slight suppression.  This is contrary to the
measurement of Belle. That is, LHCb does not confirm the Belle result
of a large FB asymmetry in the low-$q^2$ region.  Thus, the jury is
still out on whether NP has already been seen in these measurements.


\appendix
\section{\boldmath Details of the $\BXsmumu$ analysis}
\label{app-bxsmumu}

The differential branching ratio for $\BXsmumu$ in SM can be written as
\begin{eqnarray}
\Bigg(\frac{\text{d}{B}}{\text{d}z}\Bigg)_{\text{SM}}&=&
B_{0} \frac{8}{3} (1-z)^2 \sqrt{1-\frac{4 t^2}{z}} \times \nonumber \\
&&\Bigg[(2 z+1) \left(\frac{2 t^2}{z}+1\right) |C_9^{\text{eff}}|^2
+ \left(\frac{2 (1-4 z) t^2}{z}+ (2z+1)\right) |C_{10}^{\text{eff}}|^2
\phantom{space} 
\nonumber \\
&&+~4 \left(\frac{2}{z}+1\right) \left(\frac{2 t^2}{z}+1\right)
|{C}_7^{\text{eff}}|^2
+ 12 \left(\frac{2 {t}^2}{{z}}+1 \right)
\text{{\rm Re}}({C}_{7}^{\text{eff}} {C}_{9}^{\text{eff}*})\Bigg], 
\end{eqnarray}
Here ${t} \equiv {m}_\mu/{m}_{{b}}^{\text{pole}}$ and
${z} \equiv {q}^2/({m}_{{b}}^{\text{pole}})^2$.
The normalization constant ${B}_0$ is \cite{Ali:1996bm}
\begin{equation}
{B}_0= \frac{3 \alpha_{em}^2 \, {B}({\bar {{B}}}\rightarrow {X}_{c}
{e} {\bar \nu})}
 {32 \pi^2 \, f(\hat{{m}}_{{c}}) \, \kappa(\hat{{m}}_{c})}
 \frac{|{V}_{{tb}}{V}_{{ts}}^{*}|^2}{|{V}_{{cb}}|^2}\;,
\end{equation}
where $\hat{{m}}_{c} \equiv {{m}}_{{{c}}}^{\text{pole}}/
{m}_{{b}}^{\text{pole}}$. We use $\hat{m}_{c}=0.29\pm0.02$
\cite{Ali:2002jg}, $B(\bar{B}\rightarrow X_c e \bar{\nu})=0.1061\pm
0.0017$ \cite{pdg} and $|V_{tb}V^*_{ts}|/|V_{cb}|=0.967\pm 0.009$
\cite{Charles:2004jd}.  Here $f(\hat{{m}}_{{c}})$ is the lowest-order
(i.e.\ parton-model) phase-space factor in ${B}({\bar
  {{B}}}\rightarrow {X}_{c} {e} {\bar \nu})$:
\bea
f(\hat{{m}}_{{c}}) = 1 - 8\hat{{m}}^2_c + 8\hat{{m}}_c^6 - \hat{{m}}_c^8
- 24\hat{{m}}_c^4 \ln \hat{{m}}_c \; ,
\eea
and the function $\kappa(\hat{{m}_{c}})$ includes both the $O(\alpha_{s})$ QCD
corrections and the leading-order $(1/{m}_{b}^2)$ power correction to
${B}({\bar {{B}}}\rightarrow {X}_{c} {e} {\bar \nu})$ :
\begin{eqnarray}
\kappa(\hat{{m}}_{{c}}) = 1 - \frac{2 \alpha_{s}({m}_{b})}{3 \pi} g(\mc)
+ \frac{h(\mc)}{2 {m}_{{b}}^2} \; .
\end{eqnarray}
Here the two functions are
\begin{eqnarray}
g(\mc) &=& (\pi^2-\frac{31}{4})(1-\mc)^2 + \frac{3}{2} \; , \nonumber\\
h(\mc) &=& \lambda_1 + \frac{\lambda_2}{f(\mc)} \left[ -9 +24 \mc^2
-72\mc^4 + 72\mc^6 -15\mc^8 -72 \mc^4 \ln \mc \right].
\end{eqnarray}
After including all the NP interactions, and neglecting terms suppressed
by $m_\mu/m_b$ and $m_s/m_b$,
the total differential branching ratio $\text{d}{B}/{\text{d}z}$
can be written in the form
\bea
\Bigg(\frac{\text{d}{B}}{\text{d}z}\Bigg)_{\text{Total}}~=~
\Bigg(\frac{\text{d}{B}}{\text{d}z}\Bigg)_{\text{SM}} + {B}_0
\Bigg[{B}_{SM{\hbox{-}}VA} +  {B}_{VA} +  {B}_{SP} + {B}_{T}\Bigg] \; ,
\eea

where
\bea
{B}_{SM{\hbox{-}}VA} &~=~& \frac{16}{3}(1-z)^2(1+2z)
\left[{\rm Re}(C_9^{\rm eff}R^*_V)
+ {\rm Re}(C_{10} R_A^*) \right]
\label{bsmva}
\nn\\ &&
+~32\,(1-z)^2\,{\rm Re}(C_7^{\rm eff} R_V^*)\;,
\\
{B}_{VA} &~=~& \frac{8}{3}(1-z)^2(1+2z)\Big[ |R_V|^2 + |R_A|^2 +
|{R'_V}|^{\hskip-0.2truemm 2} + |{R'_A}|^{\hskip-0.2truemm 2}\Big]\;,
\\
{B}_{SP} &~=~& 4\,(1-z)^2\,z\Big[ |R_S|^2 + |R_P|^2 + |{R'_S}|^{\hskip-0.2truemm 2} + |{R'_P}|^{\hskip-0.2truemm 2}\Big]\;,
\\
{B}_{T} &~=~& \frac{128}{3}(1-z)^2(1+2z)\Big[ |C_T|^2 + 4|C_{TE}|^2\Big]\;.
\eea
Note that here we have separated the contribution of the SM VA
operators (subscript $SM{\hbox{-}}VA$) from that of the NP VA operators
(subscript $VA$), for clarity.

The forward-backward asymmetry in $\BXsmumu$ is
\bea
\label{AFB-Xsmumu}
A_{FB}(q^2) &=&\frac{\int^1_0 d\cos{\theta_\mu}
\frac{d^2B}{dq^2d\cos{\theta_\mu}  }-\int^0_{-1}
d\cos{\theta_\mu}\frac{d^2B}{dq^2d\cos{\theta_\mu}  }}
{\int^1_0 d\cos{\theta_\mu} \frac{d^2B}{dq^2d\cos{\theta_\mu}  }
+\int^0_{-1} d\cos{\theta_\mu}\frac{d^2B}{dq^2d\cos{\theta_\mu}  }} \; ,
\eea
where $\theta_\mu$ is the angle between the $\mu^+$ and the $\bar{B}^0$ in the
dimuon center-of-mass frame.
We can write $\AFB$ in the form
\begin{equation}
A_{FB}(q^2)=\frac{N(z)}{\text{d} {B}/ \text{d}z}\;,
\end{equation}
where the numerator is given by
\bea
N(z) &~=~& B_0
\Bigg[N_{SM} + N_{SM{\hbox{-}}VA} +  N_{VA} + N_{SP{\hbox{-}}T}
\Bigg] \;,
\label{N-Xsmumu}
\eea
with
\bea
N_{SM} &~=~& -8\,C_{10}\,(1-z)^2\Big[2C^{\rm eff}_7 + z \, {\rm Re}(C_9^{\rm eff})\Big]\;,
\\
N_{SM{\hbox{-}}VA} &~=~& -8\,(1-z)^2\Big[z\,{\rm Re}\Big(C_{10}R^*_V+C_9^{\rm eff}R^*_A\Big)+2C^{\rm eff}_7{\rm Re}(R^*_A)\Big]\;,
\\
N_{VA} &~=~& -8\,z\,(1-z)^2\Big[{\rm Re}(R_V R^*_{A})-{\rm Re}(R'_V{R'_A}^{\hskip-1.2truemm *}) \Big]\;,
\\
N_{SP{\hbox{-}}T} &~=~& -8\,z\,(1-z)^2\Big[{\rm Re}\Big\{(R_S+R_P)\,(C^*_{T}-2C^*_{TE})\Big\}
\nn\\ &&
+~ {\rm Re}\Big\{(R'_S-R'_P)\,(C^*_{T}+2C^*_{TE})\Big\}\Big]
\label{nspt}\;.
\eea
The expressions of Eqs.~(\ref{bsmva})-(\ref{nspt}) are in agreement
with Ref.~\cite{Fukae:1998qy}.

The polarization fractions $f_L$ and $f_T$ are defined as
\be
f_L=\frac{H_L(z)}{H_L(z)+H_T(z)} ~, \qquad \qquad 
f_T=\frac{H_T(z)}{H_L(z)+H_T(z)} ~,
\ee
where 
\begin{equation}
H_L(z)=H^{SM}_L(z)+H^{SM-VA}_L(z)+H^{VA}_L(z)+H^{SP}_L(z)+H^{T}_L(z)\;,
\end{equation}
\begin{equation}
H_T(z)=H^{SM}_T(z)+H^{SM-VA}_T(z)+H^{VA}_T(z)+H^{SP}_T(z)+H^{T}_T(z)\;.
\end{equation}
The components of $H_L$ and $H_T$ functions are 
\bea
H^{SM}_L(z) &~=~& \frac{8\,B'_0}{3} (1-z)^2 \Big[ \left|C^{\rm eff}_9+2C^{\rm eff}_7\right|^2 + |C_{10}|^2\Big]\;,
\\
H^{SM}_T(z) &~=~& \frac{16\,B'_0}{3} z(1-z)^2 \Big[ \left|C^{\rm eff}_9+\frac{2}{z}C^{\rm eff}_7\right|^2 + |C_{10}|^2\Big]\;,
\\
H^{SM{\hbox{-}}VA}_L(z) &~=~& \frac{16\,B'_0}{3} (1-z)^2 \Big[ {\rm Re}\left(C^{\rm eff}_9\,R^*_V + C_{10} R^*_A\right) + 2{\rm Re}(C^{\rm eff}_7\,R^*_V )\Big]\;,
\\
H^{SM{\hbox{-}}VA}_T(z) &~=~& \frac{32\,B'_0}{3} (1-z)^2 \Big[ z{\rm Re}\left(C^{\rm eff}_9\,R^*_V + C_{10} R^*_A\right) + 2{\rm Re}(C^{\rm eff}_7\,R^*_V )\Big],
\\
H^{VA}_L(z) &~=~& \frac{8\,B'_0}{3} (1-z)^2 \Big[ |R_V|^2 + |R_A|^2 +|{R'_V}|^{\hskip-0.2truemm 2} + |{R'_A}|^{\hskip-0.2truemm 2}\Big]\;,
\\
H^{VA}_T(z) &~=~& \frac{16\,B'_0}{3} z(1-z)^2 \Big[ |R_V|^2 + |R_A|^2 +|{R'_V}|^{\hskip-0.2truemm 2} + |{R'_A}|^{\hskip-0.2truemm 2}\Big]\;,
\\
H^{SP}_L(z) &~=~& \frac{4\,B'_0}{3} z(1-z)^2 \Big[ |R_S|^2 + |R_P|^2 +|{R'_S}|^{\hskip-0.2truemm 2} + |{R'_P}|^{\hskip-0.2truemm 2}\Big]\;,
\\
H^{SP}_T(z) &~=~& \frac{8\,B'_0}{3} z(1-z)^2 \Big[ |R_S|^2 + |R_P|^2 +|{R'_S}|^{\hskip-0.2truemm 2} + |{R'_P}|^{\hskip-0.2truemm 2}\Big]\;,
\\
H^{T}_L(z) &~=~& \frac{64\,B'_0}{3} (2-z)(1-z)^2 \Big[ |C_T|^2 + 4|C_{TE}|^2\Big]\;,
\\
H^{T}_T(z) &~=~& \frac{128\,B'_0}{3} z(1-z)^2 \Big[ |C_T|^2 + 4|C_{TE}|^2\Big]\;.
\eea

\section{\boldmath Details of the $\Bsmumugamma$ analysis}
\label{app-Bsmumugamma}

The transition amplitude for $\Bsmumugamma$ is
\barr
& & i{\cal M}(\bsbar \to \mu^+ \mu^- \gamma)  =
(-i)\frac{1}{2}\Bigg[-\frac{4 G_F}{\sqrt{2}}
\frac{\alpha_{em}}{4 \pi} (V_{ts}^* V_{tb})\Bigg] \times \nn \\
&\Bigg\{ &
\langle \gamma(k) | \bar{s} \gamma_\mu b | \bsbar (p_B) \rangle
\left[ (C_9^{\rm eff} + R_V+R'_V) L^\mu +
(C_{10} + R_A+R'_A) L^{\mu 5} \right] \nn \\
&& + \left< \gamma(k) | \bar{s} \gamma_\mu \gamma_5 b | \bsbar (p_B) \right>
\left[-(C_9^{\rm eff} + R_V-R'_V) L^\mu -
(C_{10} + R_A-R'_A) L^{\mu 5} \right] \nn \\
&& +
\left< \gamma(k) | \bar{s} i \sigma_{\mu \nu} q^\nu b | \bsbar (p_B) \right>
[-2 m_b \frac{C_7^{\rm eff}}{q^2} L^\mu]  \nn \\
&& + \left< \gamma(k) | \bar{s} i \sigma_{\mu \nu} \gamma_5 q^\nu b | \bsbar (p_B) \right>
[-2 m_b \frac{C_7^{\rm eff}}{q^2} L^\mu]  \nn \\
&& +  \left< \gamma(k) | \bar{s} \sigma_{\mu \nu} b | \bsbar (p_B) \right>
[  2C_T L^{\mu\nu}+  2 i C_{TE} \epsilon^{\mu\nu\alpha\beta}
L_{\alpha \beta} ] \;  \Bigg\} ~,
\label{M-mumugamma}
\earr
where the $L$'s are defined in Eq.~(\ref{Ldefs}).

In order to calculate the DBR, one needs the $\bsbar \to \gamma$
matrix elements and form factors. The matrix elements are given in
Ref.~\cite{Kruger:2002gf}\footnote{We use the convention
  $\epsilon^{0123}= + 1$.}:
\begin{equation}
\label{ff:bsg}
\begin{split}
\left< \gamma(k) | \bar{s} \gamma_\mu b | \bar \bs (p_B) \right>
&=
- e \, \epsilon_{\mu \nu \rho \sigma} \varepsilon^{*\nu} q^\rho k^\sigma \frac{f_V( q^2)}{m_{B_s}}\, , \\
\left< \gamma(k) | \bar{s} \gamma_\mu \gamma_5 b | \bar \bs (p_B) \right>
&=
i e \biggl[ \varepsilon_\mu^* k \cdot q - \varepsilon^* \cdot q k_\mu \biggr] \frac{f_A(q^2)}{m_{B_s}}\,, \\
\left< \gamma(k) | \bar{s} i \sigma_{\mu \nu} q^\nu b | \bar \bs (p_B) \right>
&=
e \, \epsilon_{\mu \nu \rho \sigma} \varepsilon^{* \nu} q^\rho k^\sigma f_{TV}(q^2)\, ,\\
\left< \gamma(k) | \bar{s} i \sigma_{\mu \nu} \gamma_5 q^\nu b | \bar \bs (p_B) \right>
&=
i e \biggl[ \varepsilon^*_\mu k \cdot q - \varepsilon^* \cdot q k_\mu \biggr] f_{TA} (q^2)\,,
\end{split} \nn
\end{equation}
\vspace{-4mm}
\bea
\left< \gamma(k) | \bar{s} \sigma_{\mu \nu} b | \bar \bs (p_B) \right>
&=&
-i e \, \epsilon_{\mu \nu \rho \sigma} \Big[  \frac{\left\{f_{TV}(q^2)-f_{TA}(q^2)\right\}}{q^2} \Big\{ (q \cdot k)\, \varepsilon^{*\rho}\,q^\sigma + (\varepsilon^{*} \cdot q) \, q^\rho\, k^\sigma \Big\}
\nonumber \\
&-& f_{TV}(q^2)\,\varepsilon^{*\rho}\,k^\sigma\Big]\;.
\eea
Here $\varepsilon_\mu$ is the four-vector polarization of the photon
and $q=p_B-k$.  For the $\Bsmumugamma$ form factors $f_i ~(i =
V,A,TA,TV)$, we use the parameterization \cite{Kruger:2002gf}
\begin{equation}
f_{i}(q^{2})=\beta_{i}\frac{f_{B_{s}}m_{B_{s}}}{
\Delta_{i}+0.5 m_{B_s}
\left(1-q^2/m_{B_s}^2\right)} ~,
\label{ff}
\end{equation}
where the parameters $\beta_i$ and $\Delta_i$ are given in
Table~\ref{ffparams}. These values of parameters ensure that
the large energy effective theory (LEET) relations
between form factors are satisfied to a 10\% accuracy \cite{Kruger:2002gf}.
In our numerical analysis we take the errors in these form factors 
to be $\pm 10\%$.

\TABLE[hbt]{
\begin{tabular}{|c|c|c|c|c|}
\hline
Parameter&
$f_{V}$&
$f_{TV}$&
$f_{A}$&
$f_{TA}$\tabularnewline
\hline
\hline
~~$\beta(\gev^{-1})~~$&
~~0.28~~&
~~0.30~~&
~~0.26~~&
~~0.33~~ \tabularnewline
\hline
$\Delta(\gev)$&
0.04&
0.04&
0.30&
0.30\tabularnewline
\hline
\end{tabular}
\label{ffparams}
\caption{The parameters for $\bsbar \to \gamma$ form factors, as
defined in Eq.~(\ref{ff}). }
}

In terms of the dimensionless parameter $x_\gamma=2 E_\gamma/m_{B_s}$,
where $E_\gamma$ is the photon energy in the $\bsbar$ rest frame,
one can calculate the double differential decay rate to be
\begin{eqnarray}
\frac{\text{d}^2\Gamma}{\text{d}x_{\gamma}\text{d}(\cos\theta_\mu)} =
\frac{1}{2 m_{B_s}} \dfrac{2 v\, m_{B_s}^2 x_{\gamma}}{(8\pi)^3}
{\cal M}^{\dagger}{\cal M} \; ,
\label{ddbr-mumugamma}
\end{eqnarray}
where
$v \equiv \sqrt{1- 4 m_{\mu}^2/[m_{B_s}^2(1-x_\gamma)]}$.
From Eq.~(\ref{ddbr-mumugamma}) we get the DBR to be
\begin{eqnarray}
\frac{\text{d}B}{\text{d} x_{\gamma}} &=&
\tau_{B_s}\int_{-1}^{1} \frac{\text{d}^2\Gamma}
{\text{d} x_{\gamma}\text{d}(\cos\theta_\mu)}\,
\text{d}\cos\theta_\mu  \nn \\
&=&\tau_{B_s}\Bigg[\frac{1}{2 m_{B_s}}
\dfrac{2 v m_{B_s}^2}{(8\pi)^3}\Bigg]
\Bigg[\frac{1}{4} ~ \frac{16 G_F^2}{2}
\frac{\alpha_{em}^2}{16 \pi^2} |V_{tb}V_{ts}^*|^2 e^2\Bigg] \Theta \; .
\label{dbr:bsg:main}
\end{eqnarray}
Here the quantity $\Theta$ has the form
\beq
\Theta = \frac{2}{3}~m_{B_s}^4~x^3_{\gamma}
\Big[X_{VA}+X_{T}+X_{VA{\hbox{-}}T}\Big] \; ,
\label{mumugamma-theta-expansion}
\eeq
where the $X$ terms are
\begin{eqnarray}
X_{VA}&=&
\Big(|A|^2+|B|^2\Big)
m_{B_s}^2 \left(3-v^2\right)(1-x_{\gamma})+
\Big(|C|^2+|D|^2\Big)
2 m_{B_s}^2 v^2 (1-x_{\gamma}) \; , \nn \\
X_T & = & 4 |E|^2 (3-v^2) + 4|F|^2 m_{B_s}^4 v^2 (1-x_{\gamma})^2 \nn \\
&&+ 16 |G|^2 \left(3-v^2\right) +16 |H|^2 m_{B_s}^4 \left(3 - 2 v^2\right)
(1-x_{\gamma})^2  \nn \\
&& + 8 m_{B_s}^2 v^2 (1-x_{\gamma})\text{{\rm Re}}\left(E^* F\right)
+ 32 m_{B_s}^2 (3-2v^2)(1-x_{\gamma}) \text{{\rm Re}}\left(G^* H\right) \; ,\nn \\
X_{VA{\hbox{-}}T} & = & -24 m_{\mu}\text{{\rm Re}}\left(A^* E\right)
- 48 m_{\mu}\text{{\rm Re}}\left(B^* G\right)
- 48  m_{\mu} m_{B_s}^2 (1-x_{\gamma})\text{{\rm Re}}\left(B^*H\right) \, .
\end{eqnarray}
Note that here, the $VA$ subscript includes the SM operators.
The parameters $A$--$H$ are combinations of the Wilson coefficients,
form factors and NP parameters, and are given by
\bea
A &=& (C^{\rm eff}_9 +R_V+R'_V) \, \frac{f_V(q^2)}{m_{B_s}}  + \frac{2 \,m_b C^{\rm eff}_7}{q^2} \, f_{TV}(q^2) \,,\nonumber \\
B &=& (C^{\rm eff}_9 + R_V - R'_V) \, \frac{f_A(q^2)}{m_{B_s}}  + \frac{2 \,m_b C^{\rm eff}_7}{q^2} \, f_{TA}(q^2) \,,\nonumber  \\
C &=& (C^{\rm eff}_{10} +R_A+R'_A) \, \frac{f_V(q^2)}{m_{B_s}} \,,  \nonumber \\
D &=& (C^{\rm eff}_{10} +R_A-R'_A) \, \frac{f_A(q^2)}{m_{B_s}} \,,  \nonumber \\
E &=& -2C_T  f_{TV}(q^2)\,,\nonumber \\
F &=& 2C_T \frac{f_{TV}(q^2)-f_{TA}(q^2)}{q^2} \,,\nonumber \\
G &=& -2C_{TE}  f_{TV}(q^2)\,,\nonumber \\
H &=& 2C_{TE} \frac{f_{TV}(q^2)-f_{TA}(q^2)}{q^2}~~. \\ \nonumber
\eea

The normalized forward-backward asymmetry of muons in
$\Bsmumugamma$ is defined as
\beq
\AFB(q^2) = \frac {\displaystyle \int_{0}^{1} d\cos\theta_\mu
\frac{d^2B}{dq^2 d\cos\theta_\mu} - \int_{-1}^{0} d\cos\theta_\mu
\frac{d^2B}{dq^2 d\cos\theta_\mu} }{\displaystyle \int_{0}^{1}
d\cos\theta_\mu \frac{d^2B}{dq^2 d\cos\theta_\mu} + \int_{-1}^{0}
d\cos\theta_\mu \frac{d^2B}{dq^2 d\cos\theta_\mu} } ~,
\eeq
where $\theta_\mu$ is the angle between the three-momentum vectors of the
$\bsbar$ and the $\mu^+$ in the dimuon center-of-mass frame. The
calculation of $\AFB$ gives
\begin{eqnarray}
A_{FB}(q^2) &=& \frac{1}{\Theta}~
\Bigg(2 m^4_{B_s} v ~ x_{\gamma}^3\Bigg)\Bigg[ Y_{VA} + Y_{VA{\hbox{-}}T} \Bigg] \; ,
\end{eqnarray}
with the $Y$ terms given by
\begin{eqnarray}
Y_{VA} & = & \Big(\text{{\rm Re}}\left(A^* D\right)+
\text{{\rm Re}}\left(B^* C\right)\Big)m_{B_s}^2 (1 - x_{\gamma})\; ,\nn \\
Y_{VA{\hbox{-}}T} & = & - 4 m_{\mu} \Bigg( 2 \text{{\rm Re}}\left(C^* G\right)
+ 2 m_{B_s}^2 (1 - x_{\gamma})\text{{\rm Re}}\left(C^* H\right)
+ \text{{\rm Re}}\left(D^* E\right) \Bigg)  \;.\\ \nn
\label{afb:bsg}
\end{eqnarray}

\section{\boldmath Details of the $\BKmumu$ analysis}
\label{app-bkmumu}

The transition matrix element for $\BKmumu$ is given by
\begin{eqnarray}
&& i {\cal M}\,(\BKmumu) =  (-i)~\frac{1}{2}~\Bigg[-\frac{4~G_F}{\sqrt{2}}
\frac{\alpha_{em}}{4 \pi} (V_{ts}^* V_{tb})\Bigg] \times
\nonumber \\
& \Bigg\{ &
\left< K(p_2) \left|\bar{s}\gamma_{\mu}b\right|B(p_1)\right>
[ (C_{9}^{\rm eff}+R_V+R'_V)  L^\mu
+ (C_{10} + R_A + R'_A)  L^{\mu 5} ]
\nonumber \\
&& + \left< K(p_2)\left|\bar{s}b\right|B(p_1)\right>
[(R_S+R'_S) L + (R_P+R'_P) L^5] \nonumber\\
&&
+ \left< K(p_2)\left|\bar{s}i\sigma_{\mu\nu}q^{\nu}b\right|B(p_1)\right>
[- 2 C^{\rm eff}_7 (m_b/q^2)  L^\mu]
\nonumber \\
&& + \left< K(p_2)\left|\bar{s}\sigma_{\mu\nu}b\right|B(p_1)\right>\;
[2 C_T L^{\mu\nu} +  2iC_{TE}\epsilon^{\mu \nu \alpha\beta}
L_{\alpha \beta} ] \; \Bigg\} \; ,
\label{matrix}
\end{eqnarray}
where the $L$'s are defined in Eq.~(\ref{Ldefs}).

The $\bdbar \to \Kbar$ matrix elements needed to calculate the
decay rate and asymmetry in $\BKmumu$ are \cite{ali-00}
\begin{eqnarray}
\left< \bar{K}(p_2) \left|\bar{s}\gamma_{\mu}b\right|\bdbar(p_1)\right> &=&
(2p_1-q)_{\mu}f_{+}(z)+(\frac{1-k^2}{z})\, q_{\mu}[f_{0}(z)-f_{+}(z)]\;, \nn \\
\left< \bar{K}(p_2)\left|\bar{s}i\sigma_{\mu\nu}q^{\nu}b\right|
\bdbar(p_1)\right> &=& -\Big[(2p_1-q)_{\mu}q^2-(m_{B}^{2}-m_{K}^{2})
q_{\mu}\Big]\,\frac{f_{T}(z)}{m_B+m_{K}}\;, \nn \\
\left< \bar{K}(p_2)\left|\bar{s}b\right|\bdbar(p_1)\right> &=&
\frac{m_B(1-k^2)}{\hat{m}_b}f_0(z)\;, \nn \\
\left< \bar{K}(p_2)\left|\bar{s}\sigma_{\mu\nu}b\right|\bdbar(p_1)\right>
&=& i\Big[(2p_1-q)_{\mu}q_{\nu}-(2p_1-q)_{\nu}q_{\mu}\Big]\,
\frac{f_T(z)}{m_B+m_{K}}\; ,
\end{eqnarray}
where $k \equiv m_K/m_B$, $\hat{m}_b \equiv m_b/m_B$, $q_\mu =
(p_1-p_2)_\mu = (p_+ + p_-)_\mu$, and $z \equiv q^2/m^2_{B}$.  
The form factors $f_{+,\,0,\,T}$ were calculated in the framework of QCD
light-cone sum rules in Ref.~\cite{ali-00}.  The $z$ dependence
of these is parametrized by
\begin{eqnarray}
f(z)=f(0)\,\exp(c_1z+c_2z^2+c_3z^3)\;,
\end{eqnarray}
where the parameters $f(0), c_1$, $c_2$ and $c_3$ for each form factor
are taken from Tables III, IV and V of Ref.~\cite{ali-00}.
Using these, the differential branching ratio is given by
\begin{eqnarray}
\frac{dB}{dz} & = & B'_0\, \phi^{1/2}\,\beta_\mu
\Bigg[X'_{VA} + X'_{SP} + X'_T + X'_{VA{\hbox{-}}SP} + X'_{VA{\hbox{-}}T} \Bigg] \; ,
\end{eqnarray}
where $B_0'$ is the normalization factor:
\beq
B'_0  = \frac{G_F^2\alpha^2 \tau_B}{2^{12}\pi^5}|V_{tb}V^*_{ts}|^2m_B^5\;,
\eeq
the phase factor $\phi$ is
\beq
\phi  \equiv  1+k^{4}+z^{2}-2(k^{2}+k^{2}z+z)\;,
\eeq
and the $X'$ terms are given by
\begin{eqnarray}
X'_{VA} & = & \phi\left(1-\frac{1}{3}\beta_\mu^2\right)\,(|A'|^2+|B'|^2)\, +\,
  4\,\hat{m}_{\mu}^2\,|B'|^2\,(2+2k^2-z) \nn \\
&& +\, 4\,\hat{m}_{\mu}^2\,z\,|C'|^2\,
  +\,8\,\hat{m}_{\mu}^2\,(1-k^2)\,{\rm Re}
  (B'C'^*)\; , \nn \\
X'_{SP} & = & \frac{z}{m_B^2}\,(|E'|^2+\beta_\mu^2\,|D'|^2) \; , \nn \\
X'_T & = & \frac{4}{3}\,\phi\,z\,m_B^2\,
\Big[3|F'|^2\,+\,2\,\beta_\mu^2\,(2|G'|^2-|F'|^2)\,\Big] \; , \nn \\
X'_{VA{\hbox{-}}SP} & = &
\frac{4\hat{m}_\mu}{m_B}\,(1-k^2)\,{\rm
    Re}(B'E'^*)+\frac{4\hat{m}_\mu}{m_B}\, z\, {\rm Re}(C'E'^*) \; ,
  \nn \\
X'_{VA{\hbox{-}}T} & = &
8\hat{m}_{\mu}\,m_B\phi \,{\rm Re}(A'F'^*) \; .
\end{eqnarray}
Here $\hat{m}_\mu \equiv m_\mu/m_B $ and
$\beta_\mu  \equiv \sqrt{1- 4\hat{m}_{\mu}^2/z}$.
The parameters $A'$--$G'$ are combinations of the Wilson coefficients,
form factors and NP parameters, and are given by
\begin{eqnarray}
A' & \equiv & 2(C^{eff}_9+R_V+R'_V)\,f_{+}(z)
+ 4C^{eff}_7\hat{m}_b \frac{f_{T}(z)}{1+k}\;,
\nn\\
B' & \equiv  & 2(C_{10}+R_A+R'_A)\, f_{+}(z)\;,
\nn\\
C' & \equiv  & 2(C_{10}+R_A+R'_A)\,\frac{1-k^2}{z}\Big[f_{0}(z)-f_{+}(z)\Big] \;,
\nn\\
D' & \equiv &2(R_S+R'_S)\frac{m_B(1-k^2)}{\hat{m}_b}f_0(z)  \;,
\nn\\
E' & \equiv& 2(R_P+R'_P)\frac{m_B(1-k^2)}{\hat{m}_b}f_0(z)  \;,
\nn\\
F' & \equiv & 4C_T\frac{f_T(z)}{m_B(1+k)}\;,
\nn \\
G' & \equiv  & - 4C_{TE}\frac{f_T(z)}{m_B(1+k)}\; .
\end{eqnarray}
The limits on the kinematical variables $z$ and  $\cos\theta_\mu$ are
\begin{equation}
-1\leq \cos\theta_\mu \leq 1 ~~,~~~~
4\hat{m}^2_{\mu}\leq z \leq(1-k)^2\; .
\label{kine-bounds}
\end{equation}
Note that in the large energy (LEET) limit,  there are relations
between form factors that are valid
up to $\alpha_s$, $1/E_K$ and $1/m_b$ corrections \cite{Charles, Beneke}.
These are
\bea
f_+(z) &=& \zeta (m_B,E_P),\nl
f_0(z)&=& \Big(1-\frac{q^2}{m^2_B -m^2_P}\Big)\zeta (m_B,E_P),\nl
f_T(z)&=& \Big(1+\frac{m_P}{m_B}\Big)\zeta (m_B,E_P)  \; .
\label{leet_rel}
\eea
Thus, all form factors can be expressed in terms of a single universal
soft form factor
$\zeta(m_B, E_P)$  in this limit.

The normalized forward-backward asymmetry for the muons in $\BKmumu$
is defined as
\beq
\AFB(q^2) = \frac {\displaystyle \int_{0}^{1} d\cos\theta_\mu
\frac{d^2B}{dq^2 d\cos\theta_\mu} - \int_{-1}^{0} d\cos\theta_\mu
\frac{d^2B}{dq^2 d\cos\theta_\mu} }{\displaystyle \int_{0}^{1}
d\cos\theta_\mu \frac{d^2B}{dq^2 d\cos\theta_\mu} + \int_{-1}^{0}
d\cos\theta_\mu \frac{d^2B}{dq^2 d\cos\theta_\mu} } ~,
\eeq
where $\theta_\mu$ is the angle between the three-momenta of the $\bdbar$
and the $\mu^+$ in the dimuon center-of-mass frame.
The calculation of $\AFB(q^2)$ gives
\begin{equation}
A_{FB}(q^2)=\frac{2B'_0 \, \beta_\mu \, \phi}{dB/dz}
\Bigg[Y'_{VA{\hbox{-}}SP} + Y'_{VA{\hbox{-}}T} + Y'_{SP{\hbox{-}}T} \Bigg]
\label{afb-Kmumu-main}
\end{equation}
where
\bea
Y'_{VA{\hbox{-}}SP} & = & -\frac{\hat{m}_\mu}{m_B}{\rm Re}(A'D'^*) \nn \\
Y'_{VA{\hbox{-}}T} & = & -4m_\mu(1-k^2) {\rm Re}(B'G'^*)
-4z m_\mu {\rm Re}(C'G'^*)\nn\\
Y'_{SP{\hbox{-}}T} & = & - \frac{z}{4}{\rm Re}(D'F'^*)-2z{\rm Re}(E'G'^*) \; .
\eea
Note that only $Y'_{SP{\hbox{-}}T}$ term is unsuppressed by the muon mass.

\section{\boldmath Details of the $\BKstarmumu$ angular analysis} 
\label{app-bkstarmumu}
\subsection{Matrix elements}

The full transition amplitude for
$\bar{B}(p_B)\rightarrow \bar{K}^*(p_{K^*},\epsilon^*) \mu^+(p_\mu^+)
\mu^-(p_\mu^-)$ is
\bea
& & i {\cal M}\,(\BKstarmumu) =  (-i)~\frac{1}{2}~\Bigg[-\frac{4~G_F}{\sqrt{2}}
\frac{\alpha_{em}}{4 \pi} (V_{ts}^* V_{tb})\Bigg] \times
\nonumber \\
& \Bigg\{ &
\langle K^*(p_{K^*},\epsilon)|\bar{s}\gamma_{\mu}b|B(p_B)\rangle \,
[(C_{9}^{\rm eff} + R_V + R'_V) L^\mu +
 (C_{10}+ R_A + R'_A) L^{\mu 5}]  \nn \\
& + &
\langle K^*(p_{K^*},\epsilon)|\bar{s}\gamma_{\mu}\gamma_5 b|B(p_B)\rangle
[-(C_{9}^{\rm eff} + R_V - R'_V) L^\mu -
 (C_{10}+ R_A - R'_A) L^{\mu 5}]  \nn \\
& + &
\langle K^*(p_{K^*},\epsilon) |\bar{s} i\sigma_{\mu\nu} q^{\nu} (1+\gamma_5)b|B(p_B)\rangle \;
[- 2 C^{\rm eff}_7 (m_b/q^2)L^\mu] \nn \\
& +  &
\langle K^*(p_{K^*},\epsilon) |\bar{s} b | B(p_B) \rangle\;
[(R_S + R'_S) L + (R_P + R'_P) L^5]  \nn \\
& +  &
\langle K^*(p_{K^*},\epsilon) |\bar{s} \gamma_5 b | B(p_B) \rangle\;
[(R_S - R'_S) L + (R_P - R'_P) L^5] \nn \\
& + & \langle K^*(p_{K^*},\epsilon)|\bar{s} \sigma_{\mu\nu}b | B(p_B) \rangle \;
[2 C_T L^{\mu\nu} + 2 i C_{TE} \epsilon^{\mu\nu\alpha\beta} L_{\alpha\beta}]
\; \Bigg\} \; , \label{bkllampNP}
\eea
where the $L$'s are defined in Eq.~(\ref{Ldefs}). Here $q= p_B-p_{K^*}
=p_\mu^++p_\mu^-$. This can be written in the form
\bea
i{\cal M}(\BKstarmumu) & = & (-i)\frac{1}{2}~\Bigg[\frac{4~G_F}{\sqrt{2}}
\frac{\alpha_{em}}{4 \pi} (V_{ts}^* V_{tb})\Bigg] \times
\nn \\
&& \hspace{-3.5cm}
[ M_{V\mu}  L^\mu+ M_{A\mu}  L^{\mu5}+M_S  L+M_P  L^5 +M_{T\mu \nu}  L^{\mu\nu}
+i M_{E\mu \nu}  L_{\alpha \beta} \epsilon^{\mu\nu\alpha\beta}] \; ,
\eea
with
\bea
\label{VecMesonME}
M_{V\mu}&=& -A^{\prime \prime} \epsilon_{\mu\nu\alpha\beta}\varepsilon^{*\nu}p_{K^*}^\alpha q^\beta +iB^{\prime \prime}\varepsilon^{*}_\mu + i C^{\prime \prime}\varepsilon^{*}.q (p_B+p_{K^*})_\mu +i D^{\prime \prime} \varepsilon^{*}.q q_\mu ,\nl
M_{A\mu}&=& -E^{\prime \prime} \epsilon_{\mu\nu\alpha\beta}\varepsilon^{*\nu}p_{K^*}^\alpha q^\beta +i F^{\prime \prime}\varepsilon^{*}_\mu + i G^{\prime \prime} \varepsilon^{*}.q (p_B+p_{K^*})_\mu + i H^{\prime \prime} \varepsilon^{*}.q q_\mu,\nl
M_S &=& i S^{\prime \prime} \varepsilon^{*}.q,\nl
M_P & =& i P^{\prime \prime} \varepsilon^{*}.q,\nl
M_{T\mu \nu} & =& C_T (i T^{\prime \prime}_1  \epsilon_{\mu\nu\alpha\beta}\varepsilon^{*\alpha}(p_B+p_{K^*})^\beta+i T^{\prime \prime}_2  \epsilon_{\mu\nu\alpha\beta}\varepsilon^{*\alpha}q^\beta-i T^{\prime \prime}_3  \epsilon_{\mu\nu\alpha\beta}\varepsilon^{*}.q p_{K^*}^\alpha q^\beta),\nl
M_{E\mu \nu}  & =& C_{TE}( i T^{\prime \prime}_1  \epsilon_{\mu\nu\alpha\beta}\varepsilon^{*\alpha}(p_B+p_{K^*})^\beta+i T^{\prime \prime}_2 \epsilon_{\mu\nu\alpha\beta}\varepsilon^{*\alpha}q^\beta-i T^{\prime \prime}_3 \epsilon_{\mu\nu\alpha\beta}\varepsilon^{*}.q p_{K^*}^\alpha q^\beta ) ~.
\eea
The quantities $A^{\prime \prime}$, $B^{\prime \prime}$, $C^{\prime
  \prime}$, $D^{\prime \prime}$,$ E^{\prime \prime}$, $F^{\prime
  \prime}$, $G^{\prime \prime}$, $S^{\prime \prime}$, $P^{\prime
  \prime}$, and $T^{\prime \prime}_i$ (1=1,2,3) are related to the
$\bar{B}\rightarrow \bar{K}^*$ form factors which are given below.
The contribution to the transition amplitudes
from the quantity $D^{\prime \prime}(q^2)$ vanishes and that from $H^{\prime
  \prime}(q^2)$ is suppressed because of the equation of motion of the
muons.
\subsection{Form factors}

The form factors for the decay amplitude for $\BKstarmumu$ [Eq.~(\ref{bkllampNP})] 
in terms of matrix elements of the quark
operators are given by 
\cite{ali-00}
\bea
\langle K^*(p_{K^*},\epsilon)|\bar{s}\gamma_\mu (1 \pm
\gamma_5)b|B(p_B) \rangle & ~=~ & \mp~iq_{\mu}
\frac{2m_{K^*}}{q^2} \, \epsilon^{*} \cdot q \, \bigg[
A_3(q^2)-A_0(q^2) \bigg] \nn \\
&& \hskip-4truecm \pm~i\epsilon_{\mu}^{*}(m_B+m_{K^*})
A_1(q^2) \mp~i(p_B+p_{K^*})_{\mu} \, \epsilon^{*} \cdot q
\, \frac{A_2(q^2)}{(m_B+m_{K^*})} \nn \\
&& \hskip-4truecm
-~\epsilon_{\mu\nu\lambda\sigma} \epsilon^{*\nu}
p^{\lambda}_{K^*} q^{\sigma} \frac{2V(q^2)}{(m_B+m_{K^*})} ~,
\label{me1}
\eea
where
\beq
A_3(q^2)~=~\frac{m_B+m_{K^*}}{2m_{K^*}}A_1(q^2)-\frac{m_B-m_{K^*}}{2m_{K^*}}A_2(q^2)\;.
\eeq
\bea
\langle K^*(p_{K^*},\epsilon)|\bar{s}\sigma_{\mu\nu}b|B(p_B)
\rangle & ~=~ & i\epsilon_{\mu\nu\lambda\sigma} \bigg\{ -
T_1(q^2)\epsilon^{*\lambda} (p_B+p_{K^*})^{\sigma}  \nn \\
&& \hskip-2truecm +~\frac{ (m_B^2-m_{K^*}^2) }{q^2}
\bigg(T_1(q^2)-T_2(q^2) \bigg)\epsilon^{*\lambda} q^{\sigma}
 \nn \\
&& \hskip-4truecm -~\frac{2}{q^2} \bigg(T_1(q^2)-T_2(q^2)
-\frac{q^2}{(m_B^2-m_{K^*}^2)}~T_3(q^2) \bigg) \epsilon^{*}
\cdot q \, p^{\lambda}_{K^*} q^{\sigma} \bigg\}\; .
\label{me2}
\eea
\bea
\langle K^*(p_{K^*},\epsilon)|\bar{s}i\sigma_{\mu\nu}q^\nu (1
\pm \gamma_5)b|B(p_B) \rangle & ~=~ & 2
\epsilon_{\mu\nu\lambda\sigma} \epsilon^{*\nu}
p^\lambda_{K^*} q^\sigma ~ T_1 (q^2) \nn \\
&& \hskip-2truecm \pm~i \bigg\{
\epsilon^*_{\mu}(m_B^2-m_{K^*}^2)-(p_B+p_{K^*})_\mu
\, \epsilon^{*} \cdot q
\,  \bigg\}~T_2(q^2) \nn \\
&& \hskip-2truecm \pm~i \, \epsilon^{*} \cdot q
\, \bigg\{ q_\mu -
\frac{(p_B + p_{K^*})_\mu q^2}{(m_B^2-m_{K^*}^2)} \bigg\} ~
T_3(q^2)\; .
\label{eq:4}
\eea
\bea
\langle K^*(p_{K^*},\epsilon)|\bar{s}(1\pm\gamma_5)b|B(p_B)
\rangle & ~=~ & \mp~2i \frac{m_{K^*}}{m_b} \,
\epsilon^{*} \cdot q \,A_0(q^2) \;.
\label{eq:5}
\eea
Here we have neglected the strange-quark mass. The matrix
elements are functions of 7 unknown form factors:
$A_{0,1,2}(q^2)$, $V(q^2)$, $T_{1,2,3}(q^2)$.

The matrix elements $M_{V,A,S,P,T,E}$ appearing in Eq.~(\ref{VecMesonME})
can be written in terms of these 7 form factors, coupling constants
and kinematic variables as
\bea
\label{vecMeFF}
A^{\prime\prime} &=& \Big[\frac{2 V(q^2)(C^{eff}_9+R_V+R^\prime_V)}{m_B+m_{K^*}}+\frac{4 m_b}{q^2}C^{eff}_7 T_1(q^2)\Big],\nl
B^{\prime\prime} &=& -\Big[(m_B+m_{K^*})A_1(q^2)(C^{eff}_9+R_V-R^\prime_V)
+\frac{2 m_b}{q^2}C^{eff}_7 T_2(q^2)(m^2_B-m^2_{K^{*}}) \Big],\nl
C^{\prime\prime} &=& \Big[\frac{A_2(q^2)}{m_B+m_{K^*}}(C^{eff}_9+R_V-R^\prime_V)+\frac{2 m_b}{q^2}C^{eff}_7 \Big(T_2(q^2)+\frac{q^2 T_3(q^2)}{(m^2_B-m^2_{K^{*}})}\Big)\Big],\nl
D^{\prime\prime} &=& \Big[\frac{2 m_{K^*}}{q^2}(C^{eff}_9+R_V-R^\prime_V)(A_3(q^2)-A_0(q^2))-\frac{2 m_b}{q^2}C^{eff}_7 T_3(q^2) \Big],\nl
E^{\prime\prime} &=& \Big[\frac{2 V(q^2)(C_{10}+R_A+R^\prime_A)}{m_B+m_{K^*}}\Big],\nl
F^{\prime\prime} &=& -\Big[(m_B+m_{K^*})A_1(q^2)(C_{10}+R_A-R^\prime_A)\Big],\nl
G^{\prime\prime} &=& \Big[\frac{A_2(q^2)}{m_B+m_{K^*}}(C_{10}+R_A-R^\prime_A)\Big],\nl
H^{\prime\prime} &=& \Big[\frac{2 m_{K^*}}{q^2}(C_{10}+R_A-R^\prime_A)(A_3(q^2)-A_0(q^2))\Big],\nl
S^{\prime\prime} &=& \Big[-2 (R_S-R^\prime_S)\frac{m_{K^*}}{m_b}A_0(q^2)\Big],\nl
P^{\prime\prime} &=& \Big[-2 (R_P-R^\prime_P)\frac{m_{K^*}}{m_b}A_0(q^2)\Big],\nl
T^{\prime\prime}_1 &=& -2 T_1(q^2) ,\nl
T^{\prime\prime}_2  &=& \Big[\frac{2(m^2_B-m^2_{K^{*}})}{q^2}( T_1(q^2)- T_2(q^2))\Big],\nl
T^{\prime\prime}_3 &=& \Big[\frac{4 }{q^2}\Big(T_1(q^2)- T_2(q^2)-\frac{q^2  T_3(q^2)}{m^2_B-m^2_{K^{*}}}\Big)\Big].
\eea
Also, we define
\bea
\label{FFTs}
T_0 &=& \frac{1}{m_{K^*}}\Big(\sqrt{q^2}(E_{K^*} \sqrt{q^2}+2 m^2_{K^*})T^{\prime \prime} _1 + q^2 (E_{K^*} T^{\prime \prime} _2 -|\vec{p}_{K^{*}}|^2  \sqrt{q^2}  T^{\prime \prime} _3 )\Big)~,\nl
T_+ &=&  (q^2+2 E_{K^*} \sqrt{q^2}) T^{\prime \prime} _1 +q^2  T^{\prime \prime} _2~,\quad
T_- = 2 |\vec{p}_{K^{*}}| \sqrt{q^2} T^{\prime \prime} _1~.
\eea

\subsection{Transversity amplitudes}

We summarize the various transversity amplitudes that appear in the
$\BKstarmumu$ angular distribution. The decay amplitude of
$\BKstarmumu$ depends on the $K^*$ polarization vector $\varepsilon
(\lambda)$ with helicity $\lambda~(0,\pm 1)$. Hence, the decay
amplitude can be decomposed into three components. Below we define the
helicity amplitudes of various operators with different Lorentz
structures (V, A, S, P, T, TE) in Eq.~(\ref{bkllampNP}).
\bea
\label{Ksthelamp}
A^0_{V} &=& \sqrt{q^2}\Big(\frac{E_{K^*}}{m_{K^*}} B^{\prime \prime} +\frac{2 |\vec{p}_{K^{*}}|^2 \sqrt{q^2}}{m_{K^*}} C^{\prime \prime} \Big) ~,\quad 
A^{\pm}_{V} = \sqrt{q^2} (\pm |\vec{p}_{K^{*}}|\sqrt{q^2} A^{\prime \prime} + B^{\prime \prime} ) ~,\nl
A^0_{A}&=& \sqrt{q^2}\Big(\frac{E_{K^*}}{m_{K^*}} F^{\prime \prime}  +\frac{2 |\vec{p}_{K^{*}}|^2 \sqrt{q^2}}{m_{K^*}} G^{\prime \prime} \Big) ~,\quad 
A^{\pm}_{A} = \sqrt{q^2} (\pm |\vec{p}_{K^{*}}|\sqrt{q^2} E^{\prime \prime} + F^{\prime \prime} ) ~,\nl
A_{S} &=&  \frac{2 |\vec{p}_{K^{*}}|q^2 }{m_{K^*}} S^{\prime \prime} ~,\hspace{4.2cm}
A_{P} = \frac{2 |\vec{p}_{K^{*}}|q^2 }{m_{K^*}} P^{\prime \prime} ~,\nl
A^0_{T} &=& T_0 C_T ~,\hspace{5.25cm}
A^{\pm}_{T} = T_{\pm} C_T ~,\nl
A^0_{TE} &=&  2 T_0  C_{TE} ~,\hspace{4.6cm}
A^{\pm}_{TE} = 2 T_{\pm}  C_{TE}  ~,\nl
A_{vt} &=& -2  |\vec{p}_{K^{*}}|  \sqrt{q^2} (C_{10} + R_{A} - R_{A^\prime}) A_0~,
\eea
where the amplitude $A_{vt}$ is related to the time-like component of the virtual $K^*$. In the transversity basis, the positive and negative helicity amplitudes are replaced by the  transversity amplitudes as
\bea
\label{Kstrasamp1}
A^i_{\parallel}=\frac{1}{\sqrt{2}}(A^{+}_{i}  + A^{-}_{i})~,\quad A^i_{\perp}=\frac{1}{\sqrt{2}}(A^{+}_{i}  - A^{-}_{i})~,\quad \quad  \mathrm{i = V,A,T,TE}.
\eea
The left and right component of the transversity amplitudes of vector and axial-vector currents in \cite{Altmannshofer:2008dz} can be  written as
\bea
\label{Kstrasamp2}
A^{L,R}_{0,VA} &=& A^0_{V} \mp A^0_{A}~, A^{L,R}_{\parallel,VA} =( A^V_{\parallel}\mp A^A_{\parallel})~,A^{L,R}_{\perp,VA} = ( A^V_{\perp}\mp A^A_{\perp})~.
\eea

Note that in the notation of Ref.~\cite{Altmannshofer:2008dz}, we have
the correspondence $A^{L,R}_{(0,\parallel,\perp), VA} =
({\sqrt{q^2}}/{N}) A^{L,R }_{(0,\parallel,\perp)} $, $A_S = ({\sqrt{q^2}}/{N}) A_{S}$, and $(A_{vt} + \frac{\sqrt{q^2}}{4 m_\mu} A_P) = -{\sqrt{q^2}}/{(2 N)} A_{t}$. 

\subsection{Angular coefficients}

The expressions for the twelve angular coefficients ($I$'s) in the
$\BKstarmumu$ angular distribution are summarized here according to
$K^*$ helicity combinations $\lambda_1 \lambda_2$.
The longitudinal $I^0$'s ($\lambda_1 \lambda_2 = 0 0$) are given by
\bea
\label{eq:I0}
I^0_1 &=& 2  \Big[\frac{1}{2} (|A^{L}_{0,VA}|^2+|A^{R}_{0,VA}|^2) + \frac{1}{2} \beta^2_\mu |A_S|^2 +  \frac{4 m^2_\mu}{q^2} \Big({\rm Re}[A^{L}_{0,VA}A^{R*}_{0,VA}] \nl &&  + 2 |A_{vt} + \frac{\sqrt{q^2}}{4 m_\mu} A_P |^2  + 8 |A^0_{TE}|^2\Big) +  4 \beta^2_\mu  (|A^0_T|^2 + |A^0_{TE}|^2)\nl && - \frac{8 m_\mu}{\sqrt{q^2}} {\rm Re}[(A^{L}_{0,VA}+A^{R}_{0,VA})A^{0*}_{TE}]\Big]~,\nl
I^0_2 &=& \beta^2_\mu \Big[-(|A^{L}_{0,VA}|^2+|A^{R}_{0,VA}|^2)+ 8   (|A^0_T|^2 + |A^0_{TE}|^2)\Big]~,\nl
I^0_3 &=& 2 \beta_\mu  {\rm Re}\Big[-4 A^{0}_{TE} A^*_S + \frac{4 m_\mu}{\sqrt{q^2}} \Big(\frac{1}{2}(A^{L}_{0,VA}+A^{R}_{0,VA})A^*_S  \nl && + 4 (A_{vt} + \frac{\sqrt{q^2}}{4 m_\mu} A_P) A^{0*}_T \Big)\Big]
\eea
The transverse $I^T$'s ($\lambda_1 \lambda_2 =++,--,+-,-+$) are given by
\bea
 \label{eq:IT}
I^T_1 &=& \Big[\frac{2+\beta^2_\mu }{2}\Big(|A^V_\parallel|^2+|A^V_\perp|^2 + |A^A_\parallel|^2+|A^A_\perp|^2\Big)-4(-2+\beta^2_\mu) \Big(|A^T_\parallel|^2+|A^T_\perp|^2 + |A^{TE}_\parallel|^2 \nl &&+|A^{TE}_\perp|^2\Big) + \frac{4 m^2_\mu}{q^2} \Big(|A^V_\parallel|^2+|A^V_\perp|^2- |A^A_\parallel|^2-|A^A_\perp|^2-16{\rm Re}[ (A^T_\parallel A^{T*}_\perp -A^{TE}_\parallel A^{TE *}_\perp)]\Big) \nl && - 16 \frac{ m_\mu}{\sqrt{q^2}} \Big({\rm Re}[A^V_\perp(A^{T*}_\parallel- A^{T*}_\perp)] + A^V_\parallel (A^{TE*}_\parallel + A^{TE*}_\perp)]\Big) \Big]~,\nl
I^T_2 &=& \beta^2_\mu\Big[\frac{1 }{2}\Big(|A^V_\parallel|^2+|A^V_\perp|^2 + |A^A_\parallel|^2+|A^A_\perp|^2\Big)-4\Big(|A^T_\parallel|^2+|A^T_\perp|^2 + |A^{TE}_\parallel|^2 +|A^{TE}_\perp|^2\Big) \Big]~,\nl
I^T_3 &=& -4 \beta_\mu \Big[{\rm Re}[A^V_\perp A^{A*}_\parallel + A^V_\parallel A^{A*}_\perp ] - 4 \frac{ m_\mu}{\sqrt{q^2}}  {\rm Re}[A^A_\parallel(A^{T*}_\parallel- A^{T*}_\perp)+ A^A_\perp (A^{TE*}_\parallel+ A^{TE*}_\perp)]\Big]~,\nl
I^T_4 &=& \beta^2_\mu \Big[ \Big(|A^V_\perp|^2-|A^V_\parallel|^2 + |A^A_\perp|^2-|A^A_\parallel|^2\Big) + 16 {\rm Re}[A^{T}_\parallel A^{T*}_\perp + A^{TE}_\parallel A^{TE*}_\perp ]\Big]~,\nl
I^T_5 &=&  2 \beta^2_\mu {\rm Im}[A^{V*}_\parallel A^{V}_\perp + A^{A*}_\parallel A^{A}_\perp]~.
\eea
The mixed $I^{LT}$'s ($\lambda_1 \lambda_2 = 0\pm,\pm0$) are given by
\bea
\label{eq:ILT}
I^{LT}_1 &=& \beta^2_\mu {\rm Re}\Big[\frac{1}{\sqrt{2}}\Big(A^{R}_{0,VA}(A^{V*}_\parallel + A^{A*}_\parallel) + A^{L}_{0,VA}(A^{V*}_\parallel - A^{A*}_\parallel)\Big) \nl && - 4 \sqrt{2}\Big( A^0_T(A^{T*}_\parallel + A^{T*}_\perp) + A^0_{TE}(A^{TE*}_\parallel+ A^{TE*}_\perp)\Big)\Big]~,\nl
I^{LT}_2 &=& \frac{1}{\sqrt{2}} \beta^2_\mu  {\rm Im}[A^{R}_{0,VA}(A^{V*}_\perp + A^{A*}_\perp) + A^{L}_{0,VA}(A^{V*}_\perp - A^{A*}_\perp)]~,\nl
I^{LT}_3 &=& \sqrt{2}  \beta_\mu {\rm Re} \Big[  A^{L}_{0,VA}(A^{V*}_\perp - A^{A*}_\perp)-A^{R}_{0,VA}(A^{V*}_\perp + A^{A*}_\perp)+ 2 (A^{TE}_\parallel+ A^{TE}_\perp)A^*_S\nl &&  -2 \frac{ m_\mu}{\sqrt{q^2}} \Big(A^{V}_\parallel A^*_S + 4(A^{T}_\parallel + A^{T}_\perp)(A^*_{vt} + \frac{\sqrt{q^2}}{4 m_\mu} A^*_P)\nl && + 2(A^{T}_\parallel - A^{T}_\perp)(A^{L*}_{0,VA} - A^{R*}_{0,VA}) -4  A^{A}_\perp A^{0*}_{TE} \Big) \Big]~,\nl
I^{LT}_4 &=& \sqrt{2}  \beta_\mu  {\rm Im}\Big[ A^{L}_{0,VA}(A^{V*}_\parallel - A^{A*}_\parallel)-A^{R}_{0,VA}(A^{V*}_\parallel + A^{A*}_\parallel)  + 2 (A^{T}_\parallel - A^{T}_\perp)A^*_S \nl &&  + 2 \frac{ m_\mu}{\sqrt{q^2}}\Big( A^{V}_\perp A^*_S + 4 (A^{TE}_\parallel - A^{TE}_\perp)(A^*_{vt} + \frac{\sqrt{q^2}}{4 m_\mu} A^*_P) \nl && + 2 (A^{TE}_\parallel + A^{TE}_\perp)(A^{L*}_{0,VA} - A^{R*}_{0,VA})- 4  A^{A}_\parallel  A^{0*}_{TE}\Big) \Big]~.\nl
\eea


\end{document}